\documentclass[12pt]{article}

\usepackage{mathptmx} %for springer PS fonts 
\usepackage{bm}

\usepackage{amsfonts} %pour utiliser \mathbb 
\usepackage{amsmath} %pour utiliser l'environnement equation, ou \text
\usepackage{enumerate} %permet de choisir le style du compteur (a), b) c),...)
\usepackage{dsfont} %permet de faire de joli indicatrice d'intervalle \mathds{1}

\usepackage{cite} %pour trier les citations

\usepackage{graphicx}
\usepackage{subfigure} %pour faire un tableau de dessins dans une seule figure
\usepackage{psfrag} %permet de transformer des caracteres dans un .eps en texte LateX
\usepackage[usenames]{color}

\title{All speed scheme for the low mach number limit of
the Isentropic Euler equation}

\author{Pierre Degond$^{1,2}$, Min Tang$^{1,2}$}

\date{}

\begin{document}

\maketitle

\vspace{0.5cm}

\begin{center}
1-Universit\'e de Toulouse; UPS, INSA, UT1, UTM ;\\ 
Institut de Math\'ematiques de Toulouse ; \\
F-31062 Toulouse, France. \\
2-CNRS; Institut de Math\'ematiques de Toulouse UMR 5219 ;\\ 
F-31062 Toulouse, France.\\
email: pierre.degond@math.univ-toulouse.fr, tangmin1002@gmail.com
\end{center}

%%%%% Begin Abstract %%%%%%%%%%%
\begin{abstract}
An all speed scheme for the Isentropic Euler equation
 is presented in this paper. When the Mach number tends to zero, the
 compressible Euler equation converges to its incompressible counterpart, in which the
 density becomes a constant. Increasing approximation errors and
 severe stability constraints are the main difficulty in the low
 Mach regime. The key idea of our all speed scheme is the special semi-implicit time
   discretization, in which the low Mach number
   stiff term is divided into two parts, one being treated explicitly and the
   other one implicitly. Moreover, the flux of the density equation is also treated
   implicitly and an elliptic type equation is derived to obtain the density.
   In this way, the correct limit can be captured
   without requesting the mesh size and time step to be smaller than the Mach
   number. Compared with previous semi-implicit methods \cite{Jeff,HW,HWconserve},
    nonphysical oscillations can be suppressed. We develop this semi-implicit time
    discretization in the framework of
   a first order local Lax-Friedrich (LLF) scheme and numerical tests are
   displayed to demonstrate its performances.
\end{abstract}
%%%%% end %%%%%%%%%%%

%%%%% AMS/PACs/Keywords %%%%%%%%%%%
%\pac{}

\medskip


%%%% maketitle %%%%%
\maketitle

%%%% Start %%%%%%
\section{Introduction}
\label{sec1} Singular limit problems in fluid mechanics have drawn
great attentions in the past years, like low-Mach number flows,
magneto-hydrodynamics at small Mach and Alfven numbers and
multiple-scale atmospheric flows. As mentioned in \cite{Klein}, the
singular limit regime induces severe stiffness and stability
problems for standard computational techniques. In this paper we
focus on the simplest Isentropic Euler equation and propose a
numerical scheme that is uniformly applicable and efficient for all
ranges of Mach numbers.

The problem under study is the Isentropic Euler equation
\begin{equation}
\left\{\begin{array}{l}\partial_t\rho_\epsilon+\nabla\cdot
(\rho_\epsilon\mathbf{u}_\epsilon)=0,\\
\partial_t(\rho_\epsilon\mathbf{u}_\epsilon)+\nabla\Big(\rho_\epsilon\mathbf{u}_\epsilon\otimes
\mathbf{u}_\epsilon\Big)+\frac{1}{\epsilon^2}\nabla p_\epsilon=0.
\end{array}\right.
\label{hype}
\end{equation} where $\rho_\epsilon,\rho_\epsilon\mathbf{u}_\epsilon$
is the density and momentum of the fluid respectively and $\epsilon$
is the scaled Mach number. This is one of the most studied nonlinear
hyperbolic systems. For standard applications, the equation of state
takes the form
\begin{equation}\label{P}p(\rho)=\Lambda\rho^\gamma,\end{equation}
where $\Lambda,\gamma$ are constants depending on the physical
problem.

It is rigorously proved by Klainerman and Majda \cite{KM1, KM2} that
when $\epsilon\to 0$, i.e. when the fluid velocity is small compared
with the speed of sound \cite{Constantin}, the solution of
(\ref{hype}) converges to its incompressible counterpart. Formally,
this can be obtained by inserting the expansion
\begin{equation}\begin{array}{c}\rho_\epsilon=\rho_0+\epsilon^2\rho_{(2)}+\cdots,\\
\mathbf{u}_\epsilon=\mathbf{u}_0+\epsilon^2\mathbf{u}_{(2)}+\cdots,\end{array}
\label{assumption}\end{equation} into (\ref{hype}) and equate the
same order of $\epsilon$. The limit reads as follows
\cite{KM1,KBSM}:
\begin{subequations}\label{lim}
\begin{eqnarray}
&&\rho=\rho_0,\\
&&\nabla\cdot\mathbf{u}_0=0,\\
&&\partial_t\mathbf{u}_0+\nabla(\mathbf{u}_0\otimes\mathbf{u}_0)+\nabla
p_{0}=0.
\end{eqnarray}
\end{subequations}
Here $p_0$ is a scalar pressure that can be viewed as the Lagrange
multiplier which enforces the incompressibility constraint.
Physically, this limit means that in slow flows (compared with speed
of sound), the factor $1/\epsilon^2$ in the momentum equation in
front of the pressure gradient generates fast pressure waves, which
makes the pressure and therefore, the density, uniform in the
domain\cite{PM, MRKG}.

For atmosphere-ocean computing or fluid flows in engineering
devices, when $\epsilon$ is small in (\ref{hype}), standard
numerical methods become unacceptably expensive. Indeed,
(\ref{hype}) has wave speeds of the form
\[\lambda=\mathbf{u}_\epsilon\pm\frac{1}{\epsilon}\sqrt{p'(\rho_\epsilon)},\]
where $p'(\rho_\epsilon)$ is the derivative with respect to
$\rho_\epsilon$. If a standard hyperbolic solver is used, the CFL
requirement is $\Delta t=O(\epsilon\Delta x)$. Moreover in order to
maintain stability, the numerical dissipation required by the
hyperbolic solver is proportional to $|\lambda|$. If
$|\lambda|=O(\frac{1}{\epsilon})$, in order to control the
diffusion, we need to have $\Delta x=o(\epsilon^r)$, where $r$ is
some appropriate constant. Thus the stability and accuracy highly
depend on $\epsilon$.

Our aim is to design a method whose stability and accuracy is
independent of $\epsilon$. The idea is to find an asymptotic
preserving (AP) method, i.e. a method which gives a consistent
discretization of the isentropic Euler equations (\ref{hype}) when
$\Delta x,\Delta t$ resolve $\epsilon$, and a consistent
discretization of the incompressible limit (\ref{lim}) when
$\epsilon\to 0$ ($\Delta x,\Delta t$ being fixed). The efficiency of
AP schemes at the low Mach number regime can be proved similarly as
in \cite{GJL}. The key idea of our all speed scheme is a specific
semi-implicit time discretization, in which the low Mach number
stiff term is divided into two parts, one part being treated
explicitly and the other one implicitly. Moreover, the flux of the
density equation is also treated implicitly. For the space
discretization, when $\epsilon$ is $O(1)$, even if the initial
condition is smooth, shocks will form due to the nonlinearity of the
$\mbox{div}\big(\rho_\epsilon\mathbf{u}_\epsilon\otimes\mathbf{u}_\epsilon\big)$
term and shock capturing methods should be employed here.

In the literature, lots of efforts have been made to find numerical
schemes for the compressible equation that can also capture the zero
Mach number limit \cite{Bijl,DJL,PM, MRKG,Jeff}. In \cite{Bijl},
Bijl and Wesseling split the pressure into thermodynamic and
hydrodynamic pressure terms and solve them separately. Similar to
this approach, the multiple pressure variable (MPV) method was
proposed by Munz et al. in \cite{PM, MRKG}. There is also some
recent work by J. Hauck, J-G. Liu and S. Jin \cite{Jeff}. Their
approach involves specific splitting of the pressure term. We avoid
using this splitting, the proper design of which seems very crucial
in some cases.

Some similar ideas can be found in the ICE method, which is designed
to adapt incompressible flow computation techniques using staggered
meshes to the simulation compressible flows. The method was first
introduced by Harlow and Amsdan in 1965 and 1971 \cite{HW,HA} and is
called Implicit Continuous-fluid Eulerian (ICE) technique. It is
used to simulate single phase fluid dynamic problems with all flow
speeds. They introduce two parameters in the continuity equation and
the momentum equation to combine information from both previous and
forward time steps. However this method is not conservative, which
leads to discrepancies in the shock speeds. Additionally it suffers
from small wiggles when there are moving contact discontinuities.
The first problem was solved by an iterative method, for example
SIMPLE \cite{P}, or PISO \cite{IGW}. In some recent work, Heul and
Wesseling also find a conservative pressure-correction method
\cite{HWconserve}. All these methods are based on the so called MAC
staggered mesh in order to be consistent with the staggered grid
difference method for the incompressible Euler equations \cite{HW}.
Specifically, if we write the simplified ICE technique presented in
\cite{Bonner} in conservative form, we are led to the semi-discrete
framework:
\begin{eqnarray}
&&\left\{\begin{array}{l}\frac{\rho^{\ast}_\epsilon-\rho^{n}_\epsilon}{\Delta
t}+\nabla\cdot
(\rho_\epsilon\mathbf{u}_\epsilon)^{n}=0,\\
\frac{(\rho_\epsilon\mathbf{u}_\epsilon)^{\ast}-(\rho_\epsilon\mathbf{u}_\epsilon)^n}{\Delta
t}+\nabla (\rho^n_\epsilon\mathbf{u}^n_\epsilon\otimes
\mathbf{u}^n_\epsilon)=0,
\end{array}\right.
\label{ICE1}\\
&&\left\{\begin{array}{l}\frac{\rho^{n+1}_\epsilon-\rho^{\ast}_\epsilon}{\Delta
t}+\nabla\cdot\big(
(\rho_\epsilon\mathbf{u}_\epsilon)^{n+1}-(\rho_\epsilon\mathbf{u}_\epsilon)^{\ast}\big)=0,\\
\frac{(\rho_\epsilon\mathbf{u}_\epsilon)^{n+1}-(\rho_\epsilon\mathbf{u}_\epsilon)^\ast}{\Delta
t}+\frac{1}{\epsilon^2}\nabla p(\rho^{n+1}_\epsilon)=0.
\end{array}\right.\label{ICE2}
\end{eqnarray}
By substituting the gradient of the second equation of (\ref{ICE2})
into its first equation and using the results of the first equation
(\ref{ICE1}), $\rho_\epsilon$ can be updated by solving an elliptic
equation which does not degenerate when $\epsilon\to 0$.

We use a similar idea in our method. However, we do not use the
predictor-corrector procedure but we rather discretize the problem
in a single step. We use standard shock capturing schemes which
allows to guarantee the conservativity and the desired artificial
viscosity. We only use implicit evaluations of the mass flux and
pressure gradient terms to ensure stability and provide an extremely
simple way to deal with the implicitness. Additionally, we propose a
modification of the implicit treatment of the pressure equation.
Indeed, using a similar idea as in \cite{Jeff}, we split the
pressure into two parts and put $\alpha p(\rho_\epsilon)$ into the
hyperbolic system. This makes the first system no longer be weakly
hyperbolic and much more stable. The numerical results show the
advantage of our method in the following sense:
\begin{itemize}
\item The method is in conservative form and can capture the right
shock speeds.
\item The non-physical oscillations \cite{HH} can be suppressed by choosing the proper value of the
parameter which determines the fraction of implicitness used in the
evaluation of the pressure gradient term. The choice of this
parameter depends on the time and space step and on the specific
problem.
\end{itemize}

In this paper we only use the first order LLF scheme. Higher order
space and time discretizations will be subject of future work. The
main objective of this work is to show that the semi-discrete time
discretization provides a framework for developing AP methods for
singular limit problems. Similar ideas can be extended to the full
Euler equation and more complicated fluid model and have also been
used in other contexts such as quasineutrality limits
\cite{DEV07,DLV08} and magnetized fluids under stong magnetic fields
\cite{DDSV09}.

The organization of this paper is as follows. Section 2 exposes the
semi-implicit scheme and its capability to capture the
incompressible limit is proved. The detailed one dimensional and two
dimensional fully discretized schemes and their AP property are
presented in section 3 and 4 respectively. In section 5, how to
choose the ad-hoc parameter is discussed and finally, some numerical
tests are given in section 6 to discuss the stability and accuracy
of our scheme. The efficiency at both the compressible and low mach
number regime are displayed. Finally, we conclude in section 6 with
some discussion.

\section{Time Semi-discrete scheme}
Let $\Delta t$ be the time step, $t^n=n\Delta t, n=0,1,\cdots$ and
let the '$n$' superscript denote the approximations at $t^n$. The
semi-discrete scheme for the $n$th time step is
\begin{eqnarray}
&&\frac{\rho^{n+1}_\epsilon-\rho^n_\epsilon}{\Delta t}+\nabla\cdot
(\rho_\epsilon\mathbf{u}_\epsilon)^{n+1}=0,\label{semidis1}\\
&&\frac{(\rho_\epsilon\mathbf{u}_\epsilon)^{n+1}-(\rho_\epsilon\mathbf{u}_\epsilon)^n}{\Delta
t}+\mbox{div} \big(\rho^n_\epsilon\mathbf{u}^n_\epsilon\otimes
\mathbf{u}^n_\epsilon+\alpha
p(\rho_\epsilon^n)\big)+\frac{1-\alpha\epsilon^2}{\epsilon^2}\nabla
p(\rho^{n+1}_\epsilon){}\nonumber\\
&&\qquad\qquad\qquad\qquad\qquad\qquad\qquad\qquad\qquad\qquad\qquad\qquad\qquad=0,
\label{semidis2}
\end{eqnarray}
where $\alpha$ is an ad-hoc parameter which satisfies $\alpha\leq
1/\epsilon^2$. The choice of $\alpha$ depends on the space and time
steps and on the fluid speed. When the shock is strong, $\alpha$
should be bigger, which means that the system should be more
explicit to follow the discontinuity more closely. We discuss the
choice of $\alpha$ for specific equations of state in this paper and
test its effect numerically. It depends on the required accuracy,
the small parameter $\epsilon$ and the shock amplitude in a
sometimes quite complex way.

Rewriting the momentum equation (\ref{semidis2}) as
\[(\rho_\epsilon\mathbf{u}_\epsilon)^{n+1}=(\rho_\epsilon\mathbf{u}_\epsilon)^n
-\Delta t\nabla\Big(\rho^n_\epsilon\mathbf{u}^n_\epsilon\otimes
\mathbf{u}^n_\epsilon+\alpha p(\rho_\epsilon^n)\Big)-\Delta
t\frac{1-\alpha\epsilon^2}{\epsilon^2}\nabla
P(\rho^{n+1}_\epsilon)\] and substituting it into the density
equation, one gets
\begin{equation}\rho_\epsilon^{n+1}-\Delta
t^2\frac{1-\alpha\epsilon^2}{\epsilon^2}\Delta
P(\rho_\epsilon^{n+1})=\phi(\rho_\epsilon^n,\mathbf{u}_\epsilon^n)
\label{elliptic}\end{equation} which is an elliptic equation that
can be solved relatively easily. Here
\begin{equation}\phi(\rho_\epsilon^n,\mathbf{u}_\epsilon^n)=\rho^n_\epsilon-\Delta
t\nabla\cdot(\rho^n_\epsilon\mathbf{u}_\epsilon^n)+\Delta
t^2\nabla\cdot\nabla\big(\rho_\epsilon^n\mathbf{u}_\epsilon^n\otimes\mathbf{u}_\epsilon^n+\alpha
p(\rho_\epsilon^n)\big).\label{phi}\end{equation}

The Laplace operator in (\ref{elliptic}) can be approximated by
$\nabla\big(P'(\rho_\epsilon^n)\nabla \rho_\epsilon^{n+1}\big)$ and
(\ref{elliptic}) becomes
\begin{equation}
\rho_\epsilon^{n+1}-\Delta
t^2\frac{1-\alpha\epsilon^2}{\epsilon^2}\nabla\cdot\big(P'(\rho_\epsilon^n)\nabla
\rho_\epsilon^{n+1}\big)=\phi(\rho_\epsilon^n,\mathbf{u}_\epsilon^n),\label{ellipticapro}
\end{equation}
Though shocks will form for the original system
(\ref{semidis1})(\ref{semidis2}), we always add some numerical
diffusion terms so that
$\phi(\rho_\epsilon^n,\mathbf{u}_\epsilon^n)$ is smooth. Then so is
$\rho_\epsilon^{n+1}$. When we implement this method,
$\rho_\epsilon^{n+1}$ can be obtained from (\ref{elliptic}) first
and $\mathbf{u}_\epsilon^n$ is then updated by the momentum equation
(\ref{semidis2}) afterwards. Therefore, apart from the resolution of
the elliptic equation (\ref{ellipticapro}), the scheme only involves
explicit steps.

We now show that the scheme (\ref{semidis1})(\ref{semidis2}) is
asymptotic preserving. We introduce the formal expansion
\begin{equation}\label{nassum}\begin{array}{c}\rho^n_\epsilon(x)=\rho^n_{0c}+
\epsilon\rho^n_{(1)}(x)+\epsilon^2\rho^n_{(2)}(x)+\cdots,\\
\mathbf{u}^n_\epsilon=\mathbf{u}^n_{0}(x)+\epsilon\mathbf{u}^n_{(1)}(x)+\cdots.
\end{array}\end{equation}
In the sequel, the '$c$' in the index means that the quantity is
independent of space. When $\Delta x$, $\Delta t$ are fixed and
$\epsilon$ goes to $0$ in (\ref{elliptic}), we formally have $\Delta
P(\rho^{n+1}_0)=0$, which implies that $\rho^{n+1}_{0}$ is
independent of space, where $\rho^{n+1}_0$ is the limit of
$\rho^{n+1}_\epsilon$ when $\epsilon\to 0$. Thus we have
\begin{equation}\frac{\rho^{n+1}_{0c}-\rho^n_{0c}}{\Delta
t}+\nabla\cdot(\rho_{0c}\mathbf{u}_0)^{n+1}=0\label{order1}\end{equation}
by equating the $O(1)$ terms in the density equation
(\ref{semidis1}). Integrating (\ref{order1}) over the computational
domain, one gets
\begin{equation}|\Omega|\frac{\rho^{n+1}_{0c}-\rho^n_{0c}}{\Delta
t}=-\rho_{0c}^{n+1}\int_{\Omega}\nabla\cdot(\mathbf{u}_0)^{n+1}=-\rho_{0c}^{n+1}\int_{\partial\Omega}
\mathbf{n}\cdot\mathbf{u}_0^{n+1}.\label{order1inter}\end{equation}
As discussed in \cite{Jeff}, for wall boundary condition, periodic
boundary condition and open boundary condition, (\ref{order1inter})
gives
\begin{equation}\rho^{n+1}_{0c}=\rho^{n}_{0c},\label{rholimdis}\end{equation}that is $\rho_0$ is also independent
of time. Thus (\ref{order1}) also implies
\begin{equation}\nabla\cdot\mathbf{u}^{n+1}_0=0\label{plimdis}.\end{equation}
%From (\ref{P}),
%\begin{equation}P^{n+1}_\epsilon=P(\rho^{n+1}_\epsilon)=P^{n+1}_0+\epsilon^2P^{n+1}_{(2)}+\cdots
%=P(\rho^{n+1}_0)+\epsilon^2P^{n+1}_{(2)}+\cdots.\label{Pnplus1}\end{equation}
%Plugging (\ref{rholimdis})(\ref{plimdis})(\ref{Pnplus1}) into the
%momentum equation of (\ref{semidis}), one gets
Then, by using the fact that the curl of the gradient of any scalar
field is always zero, the curl of the $O(1)$ terms of the momentum
equation (\ref{semidis2}) becomes
\begin{equation}
\nabla\times\frac{\mathbf{u}_0^{n+1}-\mathbf{u}_0^n}{\Delta
t}+\nabla\times\nabla\big(\mathbf{u}_0^n\otimes\mathbf{u}_0^n\big)=0.
\label{curl}\end{equation} Thus
\begin{equation}\label{ulimdis}
\frac{\mathbf{u}_0^{n+1}-\mathbf{u}_0^n}{\Delta
t}+\nabla\big(\mathbf{u}_0^n\otimes\mathbf{u}_0^n\big)+\nabla
p^{n}_{(2)}=0,
\end{equation}
where $p^{n}_{(2)}$ is some scalar field.

Equations (\ref{rholimdis}), (\ref{plimdis}), (\ref{ulimdis}) are
the semi-discretization in time of (\ref{lim}) and thus the scheme
(\ref{semidis1}), (\ref{semidis2}) is consistent with the low Mach
number limit $\epsilon\to 0$ of the original compressible Euler
equation. This statement is exactly saying that the scheme is AP. We
can see that, in order to obtain the stability and AP properties, it
is crucial to treat the flux in the density equation
(\ref{semidis1}) implicitly.

Letting $U=(\rho_\epsilon,\rho_\epsilon\mathbf{u}_\epsilon)^T$, we
can write (\ref{semidis1}), (\ref{semidis2}) abstractly as
\begin{equation}\frac{U^{n+1}-U^{n}}{\Delta t}
+\nabla \cdot
F(U^{n+\frac{1}{2}})+QU^{n+1}=0,\label{onedvector}\end{equation}
where
\begin{equation}F(U^{n+1/2})=\left(\begin{array}{c}(\rho_\epsilon\mathbf{u}_\epsilon)^{n+1}\\
\rho_\epsilon^{n}\mathbf{u}^n_\epsilon\otimes\mathbf{u}^n_\epsilon+\alpha
p(\rho_\epsilon^n)\end{array}\right), \qquad
Q=\left(\begin{array}{cc}0&0\\\frac{1-\alpha\epsilon^2}{\epsilon^2}\nabla
P&0\end{array}\right). \label{FQ}
\end{equation}Here $P$ is an operator on $\rho_\epsilon$ and $U^{n+1/2}$
reminds that the flux is partly implicit and partly explicit.

This semi-discretization gives us a framework for developing AP
schemes that can capture the incompressible limit. Now we are left
with the problem of how discretizing the space variable. Because
shocks can form, considerable literature has been devoted to the
design of high resolution methods that can capture the correct shock
speed. Upwind schemes and central schemes are among the most widely
used Godunov type schemes \cite{KT, KT2,Leveque}.

In the present paper, the hyperbolic operator
\[\frac{U^{n+1}-U^{n}}{\Delta t} +\nabla \cdot
F(U^{n+\frac{1}{2}})\] is approximated by an upwind hyperbolic
solver and the stiff $1/\epsilon^2$ factor in front of the pressure
term is treated implicitly. The implicitness of the density flux is
treated by combining it with the momentum equation. For simplicity,
in the present work we only consider the first order modified
Lax-Friedrich scheme with local evaluation of the wave-speed in the
current and neighboring cell.

\section{Full time and space discretization: One dimensional case}

For simplicity, we consider the domain $\Omega=[0,1]$. Using a
uniform spatial mesh with $\Delta x=1/M$, M being an positive
integer, the grid points are defined as
\[
x_j:=j\Delta x,\qquad j=0,1,\cdots,M.
\]

The flux and Jacobian matrix of (\ref{onedvector}) become
\begin{equation}\tilde{F}(U)=\left(\begin{array}{c}\rho_\epsilon\mathbf{u}_\epsilon\\
\rho_\epsilon \mathbf{u}_\epsilon^2+\alpha p(\rho_\epsilon)
\end{array}\right),\qquad \tilde{F}'(U)=\left(\begin{array}{cc}
0&1\\-\mathbf{u}^2_\epsilon+\alpha
p'(\rho_\epsilon)&2\mathbf{u}_\epsilon
\end{array}\right),\label{fluxjacobi}\end{equation}
so, the wave speeds are
\begin{equation}\lambda=\mathbf{u}_\epsilon\pm\sqrt{\alpha p'(\rho_\epsilon)}.\label{lambda}\end{equation}Let
$U_j$ be the approximation of $U(x_j)$ and let
\begin{equation}A_{j+\frac{1}{2}}(t)=\max\big(|\lambda_{j}|,|\lambda_{j+1}|
\big).\label{A}\end{equation} These are the local maximal
wave-speeds in the current and neighboring cells. We discretize
(\ref{onedvector}) in space as follows:
\begin{equation}
\frac{U_j^{n+1}-U_j^{n}}{\Delta t}
+\frac{F_{j+\frac{1}{2}}(U^{n+\frac{1}{2}})-F_{j-\frac{1}{2}}(U^{n+\frac{1}{2}})}{\Delta
x}+Q_jU^{n+1}=0,\label{discrete2step}
\end{equation}
where $F_{j\pm1/2}(U^{n+\frac{1}{2}})$ is the numerical flux
\begin{equation}F_{j+\frac{1}{2}}(U^{n+\frac{1}{2}})=\frac{1}{2}\big(F^+_{j+\frac{1}{2}}(U^{n+\frac{1}{2}})+
F^-_{j+\frac{1}{2}}(U^{n+\frac{1}{2}})\big)\label{Fj1}\end{equation}
and
\[F^+_{j+\frac{1}{2}}(U^{n+\frac{1}{2}})=
\left(\begin{array}{c}(\rho_\epsilon\mathbf{u}_\epsilon)_{j}^{n+1}
+A_{j+\frac{1}{2}}^n\rho_{\epsilon
j}^n\\
(\rho_{\epsilon }\mathbf{u}_{\epsilon
}\otimes\mathbf{u}_\epsilon)_{j}^{n}+\alpha p(\rho_{\epsilon
j}^n)+A_{j+\frac{1}{2}}^n
(\rho_{\epsilon}\mathbf{u}_\epsilon)_{j}^n\end{array}\right),\]\[
F^-_{j+\frac{1}{2}}(U^{n+\frac{1}{2}})=
\left(\begin{array}{c}(\rho_\epsilon\mathbf{u}_\epsilon)_{j+1}^{n+1}
-A_{j+\frac{1}{2}}^n\rho_{\epsilon
j+1}^n\\
(\rho_{\epsilon }\mathbf{u}_{\epsilon
}\otimes\mathbf{u}_\epsilon)_{j+1}^{n}+\alpha p(\rho_{\epsilon
j+1}^n)-A_{j+\frac{1}{2}}^n
(\rho_{\epsilon}\mathbf{u}_\epsilon)_{j+1}^n\end{array}\right)
\]
and
\[QU_j^{n+1}=\left(\begin{array}{c}0\\\frac{1-\alpha\epsilon^2}{\epsilon^2}\frac{1}{2\Delta
x}\big(P(\rho_{\epsilon j+1}^{n+1})-P(\rho_{\epsilon
j-1}^{n+1})\big)
\end{array}\right).\]
Let
\begin{equation}\mathbf{q}=\rho\mathbf{u}\label{q},\end{equation}
and $F^{(1)}, F^{(2)}$ denote the first and second element of $F$
respectively, we can rewrite the momentum discretization in
(\ref{discrete2step}) as follows:
\begin{equation}\mathbf{q}_{\epsilon j}^{
n+1}=\mathbf{q}_{\epsilon j}^n-\Delta
tD^x_jF^{(2)}\big(\rho^n_\epsilon,\mathbf{u}^n_\epsilon\big)-\frac{1-\alpha\epsilon^2}{\epsilon^2}\frac{\Delta
t}{2\Delta x}\big(p(\rho^{n+1}_{\epsilon
j+1})-p(\rho^{n+1}_{\epsilon j-1})\big).\label{pn1}\end{equation}
Here
\[D^x_ju=\frac{u_{j+1/2}-u_{j-1/2}}{\Delta x}.\]  By substituting
(\ref{pn1}) into the density equation in (\ref{discrete2step}), one
gets
\begin{equation}\rho_{\epsilon j}^{n+1}-\frac{(1-\alpha\epsilon^2)\Delta t^2}{4\epsilon^2\Delta
x^2}\Big(p(\rho^{n+1}_{\epsilon j+2})-2p(\rho^{n+1}_{\epsilon
j})+p(\rho^{n+1}_{\epsilon j-2})\Big)
=D\phi(\rho_\epsilon^n,\mathbf{q}_\epsilon^n),\label{ellipticdis}\end{equation}
where
\begin{equation}D\phi(\rho_\epsilon^n,\mathbf{q}_\epsilon^n)=\rho_\epsilon^n-\Delta
tD_j^x F^{(1)}(\rho_\epsilon^n,\mathbf{u}_\epsilon^n)+ \frac{\Delta
t^2}{2\Delta
x}\big(D_{j+1}^x-D_{j-1}^x\big)F^{(2)}(\rho_\epsilon^n,\mathbf{u}_\epsilon^n)\label{dphi}\end{equation}
is a discretization of $\phi(\rho_\epsilon^n,\mathbf{u}_\epsilon^n)$
in (\ref{phi}). We notice that (\ref{ellipticdis}) is a
discretization of the elliptic equation (\ref{ellipticapro}). We can
update $\mathbf{q}^{n+1}_\epsilon$ through (\ref{pn1}) afterwards.

To obtain $\rho^{n+1}_\epsilon$ in (\ref{ellipticdis}), a nonlinear
system of equations needs to be solved. One possible way to simplify
it is to replace $\nabla P(\rho_\epsilon^{n+1})$ by
$P'(\rho_\epsilon^n)\nabla\rho_\epsilon^{n+1}$, so that the
following linear system is obtained:
%\begin{equation}\label{discrete2step1}
%\left\{\begin{array}{l} \frac{\rho_{\epsilon j}^{
%n+1}-\rho_{\epsilon j}^{n}}{\Delta t}+\frac{p_{\epsilon
%j+1}^{n+1}-p_{\epsilon j-1}^{n+1}}{2\Delta x}+\frac{1}{2\Delta
%x}A^n\big(-\rho_{\epsilon j-1}^n-\rho^n_{\epsilon
%j+1}+2\rho_{\epsilon j}^n\big)=0\\\frac{p_{\epsilon j}^{
%n+1}-p_{\epsilon j}^{n}}{\Delta t}+\frac{1}{2\Delta
%x}\big(\frac{p^{n2}_{\epsilon j+1}}{\rho_{\epsilon
%j+1}^n}-\frac{p^{n2}_{\epsilon j-1}}{\rho_{\epsilon j-1}^n}\big)+
%\frac{1}{2\Delta x}A^n\big(-p_{\epsilon j-1}^n-p_{\epsilon
%j+1}^n+2p_{\epsilon
%j}^n\big)\\\qquad\qquad\qquad\qquad\qquad+\frac{1}{\epsilon^2}\frac{1}{2\Delta
%x}P'(\rho^n_{\epsilon j})\big(\rho^{n+1}_{\epsilon
%j+1}-\rho^{n+1}_{\epsilon j-1}\big)=0
%\end{array}\right..
%\end{equation}
\begin{eqnarray}\rho_{\epsilon j}^{n+1}-\frac{(1-\alpha\epsilon^2)\Delta t^2}{4\epsilon^2\Delta
x^2}\Big(p'(\rho^{n}_{\epsilon j+1})\big(\rho_{\epsilon
j+2}^{n+1}-\rho_{\epsilon j}^{n+1}\big)-p'(\rho^{n}_{\epsilon
j-1})\big(\rho^{n+1}_{\epsilon j}-\rho^{n+1}_{\epsilon
j-2}\big)\Big)
&&{}\nonumber\\=D\phi(\rho_\epsilon^n,\mathbf{q}_\epsilon^n).&&\quad\label{ellipticdis1}\end{eqnarray}
This is a five point scheme which is too much diffusive, especially
near the shock. One possible improvement is that instead of
(\ref{ellipticdis1}), we use the following three points
discretization
\begin{eqnarray}\rho_{\epsilon j}^{n+1}-\frac{(1-\alpha\epsilon^2)\Delta t^2}{\epsilon^2\Delta
x^2}\Big(p'(\rho^n_{\epsilon j+1})\big(\rho^{n+1}_{\epsilon
j+1}-\rho^{n+1}_{\epsilon j}\big)-p'(\rho^n_{\epsilon
j})\big(\rho^{n+1}_{\epsilon j}-\rho^{n+1}_{\epsilon
j-1}\big)\Big)&&{}\nonumber\\
=D\phi(\rho_\epsilon^n,\mathbf{q}_\epsilon^n).&&\quad\label{ellipticdis2}\end{eqnarray}
After obtaining $\rho^{n+1}_\epsilon$, we can substitute it into
(\ref{pn1}) to get $\mathbf{q}_{\epsilon j}^{n+1}$.

To summarize, three schemes are proposed here: (\ref{ellipticdis}),
(\ref{pn1}); (\ref{ellipticdis1}), (\ref{pn1}) and
(\ref{ellipticdis2}), (\ref{pn1}). To investigate the AP property,
we take (\ref{ellipticdis1}), (\ref{pn1}) as an example. The proofs
for the other two schemes are similar. By substituting the following
expansion \begin{equation}\rho_{\epsilon
j}^n=\rho_{0c}^n+\epsilon^2\rho_{(2)j}^n+\cdots,\qquad
\mathbf{q}_{\epsilon j}^n=\mathbf{q}_{0c}^n+\epsilon
\mathbf{q}_{(2)j}^n +\cdots,\label{expansion}\end{equation} into
(\ref{ellipticdis1}), the $O(\frac{1}{\epsilon^2})$ terms give that
$\rho_{(0)j}^{n+1}=\rho_{0c}^{n+1}$ is constant in space by using
the periodic boundary condition, and thus:
\[\rho_{\epsilon
j}^{n+1}=\rho_{(0)c}^{n+1}+\epsilon^2\rho_{(2)j}^{n+1}+\cdots.\]
Summing (\ref{ellipticdis1}) over all the grid points, one gets
\begin{subequations}\label{onedlim}
\begin{equation}\rho^{n+1}_{0c}=\rho^{n}_{0c}=\rho_{0c},\end{equation}which implies that $\rho_0$ is independent
of time and space. Thus, the $O(1)$ terms of (\ref{ellipticdis1})
are
\[p'(\rho^{n}_{0c})\big(\rho^{n+1}_{(2)j+2}-\rho^{n+1}_{ (2)j}\big)-
p'(\rho^n_{0 c})\big(\rho^{n+1}_{
(2)j}-\rho^{n+1}_{(2)j-1}\big)=0,\]by recalling that the $O(1)$
terms of both $\rho_\epsilon^n$ and $\mathbf{q}_\epsilon^n$ are
constant in space. Then the periodic boundary condition gives
\begin{equation}\rho^{n+1}_{(2)j}=\rho^{n+1}_{(2)c},\end{equation}
 which gives that $\rho^{n+1}_{(2)}$ is also independent of space.
Therefore from (\ref{P}), (\ref{pn1}), \begin{equation}\mathbf{q}_{0
j}^{n+1}=\mathbf{q}_{0j}^{n}=\mathbf{q}_{0c}^{n}.\end{equation}
\end{subequations}
In one dimension, (\ref{onedlim}) is the discretization of
(\ref{rholimdis}), (\ref{plimdis}), (\ref{ulimdis}) when periodic
boundary conditions apply and thus is consistent with the
incompressible limit. In fact all the three methods proposed here
are AP.

\section{Full time and space discretization: Two dimensional case}

We consider the domain $\Omega=[0,1]\times[0,1]$. For $M_1,M_2$ two
positive integers, we use a uniform spatial mesh $\Delta x=1/M_1,
\Delta y=1/M_2$. The grid points are
\[(x_i,y_j):=(i\Delta x, j\Delta y),\qquad
i=0,\cdots,M_1;j=0,\cdots,M_2
\]

Now
$U=(\rho_\epsilon,\mathbf{q}_{\epsilon}^{(1)},\mathbf{q}_\epsilon^{(2)})^T$
and $U_{i,j}$ is the numerical approximation of $U(x_i,y_j)$. Let
\begin{equation}
G_1(U)=\left(\begin{array}{c}\rho_\epsilon\mathbf{u}_{\epsilon 1}\\
\rho_\epsilon\mathbf{u}_1^2+\alpha
p(\rho_\epsilon)\\\rho_\epsilon\mathbf{u}_1\mathbf{u}_2\end{array}\right),\quad
G_2(U)=\left(\begin{array}{c}\rho_\epsilon\mathbf{u}_{\epsilon 2}\\
\rho_\epsilon\mathbf{u}_1\mathbf{u}_2
\\\rho_\epsilon\mathbf{u}^2_2+\alpha p(\rho_\epsilon)\end{array}\right).
\end{equation}
and
\[Q=\frac{1-\alpha\epsilon^2}{\epsilon^2}
\left(\begin{array}{ccc}0&0&0\\\partial_xP&0&0\\\partial_yP&0&0\end{array}\right).\]Eq.
(\ref{onedvector}) can be written as
\[\partial_t U+\partial_xG_1(U)+\partial_yG_2(U)+QU=0.\]
Denote
\[G_{1}(U^{n+\frac{1}{2}})=
\left(\begin{array}{c}(\rho_\epsilon\mathbf{u}_{\epsilon 1})^{n+1}\\
\rho^n_\epsilon(\mathbf{u}_{\epsilon1}^n)^2+\alpha
p(\rho_\epsilon^n)
\\\rho^n_\epsilon\mathbf{u}^n_{\epsilon1}\mathbf{u}^n_2\end{array}\right),\qquad
G_{2}(U^{n+\frac{1}{2}})=
\left(\begin{array}{c}(\rho_\epsilon\mathbf{u}_{\epsilon 2})^{n+1}\\
\rho^n_\epsilon\mathbf{u}^n_{\epsilon1}\mathbf{u}^n_{\epsilon2}
\\\rho^n_\epsilon(\mathbf{u}_{\epsilon2}^n)^2+\alpha p(\rho_\epsilon^n)\\\end{array}\right),\]
\[\tilde{Q}=\left(\begin{array}{ccc}0&0&0\\
\frac{1-\alpha\epsilon^2}{\epsilon^2}D^x\hat{P}&0&0\\\frac{1-\alpha\epsilon^2}{\epsilon^2}D^y\hat{P}&0&0\end{array}\right),\]
\[D^x_{ij}u=\frac{u_{i
j+1}-u_{i j-1}}{2\Delta x},\qquad D^y_{ij}u=\frac{u_{i+1 j}-u_{i-1
j}}{2\Delta y}.\] Now the eigenvalues of the two one-dimensional
hyperbolic equations are
\[\lambda^{(1)}=\mathbf{u}_1,\mathbf{u}_1\pm\sqrt{\alpha p'(\rho_\epsilon)},\qquad
\lambda^{(2)}=\mathbf{u}_2,\mathbf{u}_2\pm\sqrt{\alpha p'(
\rho_\epsilon)}.\]
 The fully discrete scheme for the two dimensional problem is
\begin{eqnarray}\frac{U_{ij}^{ n+1}-U_{ij}^n}{\Delta
t}+D_{ij}^x{G}_{1}(U^{n+1/2})+\frac{1}{2}\big(A_{i-\frac{1}{2},j}D^x_{ij-}-A_{i+\frac{1}{2},j}
D^x_{ij+}\big)U^n &&{}\nonumber\\+D^y_{ij}G_{2}(U^{n+1/2})+
\frac{1}{2}\big(A_{i,j-\frac{1}{2}}D^y_{ij-}-A_{i,j+\frac{1}{2}}D^y_{ij+}\big)U^n
+\tilde{Q}U_{ij}^{n+1}=0,&&\label{2ddisctere}\end{eqnarray}
 where
\[D_{ij-}^xu=\frac{u_{ij}-u_{i-1 j}}{\Delta x},\quad(D_{ij+}^xu)=\frac{u_{i+1 j}-u_{ij}}{\Delta
x},\] \[D_{ij-}^yu=\frac{u_{ij}-u_{ij-1}}{\Delta y},\quad
D_{ij+}^yu=\frac{u_{ij+1}-u_{ij}}{\Delta y},\] and
\begin{equation}\begin{array}{l}A_{i+\frac{1}{2},j}^n=\max\{|\lambda^{(1)}_{
ij}|,|\lambda^{(1)}_{ i+1,j}|,|\lambda^{(2)}_{ ij}|,|\lambda^{(2)}_{
i+1,j}|\},\\A_{i,j+\frac{1}{2}}^n=\max\{|\lambda^{(1)}_{
ij}|,|\lambda^{(1)}_{ i,j+1}|,|\lambda^{(2)}_{ ij}|,|\lambda^{(2)}_{
i,j+1}|\}.\end{array}\label{A2}\end{equation}

Let $\mathbf{q}_\epsilon$ be like in (\ref{pn1}). Like in one
dimension, we can substitute the expressions of
$\mathbf{q}_{1ij}^{n+1},\mathbf{q}_{2ij}^{n+1}$ into the density
equation and get the following discretized elliptic equation,
\begin{eqnarray}
&&\rho_{\epsilon ij}^{n+1}-\frac{(1-\alpha\epsilon^2)\Delta
t^2}{4\epsilon^2}\Big(\frac{1}{\Delta x^2}\big(P(\rho_{\epsilon
i+2,j}^{n+1})-2P(\rho^{n+1}_{\epsilon i,j})+P(\rho^{n+1}_{\epsilon
i-2,j})\big)+
\label{twodelliptic}\\
&&\qquad\qquad\frac{1}{\Delta y^2}\big(P(\rho_{\epsilon
i,j+2}^{n+1})-2P(\rho^{n+1}_{\epsilon i,j})+P(\rho^{n+1}_{\epsilon
i,j-2})\big)\Big)=D\phi_{ij}(\rho^{n}_\epsilon,\mathbf{q}_{1\epsilon}^{n},\mathbf{q}^{n}_{2\epsilon}),{}\nonumber\quad
\end{eqnarray}
where
\begin{eqnarray}
&&D\phi_{ij}(\rho_\epsilon^n,\mathbf{q}_{1\epsilon}^n,\mathbf{q}_{2\epsilon}^n){}\nonumber\\
&=&\rho_\epsilon^n-\Delta t\Big(D_{ij}^x\mathbf{q}_{\epsilon
1}^n+D_{ij}^y\mathbf{q}_{\epsilon 2}^n{}\nonumber\\
&&+\frac{1}{2}\big(A_{i-\frac{1}{2}}D^x_{ij-}-A_{i+\frac{1}{2},j}D_{ij+}^x
+A_{i,j-\frac{1}{2}}D_{ij-}^y
-A_{i,j+\frac{1}{2}}D_{ij+}^y\big)\rho_\epsilon^n\Big){}\nonumber\\
&&+\Delta
t^2\Big(D_{ij}^xD_{ij}^x\big(\rho_\epsilon^n(\mathbf{u}_{\epsilon
1}^n)^2+\alpha
p(\rho_\epsilon^n)\big)+D_{ij}^yD_{ij}^y\big(\rho_\epsilon^n(\mathbf{u}_{\epsilon
2}^n)^2+\alpha p(\rho_\epsilon^n)\big){}\nonumber\\
&&+(D_{ij}^xD_{ij}^y+D_{ij}^yD_{ij}^x)\rho_\epsilon^n\mathbf{u}_{\epsilon
1}^n\mathbf{u}_{\epsilon
2}^n+\frac{1}{2}D_{ij}^x(A_{i-\frac{1}{2},j}D_{ij-}^x-A_{i+\frac{1}{2},j}D_{ij+}^x)
\mathbf{q}_{\epsilon
1}^n{}\nonumber\\
&&+\frac{1}{2}D^y_{ij}(A_{i-\frac{1}{2},j}D_{ij-}^x-A_{i+\frac{1}{2},j}D_{ij+}^x)
\mathbf{q}_{\epsilon
2}^n+\frac{1}{2}D^x_{ij}(A_{i,j-\frac{1}{2}}D_{ij-}^y-A_{i,j+\frac{1}{2}}D_{ij+}^y)
\mathbf{q}_{\epsilon
1}^n{}\nonumber\\
&&+\frac{1}{2}D_{ij}^y(A_{i,j-\frac{1}{2}}D_{ij-}^y-A_{i,j+\frac{1}{2}}D_{ij+}^y)
\mathbf{q}_{\epsilon 2}^n\Big).{}\nonumber\\\label{dphi2}
\end{eqnarray}
 After obtaining $\rho_{ij}^{n+1}$ by (\ref{twodelliptic}),
$\mathbf{q}_{1ij}^{n+1},\mathbf{q}_{2ij}^{n+1}$ can be updated by
the momentum equation afterwards.

Similar to the one-dimensional case, the modified diffusion operator
using a reduced stencil is as follows:
\begin{eqnarray}
&&\rho_{\epsilon ij}^{n+1}-\Delta
t^2\frac{1-\alpha\epsilon^2}{\epsilon^2}\times{}\nonumber\\
&&\times \bigg(\frac{1}{\Delta x^2}\Big(P'(\rho^n_{\epsilon
i,j+1})\big(\rho^{n+1}_{\epsilon i,j+1}-\rho^{n+1}_{\epsilon
i,j}\big)-P'(\rho^n_{\epsilon i,j})\big(\rho^{n+1}_{\epsilon
i,j}-\rho^{n+1}_{\epsilon i,j-1}\big)\Big){}\nonumber
\\
&&+\frac{1}{\Delta y^2}\Big(P'(\rho^n_{\epsilon
i+1,j})\big(\rho^{n+1}_{\epsilon i+1,j}-\rho^{n+1}_{\epsilon
i,j}\big)-P'(\rho^n_{\epsilon i,j})\big(\rho^{n+1}_{\epsilon
i,j}-\rho^{n+1}_{\epsilon
i-1,j}\big)\Big)\bigg){}\nonumber\\
&=&\phi(\rho^{n}_\epsilon,\mathbf{q}_{1\epsilon}^{n},\mathbf{q}^{n}_{2\epsilon}).\label{twodelliptic2}
\end{eqnarray}

Now we prove the AP property of our fully discrete scheme. Here only
well-prepared initial conditions are considered, which means that
there will be no shock forming in the solution. Then $\alpha$ can be
chosen to be $0$ to minimize the introduced numerical viscosity.
Assuming that the expansions of $\rho_\epsilon,\mathbf{u}_\epsilon$
in (\ref{nassum}) hold at time $t^n$, when $\epsilon\to 0$, the
$O(\frac{1}{\epsilon^2})$ terms of (\ref{twodelliptic2}) give
\begin{eqnarray}&&\frac{1}{\Delta x^2}\Big(P'(\rho^n_{0
i,j+1})\big(\rho^{n+1}_{0 i,j+1}-\rho^{n+1}_{0
i,j}\big)-P'(\rho^n_{0 i,j})\big(\rho^{n+1}_{0 i,j}-\rho^{n+1}_{0
i,j-1}\big)\Big){}\nonumber
\\
&&+\frac{1}{\Delta y^2}\Big(P'(\rho^n_{0 i+1,j})\big(\rho^{n+1}_{0
i+1,j}-\rho^{n+1}_{0 i,j}\big)-P'(\rho^n_{0 i,j})\big(\rho^{n+1}_{0
i,j}-\rho^{n+1}_{0 i-1,j}\big)\Big)=0.{}\nonumber\end{eqnarray} When
using periodic boundary conditions, one gets
$\rho_{0ij}^{n+1}=\rho_{0c}^{n+1}$ from (\ref{P}). The time
independence of $\rho_{0}^{n+1}$, similar to the one dimensional
case, can be obtained by summing (\ref{twodelliptic2}) over all the
grid points. Accordingly we have
\begin{equation}
\rho_{\epsilon
ij}^{n+1}=\rho_{0c}^n+\epsilon^2\rho_{(2)ij}^{n+1}+\cdots.\label{2drho}
\end{equation}
%Thereby, the $O(1)$ terms of the density equation of
%(\ref{2ddisctere}) becomes
%\begin{equation}
%D^x\mathbf{u}_{01}^{n+1}+D^y\mathbf{u}_{02}^{n+1}=0.
%\end{equation}

To prove the limiting behavior of $\mathbf{u}_{\epsilon
1},\mathbf{u}_{\epsilon 2}$, we do not want to use the density
equation because the diffusion operator with reduced stencil does
not allow us to find the corresponding density equation. Therefore,
we consider the $O(1)$ term of (\ref{twodelliptic2}),
\begin{eqnarray}
&&\rho_{0 ij}^{n+1}-\Delta t^2\times{}\nonumber\\&&\times
\bigg(\frac{1}{\Delta x^2}\Big(P'(\rho^n_{0 c})\big(\rho^{n+1}_{(2)
i,j+1}-\rho^{n+1}_{(2) i,j}\big)-P'(\rho^n_{0c})\big(\rho^{n+1}_{(2)
i,j}-\rho^{n+1}_{(2) i,j-1}\big)\Big){}\nonumber
\\
&&+\frac{1}{\Delta y^2}\Big(P'(\rho^n_{0c})\big(\rho^{n+1}_{(2)
i+1,j}-\rho^{n+1}_{(2) i,j}\big)-P'(\rho^n_{0
c})\big(\rho^{n+1}_{(2)
i,j}-\rho^{n+1}_{(2) i-1,j}\big)\Big)\bigg){}\nonumber\\
&=&\phi(\rho^{n}_0,\mathbf{q}_{10}^{n},\mathbf{q}^{n}_{20}).\label{2dorder1}
\end{eqnarray}
%From (\ref{dphi2}), $\phi(\rho^{n}_0,p_{10}^{n},p^{n}_{20})$ is a
%discretization of \[\rho_0^n-\Delta
%t\nabla\cdot\mathbf{q}^n_0+\Delta
%t^2\nabla\cdot\mbox{div}(\frac{1}{\rho_0^{n}}\mathbf{q}^n_0\otimes\mathbf{q}^n_0)
%+O(\Delta t\Delta x).\] Here the $O(\Delta t\Delta x)$ terms come
%from the diffusion terms. Then (\ref{2dorder1}) is in fact a
%discretization of
%\begin{equation}\label{2dulimdis}
%-\nabla\cdot\mathbf{q}_0^n+ \Delta
%t\nabla\cdot\Big(\mbox{div}\big(\frac{1}{\rho^n_0}\mathbf{q}_0^n\otimes\mathbf{q}_0^n\big)\Big)+\Delta
%t\Delta p^{n+1}_{(2)}=O(\Delta x).
%\end{equation}
Moreover, noting the fact that
\begin{eqnarray}D^x_{ij}P(\rho_\epsilon^{n+1})&=&D^x_{ij}
P(\rho_{0c}^n+\epsilon^2\rho_{(2)}^{n+1}+o(\epsilon^2)){}\nonumber\\&=&
D^x_{ij}P(\rho_{0c}^n)+\epsilon^2D^x_{ij}\big(\rho_{(2)}^{n+1}P'(\rho_{0c}^n)\big)+o(\epsilon^2),
{}\nonumber\\
&=&\epsilon^2D^x_{ij}\big(\rho_{(2)}^{n+1}P'(\rho_{0c}^n)\big)+o(\epsilon^2),{}\nonumber
\end{eqnarray} and similarly,
\[D^y_{ij}P(\rho_\epsilon^{n+1})=\epsilon^2D^y_{ij}
\big(\rho_{(2)}^{n+1}P'(\rho_{0c}^n)\big)+o(\epsilon^2),\] the
$O(1)$ terms of the momentum equations of (\ref{2ddisctere}) become
\begin{subequations}\label{momentum2d}
\begin{eqnarray}
\frac{\mathbf{q}_{0 1ij}^{n+1}-\mathbf{q}_{0 1ij}^{n}}{\Delta
t}+D^x_{ij}\Big(\frac{\mathbf{q}_{0
1}^2}{\rho_0}\Big)^{n}+D_{ij}^y\Big(\frac{\mathbf{q}_{0
1}\mathbf{q}_{0
2}}{\rho_0}\Big)^{n}+\frac{1}{2}\big(A_{i-\frac{1}{2},j}D_{ij-}^x-A_{i+\frac{1}{2},j}D_{ij+}^x
&&{}\nonumber\\
+A_{i,j-\frac{1}{2}}D_{ij-}^y-A_{i,j+\frac{1}{2}}D_{ij+}^y\big)\mathbf{q}_{01}
=D^x_{ij}\big(P'(\rho^{n+1}_{0c
})\rho_{(2)}^{n+1}\big),&&\quad\label{momentum2d1}\\
\frac{\mathbf{q}_{02ij}^{n+1}-\mathbf{q}_{0 2ij}^{n}}{\Delta
t}+D^x_{ij}\Big(\frac{\mathbf{q}_{0 1}\mathbf{q}_{0
2}}{\rho_0}\Big)^{n}+D^y_{ij}\Big(\frac{\mathbf{q}^2_{0
2}}{\rho_0}\Big)^{n}+\frac{1}{2}\big(A_{i-\frac{1}{2},j}D_{ij-}^x-A_{i+\frac{1}{2},j}D_{ij+}^x
&&{}\nonumber\\
+A_{i,j-\frac{1}{2}}D_{ij-}^y-A_{i,j+\frac{1}{2}}D_{ij+}^y\big)\mathbf{q}_{02}
=D^y_{ij}\big(P'(\rho_{0c}^{n+1})\rho^{n+1}_{(2)}\big).&&\quad\label{momentum2d2}
\end{eqnarray}
\end{subequations}
%The $O(1)$ terms of $\Delta
%t*(D^x(\ref{momentum2d1})+D^y(\ref{momentum2d2}))$ gives a
%discretization of
%\begin{equation}\label{momentumd}
%\nabla\cdot\mathbf{q}_0^{n+1}-\nabla\cdot\mathbf{q}_0^n+ \Delta
%t\nabla\cdot\Big(\mbox{div}\big(\frac{1}{\rho^n_0}\mathbf{q}_0^n\otimes\mathbf{q}_0^n\big)\Big)+\Delta
%t\Delta p^{n+1}_{(2)}=O(\Delta x). \end{equation}
 Comparing
(\ref{2dorder1}) with $\Delta
t*(D_{ij}^x(\ref{momentum2d1})+D_{ij}^y(\ref{momentum2d2}))$ , one
gets
\begin{equation}D_{ij}^x\mathbf{q}^{n+1}_{01}+D_{ij}^y\mathbf{q}^{n+1}_{02}
=O(\Delta x\Delta t),\label{disq0}\end{equation} which is an
approximation of (\ref{plimdis}).
%consider the $O(1)$ terms of
%$D^x(\ref{momentum2d2})-D^y(\ref{momentum2d1})$, by noting
%(\ref{2drho}), it is a discretization of
%\begin{equation}\frac{\nabla\times\mathbf{q}_{0}^{n+1}-\nabla\times\mathbf{q}_{0}^{n}}{\Delta
%t}+\nabla\times\Big(\frac{1}{\rho^n_0}\mbox{div}\big(\mathbf{q}_0^n\otimes\mathbf{q}_0^n\big)\Big)=O(\Delta
%x),\label{curldis}\end{equation}
Moreover, it is obvious that (\ref{momentum2d}) is a discretization
of (\ref{ulimdis}). Thus we obtain a full discretization of
(\ref{lim}) in the limit $\epsilon\to 0$. Therefore, the
two-dimensional scheme is also AP.

\section{The ad-hoc parameter} In this
section we illustrate how to choose $\Delta t$ and the parameter
$\alpha$ by considering the simple state equation
$P(\rho_\epsilon)=\rho_\epsilon$. In this context, the fully
discrete scheme (\ref{discrete2step}) can be written as
\begin{equation}
\left\{\begin{array}{l}\frac{\rho^{n+1}_\epsilon-\rho^n_\epsilon}{\Delta
t}+\nabla\cdot
\mathbf{q}_\epsilon^{n+1}-\nabla\cdot\mathbf{q}_\epsilon^{n}+
\tilde{\nabla}\cdot\mathbf{q}_\epsilon^{n}=0,\\
\frac{\mathbf{q}_\epsilon^{n+1}-\mathbf{q}_\epsilon^n}{\Delta
t}+\tilde{\nabla} \big(\rho^n_\epsilon\mathbf{u}^n_\epsilon\otimes
\mathbf{u}^n_\epsilon+\alpha
\rho_\epsilon^n\big)+\frac{1-\alpha\epsilon^2}{\epsilon^2}\nabla
\rho^{n+1}_\epsilon=0.
\end{array}\right.,
\label{semidisrho}
\end{equation}
where $\nabla$ is the centered difference while $\tilde{\nabla}$
stands for the difference of fluxes. The latter is defined as
follows (in one space-dimension for simplicity):
\[(\tilde{\nabla}\cdot F(U^n))_j=\frac{F_{j+\frac{1}{2}}(U^n)-F_{j-\frac{1}{2}}(U^n)}{\Delta x},\]
where the flux $F_{j\pm1/2}$ is defined as in (\ref{Fj1}). By
substituting
$\nabla\cdot\big(\mathbf{q}_\epsilon^{n+1}-\mathbf{q}^n\big)$ from
the momentum equation of (\ref{semidisrho}) into its density
equation, one gets
\begin{equation}\label{4.2}\frac{\rho_\epsilon^{n+1}-\rho_\epsilon^n}{\Delta
t}+\tilde{\nabla}\cdot\mathbf{q}_\epsilon^n -\Delta
t\frac{1-\alpha\epsilon^2}{\epsilon^2}\Delta
\rho_\epsilon^{n+1}-\Delta t
\nabla\cdot\tilde{\nabla}\big(\rho_\epsilon^n\mathbf{u}_\epsilon^n\otimes\mathbf{u}_\epsilon^n+\alpha\rho_\epsilon^n\big)
=0.\end{equation} The $O(\Delta t)$ terms behave like a diffusion
term which suppresses the oscillations at the discontinuity.
Assuming that we use a first order explicit LLF scheme, the
diffusions needed to damp out the oscillations in the mass and
momentum equations are respectively:
%\begin{equation}-\frac{1}{24}\big(|\mathbf{u}_\epsilon|+\frac{1}{\epsilon}\big)\Delta
%x^2\partial_x^4
%\rho^n_\epsilon,\quad-\frac{1}{24}\big(|\mathbf{u}_\epsilon|+\frac{1}{\epsilon}\big)\Delta
%x^2\partial_x^4
%\mathbf{q}^n_\epsilon.\end{equation}
\begin{equation}
\frac{1}{2}(|\mathbf{u}^n_\epsilon|+\frac{1}{\epsilon})\Delta
x\Delta\rho_\epsilon^n,\qquad\frac{1}{2}(|\mathbf{u}^n_\epsilon|+\frac{1}{\epsilon})\Delta
x\Delta \mathbf{q}_\epsilon^n.\label{difneedrho}
\end{equation}
 Here in (\ref{4.2}),
besides the $O(\Delta t)$ terms,
$\tilde{\nabla}\cdot\mathbf{q}_\epsilon^n$ also includes some
numerical dissipation. By noting
\[\rho^{n+1}_\epsilon=\rho_\epsilon^n-\Delta t\nabla\cdot
\mathbf{q}_\epsilon^{n+1}+O(\Delta t\Delta x),\] the diffusion for
$\rho_\epsilon$ now is
\begin{equation}\Big(\frac{1}{2}\big(|\mathbf{u}_\epsilon|+\sqrt{\alpha}\big)\Delta
x+\frac{\Delta t}{\epsilon^2}\Big)\Delta\rho_\epsilon^n +\Delta
t\Delta(\rho_\epsilon^n\mathbf{u}_\epsilon^n\otimes\mathbf{u}_\epsilon^n)
\label{difrealrho}\end{equation}plus some higher order terms.
Moreover, the diffusion for $\mathbf{q}_\epsilon$ is
\begin{equation}\label{difrealq}
\frac{1}{2}\big(|\mathbf{u}_\epsilon|+\sqrt{\alpha}\big)\Delta
x\Delta \mathbf{q}_\epsilon^n+\Delta
t\frac{1-\alpha\epsilon^2}{\epsilon^2}\Delta\mathbf{q}_\epsilon^{n}
\end{equation}and some higher order terms.
 Comparing
(\ref{difneedrho}) and (\ref{difrealrho}), (\ref{difrealq}), in
order to suppress the oscillations at discontinuities we only need
to have
\[\frac{1}{2}\big(|\mathbf{u}_\epsilon^n|+\sqrt{\alpha}\big)\Delta x+\frac{1-\alpha\epsilon^2}
{\epsilon^2}\Delta t\geq
\frac{1}{2}\big(|\mathbf{u}_\epsilon^n|+\frac{1}{\epsilon}\big)\Delta
x,\] that is \begin{equation}\Delta
t\geq\frac{\frac{1}{2}\epsilon\Delta
x}{1+\sqrt{\alpha}\epsilon}.\label{4.5}\end{equation} Moreover the
CFL condition for the explicit part is
\begin{equation}\Delta t\leq\sigma\frac{\Delta
x}{\max\{|\mathbf{u}_\epsilon|+\sqrt{\alpha}\}},\label{CFLex}\end{equation}
where $\sigma$ is the Courant number which is less than $1$. We
usually choose $\sigma$ to be $0.5$. Then the parameter $\alpha$
should satisfy
\begin{equation}\label{alpha}
\frac{\Delta x}{2\Delta
t}-\frac{1}{\epsilon}\leq\sqrt{\alpha}\leq\frac{\sigma\Delta
x}{\Delta t}-\max\{|\mathbf{u}^n|_\epsilon\}
\end{equation}
according to (\ref{4.5}), (\ref{CFLex}). Then the following
constraint on $\Delta t$ should hold if we want the scheme to be
stable and non-oscillatory
\begin{equation}\label{deltat0}\max\{|\mathbf{u}^n_\epsilon|\}+\frac{\Delta x}{2\Delta t}\leq\frac{\sigma\Delta
x}{\Delta t}+ \frac{1}{\epsilon}.\end{equation}

The reason for the occurence of nonphysical oscillations when
$\alpha = 0$ lies in the fact that the diffusion is not large
enough. In this case, with a simple reduction of $\Delta t$, it is
likely that the diffusion can no longer suppress the oscillations.
This is why we need to introduce $\alpha$ to control the
oscillations. But, from the analysis, no matter the value of
$\alpha$, as long as it is less than $1/\epsilon^2$, the diffusion
can never be sufficient when $\Delta t\leq\frac{1}{4}\epsilon\Delta
x$. In summary there is no specific way of choosing
$\alpha<1/\epsilon$ that can guarantee that the nonphysical
oscillations will disappear in any case. For well-prepared initial
conditions in the low Mach number regime, because there is no shock
formation in the solution, it is better to choose $\alpha$ as small
as possible to get better accuracy, but if strong shocks exist in
the solution, $\alpha$ should be big enough to suppress the
oscillations. This is why the choice of $\alpha$ depends on the
considered problem.

\section{Numerical results}
Three numerical examples will allow us to test the performances of
the proposed schemes. In fact, three schemes are proposed in section
3 and 4, for example in one dimension: the scheme
(\ref{discrete2step}) without linearizing $\nabla
P(\rho^{n+1}_\epsilon)$ is denoted by "NL". We need to use Newton
iterations to solve the nonlinear system. When $\nabla
P(\rho^{n+1}_\epsilon)$ is approximated by
$P'(\rho^{n}_\epsilon)\nabla P(\rho^{n+1}_\epsilon)$, the unknowns
become a linear system. This scheme is represented by "L". "LD"
denotes the scheme with the narrower stencil (\ref{ellipticdis2}).
Here we use well-prepared initial conditions of the form
(\ref{assumption}) and $\alpha=1$ for all the test cases.

In one dimension, let the computational domain be $[a,b]$ and the
mesh size be $\Delta x$. The grid points are
\[x_j=a+(j-1)\Delta x.\]
In the following tables, the $L^2$ norm of the relative error
between the reference solutions $u$ and the numerical ones $U$
\[e(U)=\frac{\|U-u\|_{L^2}}{\|u\|_{L^2}}
=\frac{\frac{1}{M}\big(\sum_{j}|U_j-u(x_j)|^2\big)^{\frac{1}{2}}}{\frac{1}{M_e}\big(\sum_i|u(x_i)|^2\big)
^{\frac{1}{2}}}\] are displayed.

\textbf{Example 1} $P(\rho_\epsilon)$ and the initial conditions are
chosen as
\[\begin{array}{ccc}&P(\rho_\epsilon)=\rho_\epsilon^{2},&\\
\rho_\epsilon(x,0)=1,& p_\epsilon(x,0)=1-\epsilon^2/2&x\in[0,0.2]\cup[0.8,1];\\
\rho_\epsilon(x,0)=1+\epsilon^2,&
p_\epsilon(x,0)=1&x\in(0.2,0.3];\\
\rho_\epsilon(x,0)=1,&
p_\epsilon(x,0)=1+\epsilon^2/2&x\in(0.3,0.7]\\
\rho_\epsilon(x,0)=1-\epsilon^2,&
p_\epsilon(x,0)=1&x\in(0.7,0.8]\end{array}\] This example consists
of several Riemann problems. Shocks and contact discontinuities are
stronger when $\epsilon$ is bigger. We first check the difference of
the three schemes (\ref{ellipticdis}), (\ref{ellipticdis1}) and
(\ref{ellipticdis2}). The CFL condition for the linearized reduced
stencil scheme (\ref{ellipticdis2}) is discussed in (ii) and a fixed
Courant number independent of $\epsilon$ is found numerically.
Compared with the first order ICE method using local Lax-Friedrich
discretization for (\ref{ICE1}), the improvement of removing
nonphysical oscillation of our scheme is shown. We investigate the
effect of $\alpha$ for different values of $\epsilon$ in (iii). In
(iv), when $\alpha=1$, we numerically test the uniform convergence
order. Finally, the AP property and its advantages are demonstrated
in (v) by comparing with the fully explicit Lax-Fridrich scheme for
the initial Isentropic Euler equation (\ref{hype}).

\begin{figure}
\begin{center}
\includegraphics[width=0.4\textwidth]{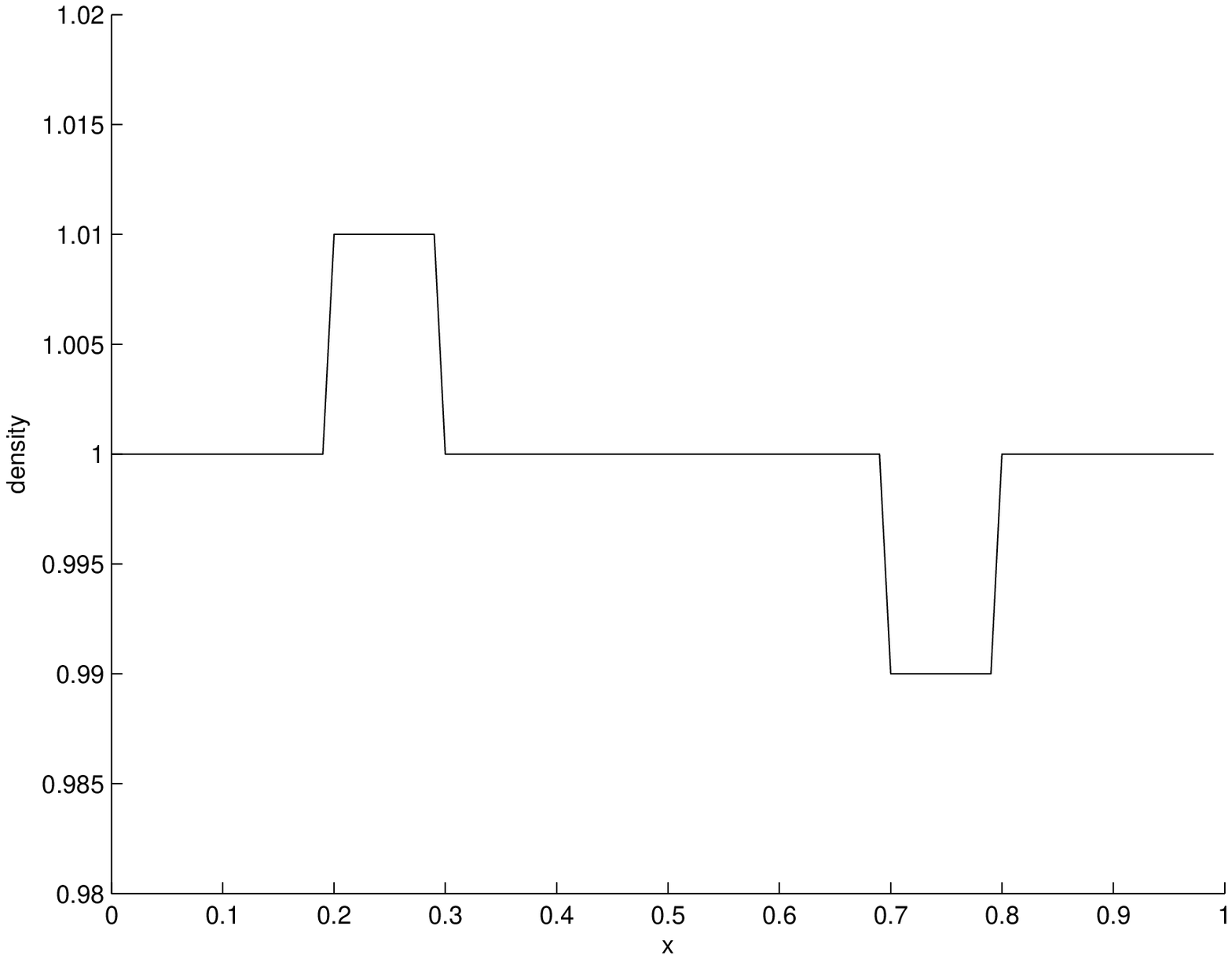}
\includegraphics[width=0.4\textwidth]{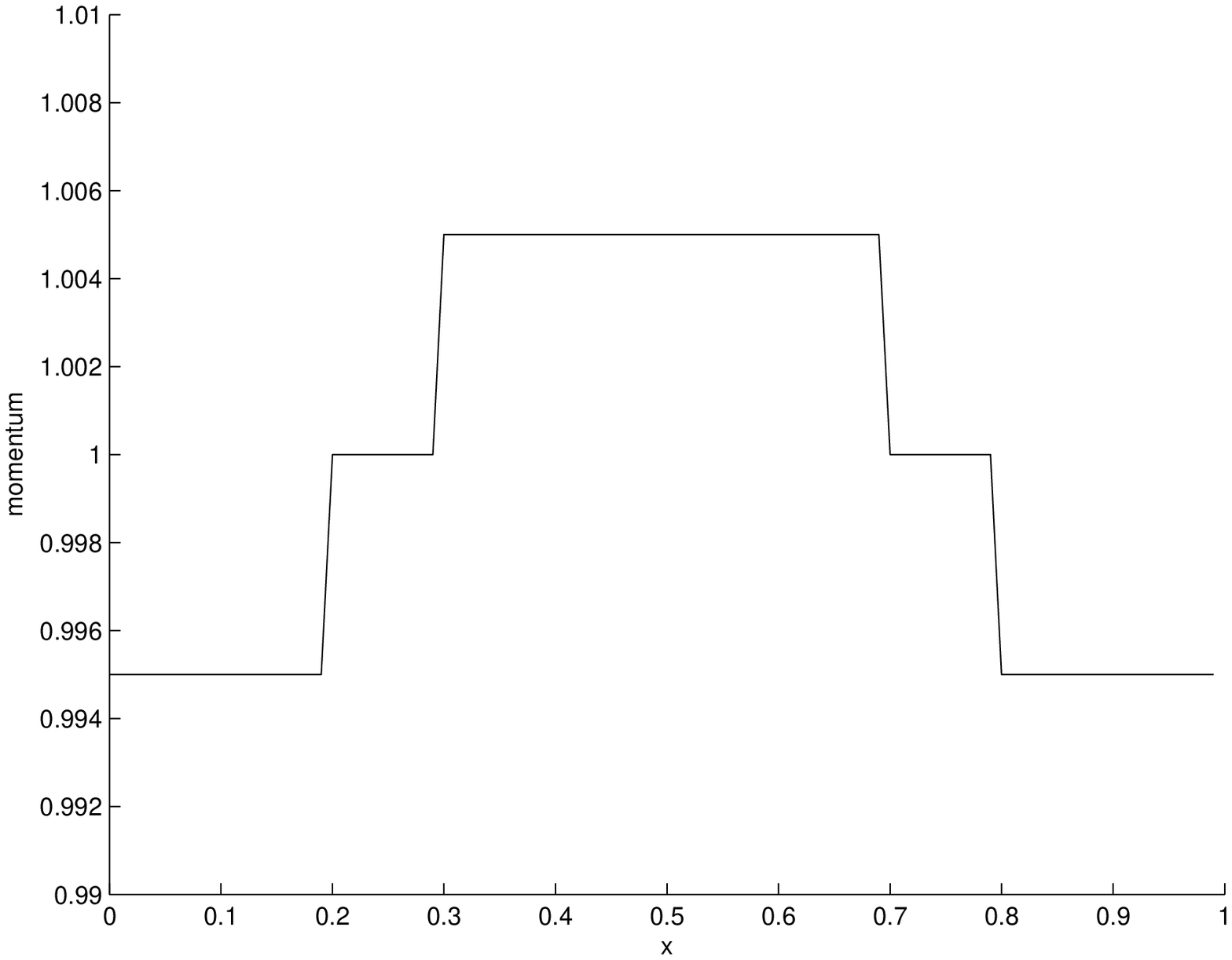}
\caption{Example 1. When $\epsilon=0.1$, the initial density and
momentum are displayed.} \label{figure1}
\end{center}
\end{figure}

When $\epsilon=0.1$, the initial density and momentum are displayed
in Figure \ref{figure1} and we can see the discontinuities clearly.

\begin{itemize}
\item[(i)] In this example, we choose $\epsilon=0.8,0.3,0.05$
corresponding to the compressible, intermediate and incompressible
regimes. The numerical results at $T=0.05$ of "NL", "L" and "LD" are
represented in Figure \ref{figure11}. Here $\Delta t$ is chosen to
make all these three schemes stable and diminishing $\Delta t$ only
will not improve much the numerical accuracy. The reference solution
is calculated by an explicit Lax-Friedrich method \cite{KT,KT2} with
$\Delta x=1/500,\Delta t=1/20000$. We can see that all these three
methods can capture the right shock speed. The results of the three
schemes are quite close, which implies that the linearization idea
does simplify the scheme but the "LD" scheme does not really
introduce less diffusion. When $\epsilon$ is small, though we can no
longer capture all the details of the waves, the error is of the
order $\Delta x$ which is the maximum information one can expect.
Numerically, for different scales of $\epsilon$, there is not much
difference between these three methods. Thus in the following one
dimensional examples, we only test the performance of the "LD"
scheme.
\begin{figure}
\begin{center}
a)\includegraphics[width=0.4\textwidth]{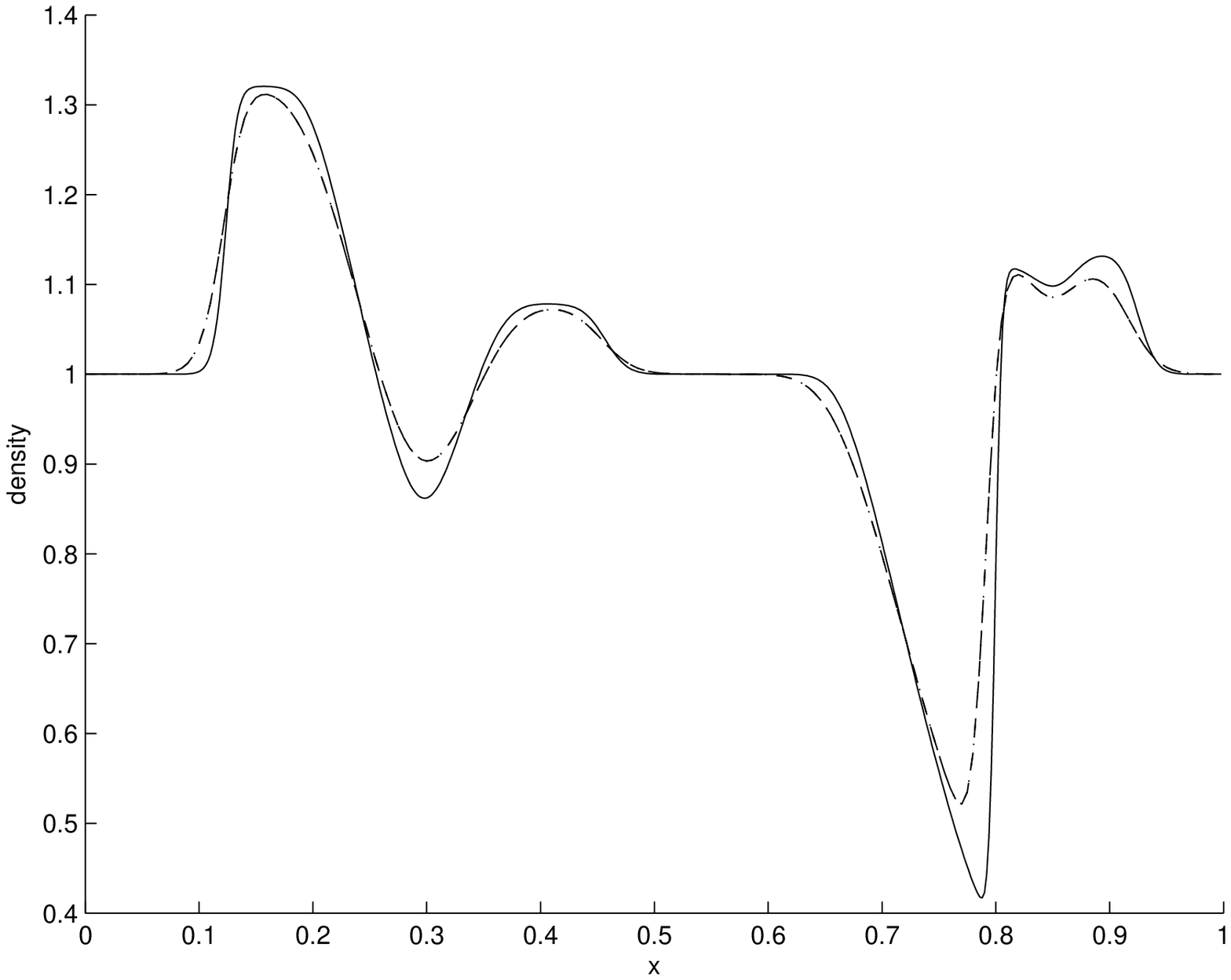}
\includegraphics[width=0.4\textwidth]{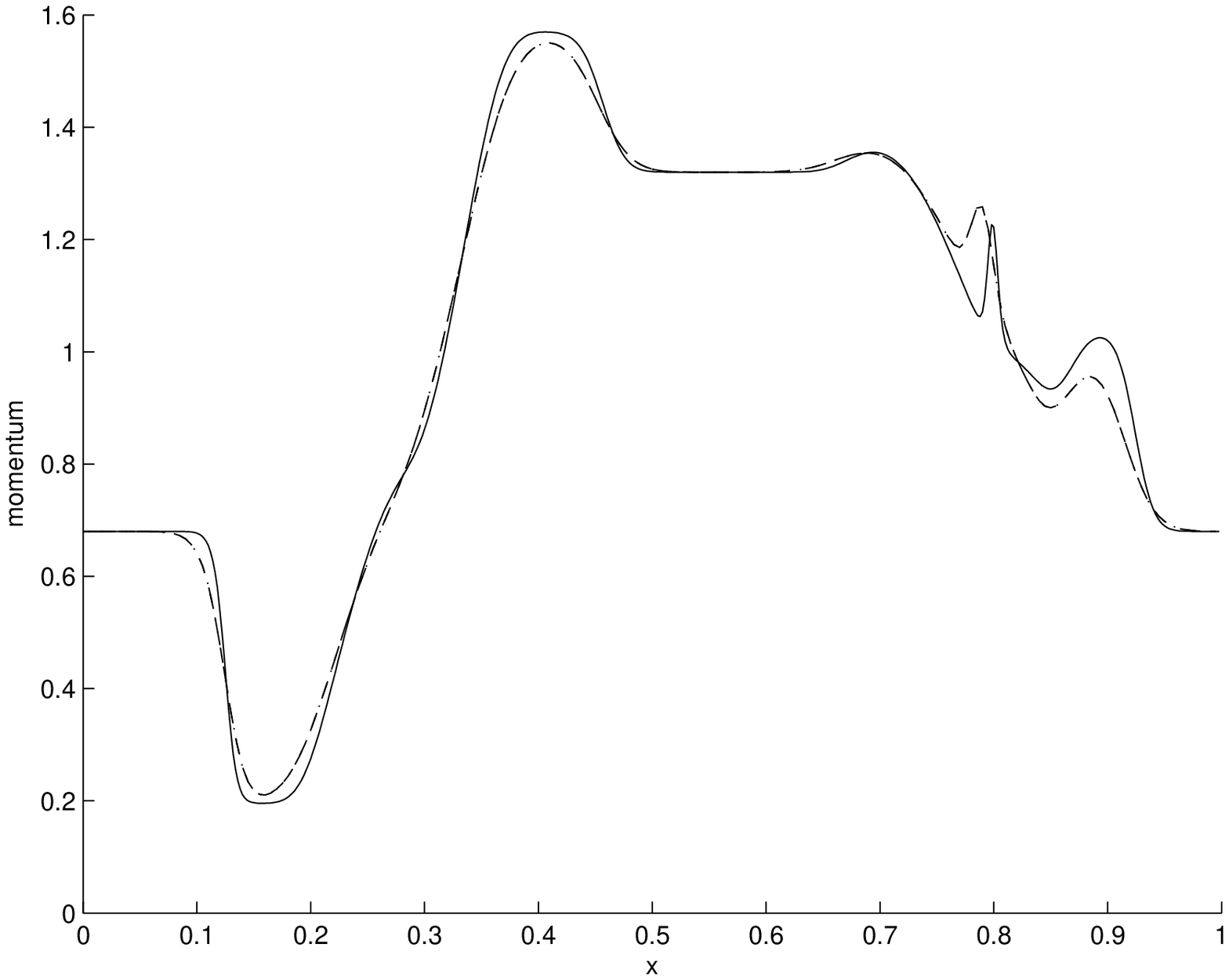}
b)\includegraphics[width=0.4\textwidth]{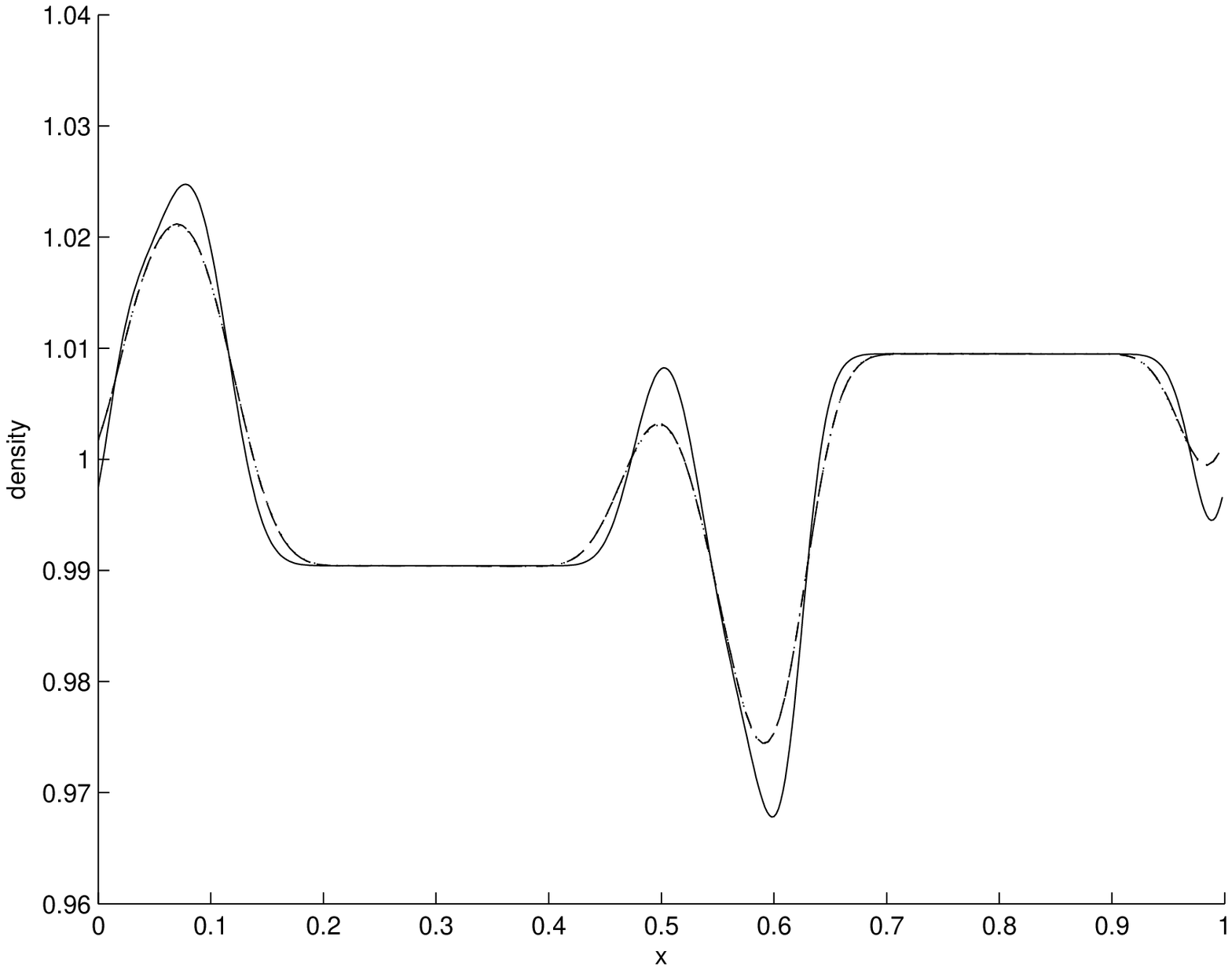}
\includegraphics[width=0.4\textwidth]{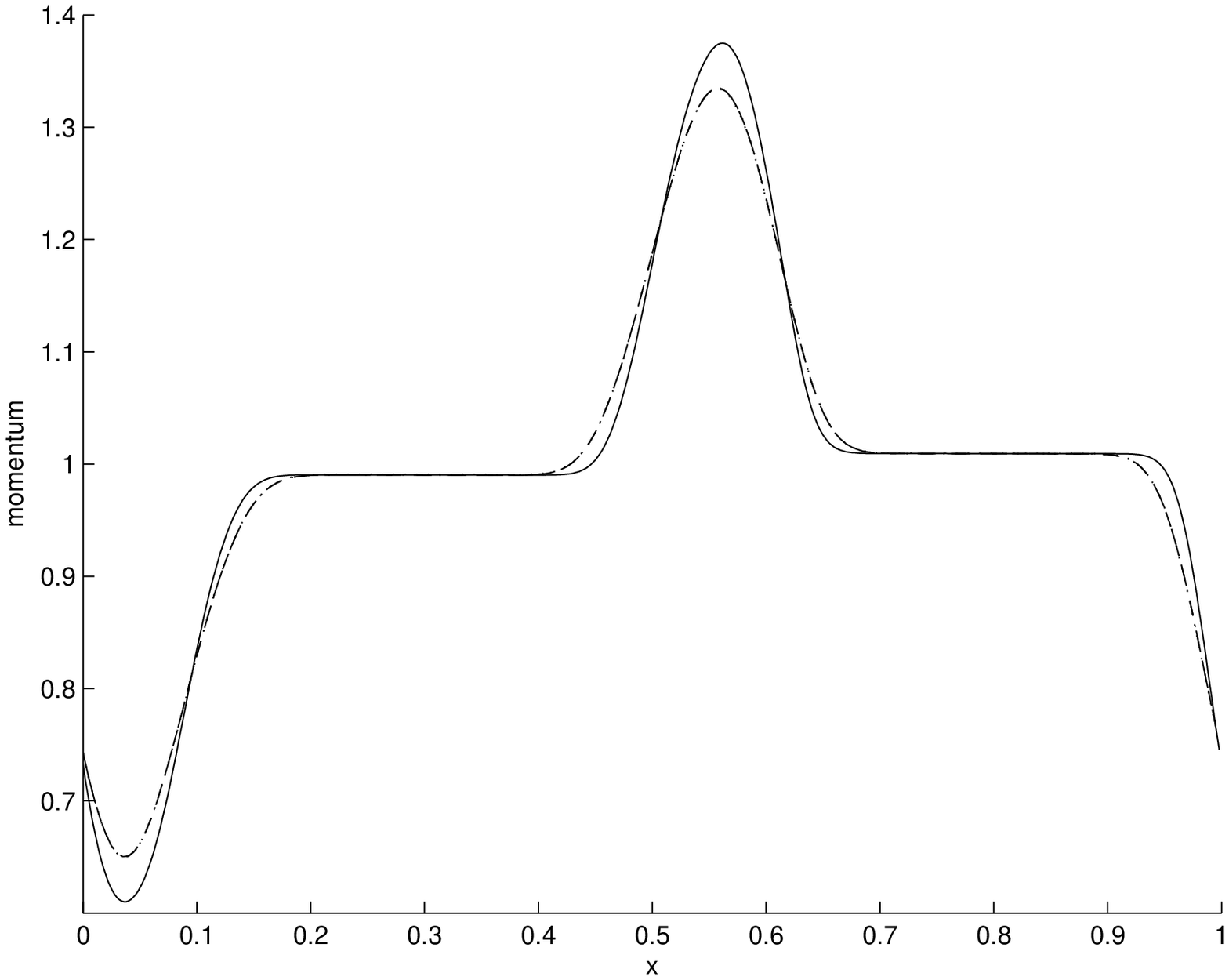}
c)\includegraphics[width=0.4\textwidth]{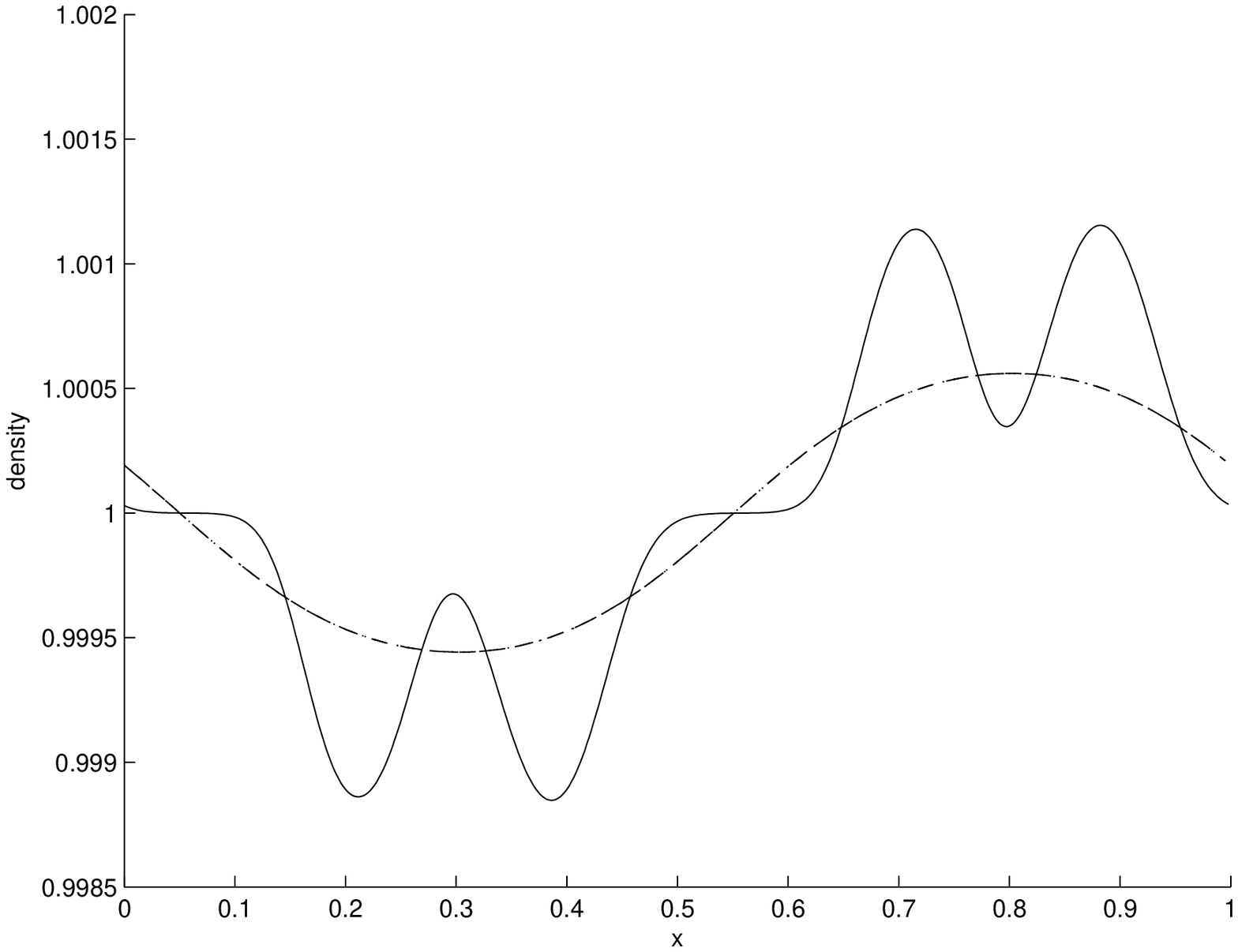}
\includegraphics[width=0.4\textwidth]{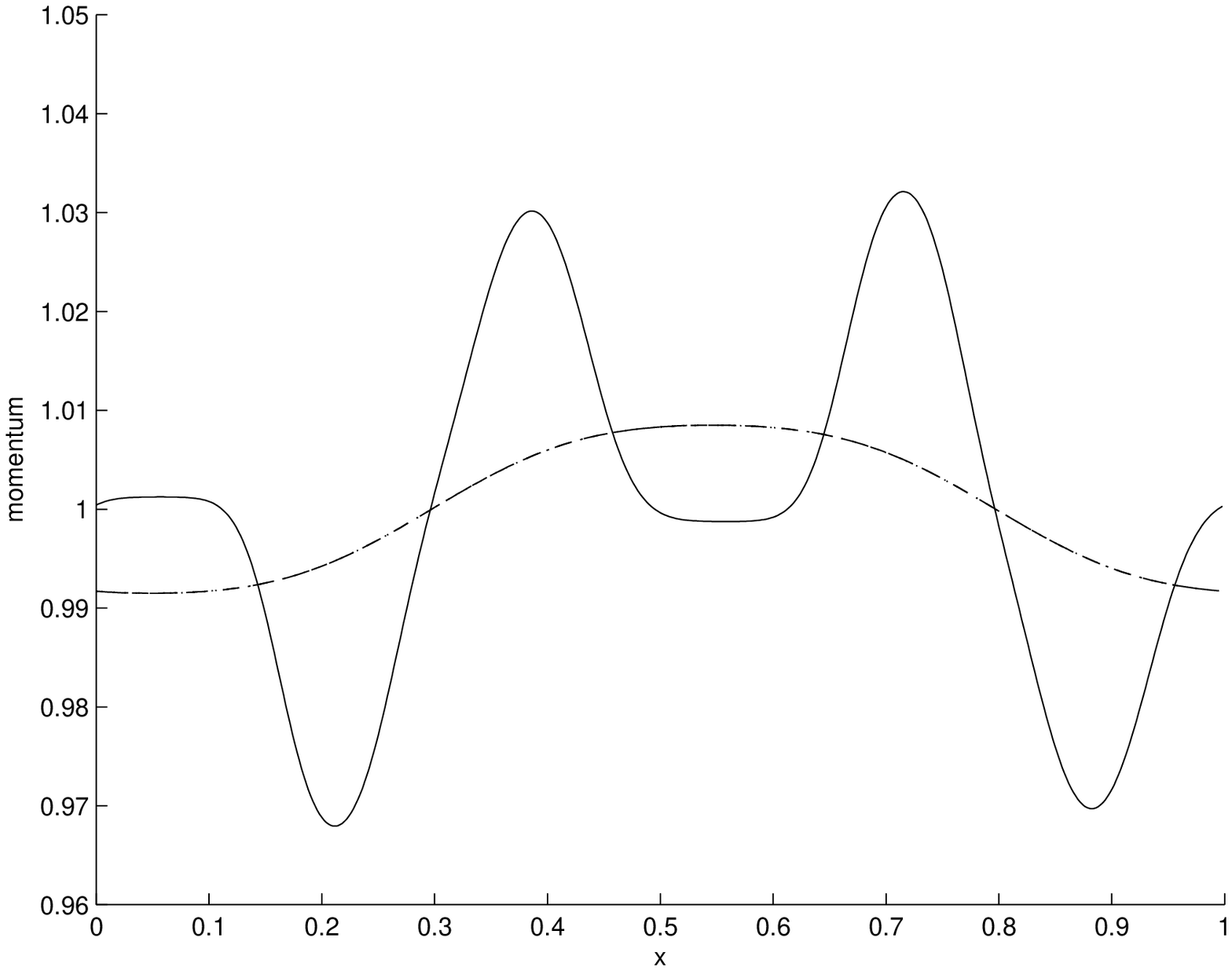}
\caption{Example 1. When $T=0.05,\Delta x=1/200,\Delta t=1/2000$,
the density and momentum of the "NL", "L" and "LD" schemes for
isentropic Euler equation are represented respectively by dashed,
dash dotted, and dotted lines. The solid line is the reference
solution calculated by an explicit Lax-Friedrich method
\cite{KT,KT2} with $\Delta x=1/500,\Delta t=1/20000$. a):
$\epsilon=0.8$; b): $\epsilon=0.3$; c): $\epsilon=0.05$. Left:
density; Right: momentum. For all $\epsilon$'s, these three lines
are so close to each other that '-.-.' and '...' are not visible in
the figure.} \label{figure11}
\end{center}
\end{figure}

\item[(ii)]Because of the explicit treatment of the flux terms in the
momentum equation, the stability of the 'LD' scheme can be only
guaranteed
 under the following CFL condition
\begin{equation}\Delta t\leq\sigma\min_i\frac{\Delta
x}{|\mathbf{u}_i|+\sqrt{\alpha
P'(\rho_\epsilon)}}.\end{equation}Here $0<\sigma<1$ is the Courant
number and is set up at initialization. Consistently with the fact
that these three methods are AP, the Courant number does not depend
on $\epsilon$. Indeed, below, we numerically verify that $\sigma$ is
independent of $\epsilon$. For $\epsilon=0.8,0.3,0.05$, the
numerical Courant numbers are displayed in Table \ref{table1_1} and
we can see numerically that the biggest allowed
$\max\{u\}\frac{\Delta t}{\Delta x}$ are close to 1 for all
 $\epsilon$'s. Therefore, $\sigma=0.9$ is enough to guarantee stability
and is numerically shown to be independent of $\epsilon$. By
contrast, the explicit local Lax-Friedrich scheme for the original
Euler equation has a stability condition which becomes more and more
restrictive as $\epsilon$ goes to zero. Thus the CFL condition of
the standard hyperbolic solver $\Delta t=O(\epsilon\Delta x)$ is
considerably improved.
\begin{table}[h]\begin{center}
\begin{tabular}{|c c c c c c|}\hline
$\epsilon$&$\max{\lambda}$&$\Delta x$&stable$\Delta t$&$\frac{\Delta x}{\Delta t}$&$u\frac{\Delta t}{\Delta x}$\\
\hline $0.8$&$4.24$&1/100&1/340&3.40&1.25\\\hline
$0.8$&$6.35$&1/200&1/970&4.85&1.31\\\hline
$0.8$&$6.58$&1/400&1/2420&6.05&1.09\\\hline
$0.8$&$6.70$&1/800&$1/5460$&6.82&0.982\\\hline
$0.3$&$2.64$&1/100&1/260&2.60&1.02\\\hline
$0.3$&$2.70$&1/200&1/510&2.55&1.06\\\hline
$0.3$&$2.76$&1/400&1/1000&2.50&1.10\\\hline
$0.3$&$2.81$&1/800&$1/2050$&2.56&1.10\\\hline
$0.05$&$2.43$&1/100&1/260&2.60&0.93\\\hline
$0.05$&$2.44$&1/200&1/490&2.45&1.00\\\hline
$0.05$&$2.45$&1/400&1/960&2.40&1.02\\\hline
$0.05$&$2.46$&1/800&$1/1920$&2.40&1.03\\\hline
\end{tabular}
\end{center}\caption{Example 1. The numerical Courant numbers for different $\epsilon$. Here $\max\{\lambda\}$
denotes the maximum of $\max\{\lambda_{j}\}$ defined in
(\ref{lambda}) until $T=0.1$ for all time steps.} \label{table1_1}
\end{table}

\item[(iii)]The classical ICE method even in its conservative form
introduces some nonphysical oscillations, no matter how small the
time step is. These oscillations cannot be diminished by decreasing
the time step. Their amplitude becomes smaller as the mesh is
refined as long as the scheme is stable. In this part we show that
our method can suppress these oscillations numerically by choosing
$\alpha=1$. When $T=0.01$, for $\epsilon=0.8,0.3,0.05$, the
numerical results of both our method with $\alpha=1$ and ICE
calculated by $\Delta x=1/200,\Delta t=1/20000$ are displayed in
Figure \ref{figure1_oscil}. The oscillations are more important for
the ICE method and smooth away when $\alpha=1$. We can see that
numerical nonphysical oscillations occur in the results of the ICE
method when $\epsilon=0.8,0.3$, but disappear when $\epsilon$
becomes small. This can also be seen from (\ref{difrealrho}),
(\ref{difrealq}). When $\epsilon$ is small the diffusion introduced
by the implicitness is bigger. These oscillations also go away as
time goes on due to dissipation.

\begin{figure}
\begin{center}
a)\includegraphics[width=0.6\textwidth]{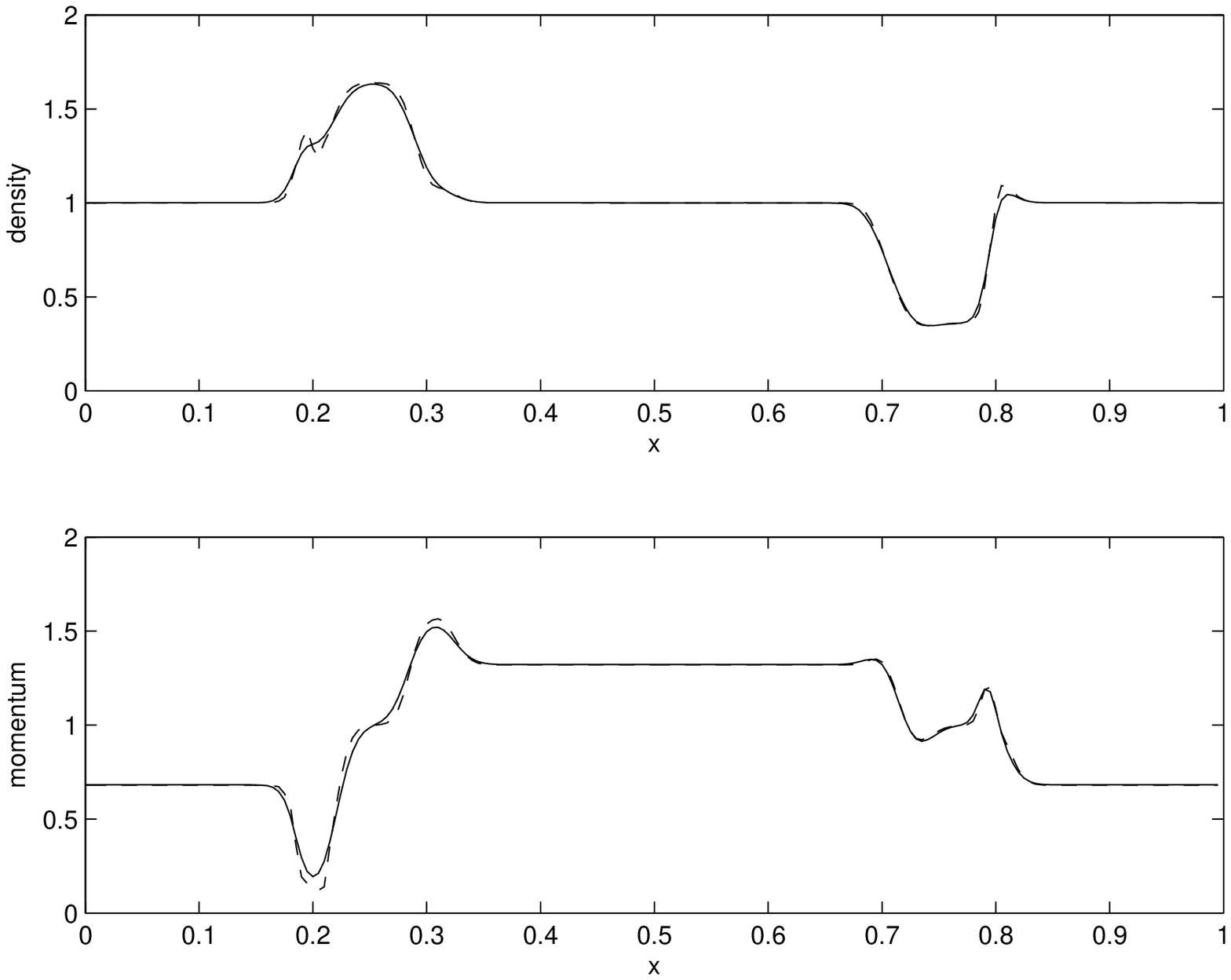}
b)\includegraphics[width=0.6\textwidth]{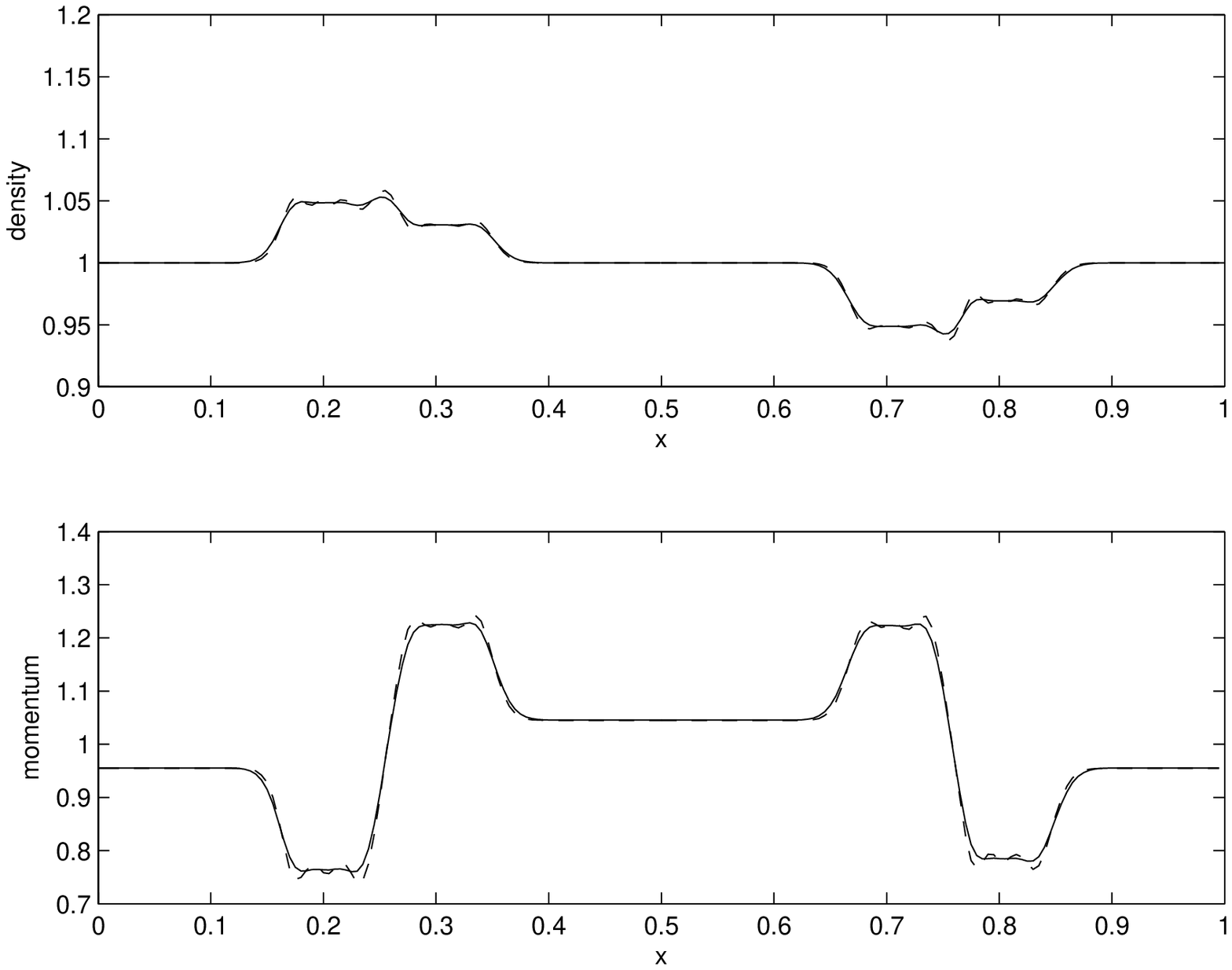}
c)\includegraphics[width=0.6\textwidth]{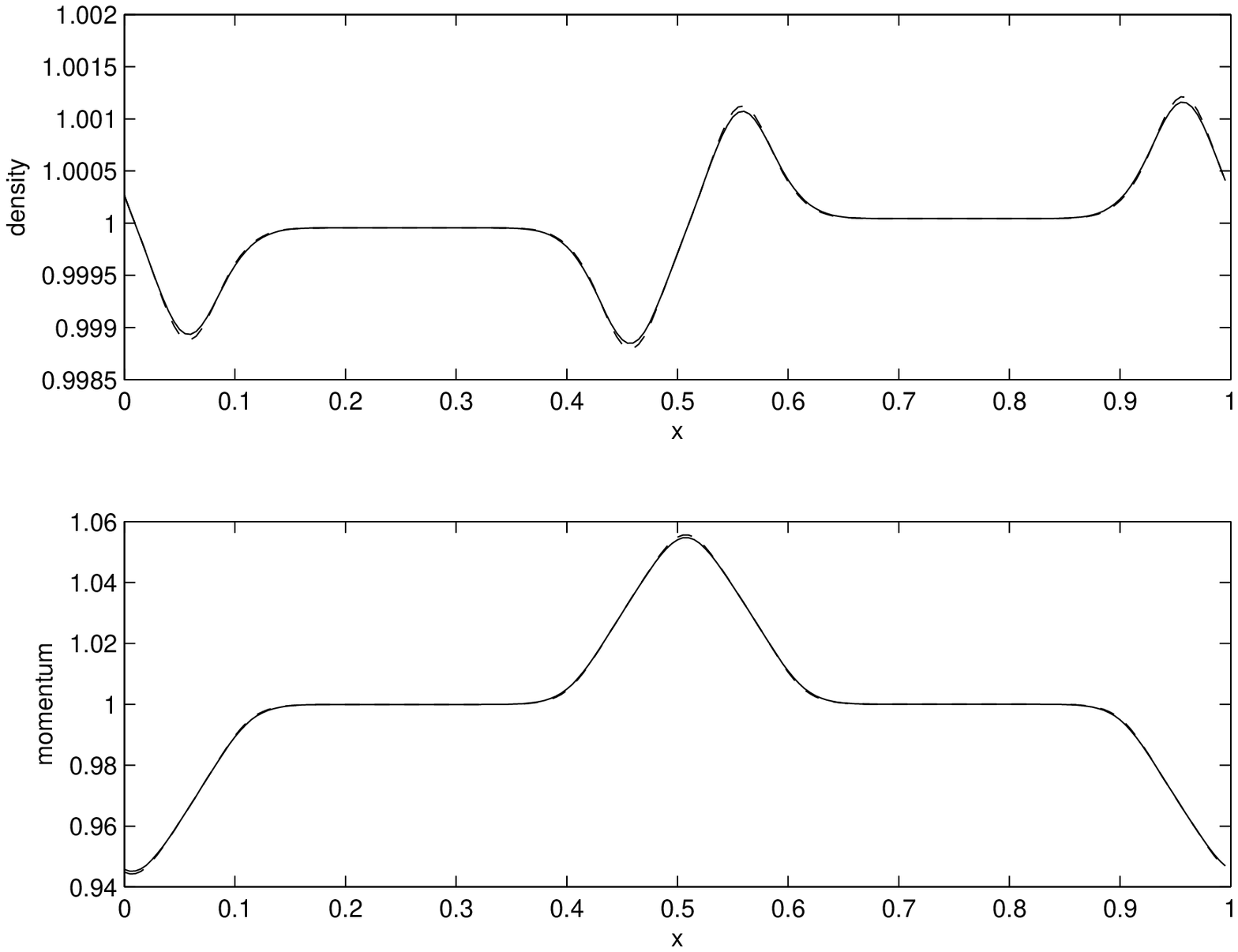}
\caption{Example 1. When $T=0.01$, the density and momentum for
different $\epsilon$ are presented. The solid and dashed lines are
the numerical results of our scheme and ICE with $\Delta
x=1/200,\Delta t=1/20000$ respectively. a) $\epsilon=0.8$; b)
$\epsilon=0.3$; c)$\epsilon=0.05$.} \label{figure1_oscil}
\end{center}
\end{figure}

\item[(iv)] When $\alpha=1$, the relative errors of the "LD" scheme for different $\Delta x$, $\Delta
t$ at time $T=0.1$ are shown in Table \ref{table1_2}. Here $\Delta
x,\Delta t$ do not need to resolve $\epsilon$ and the reference
solution is obtained by the explicit LLF scheme calculated with a
very fine mesh $\Delta x=1/1280,\Delta t=1/128000$. We can see that
good numerical approximations can be obtained without resolving the
small $\epsilon$. The convergence order is $1/2$ when $\Delta
t/\Delta x$ is fixed, uniformly with respect to $\epsilon$. This
convergence order when there are discontinuities is the same as the
explicit LLF \cite{Leveque}. We can see from Table \ref{table1_2}
that refinement in the time step does not improve the accuracy much
(provided the Courant number is appropriately small, like
$\sigma=0.7$). Take $\epsilon=0.8$ as an example. When $\Delta
x=1/320$, in order to obtain stability, $\Delta t$ should be less
than $1/1920$. It is demonstrated in Table \ref{table1_2} that the
error calculated with $\Delta x=1/320$ does not decrease much when
$\Delta t$ is changed from $1/2880$ to $1/12800$. Thus as long as
the scheme is stable, we cannot use a smaller $\Delta t$ to obtain a
better accuracy. This feature is the same as for standard hyperbolic
solvers.
\begin{table}[h]\begin{center}
\begin{tabular}{|c c c c c c c|}\hline
$\epsilon$&$\Delta x$&$\Delta t$&$e(\rho_\epsilon)$&ratio&$e(p_\epsilon)$&ratio\\
\hline $0.8$&$1/20$&$1/180$&$9.739*10^{-1}$&-&$1.197$&-\\\hline
$0.8$&$1/40$&$1/360$&$5.959*10^{-1}$&1.63&$7.484*10^{-1}$&1.16\\\hline
$0.8$&$1/80$&$1/720$&$3.467*10^{-1}$&1.72&$4.180*10^{-1}$&1.31\\\hline
$0.8$&$1/160$&$1/1440$&$1.985*10^{-1}$&1.75&$2.048*10^{-1}$&1.36\\\hline
$0.8$&$1/320$&$1/2880$&$1.126*10^{-1}$&1.76&$8.477*10^{-2}$&1.79\\\hline
$0.8$&$1/320$&$1/12800$&$ 1.126*10^{-1}$&-&$8.539*10^{-2}$&-\\\hline
$0.05$&$1/20$&$1/70$&$4.679*10^{-3}$&-&$1.355*10^{-1}$&-\\\hline
$0.05$&$1/40$&$1/140$&$3.305*10^{-3}$&1.42&$9.574*10^{-2}$&1.42\\\hline
$0.05$&$1/80$&$1/280$&$2.353*10^{-3}$&1.40&$6.758*10^{-2}$&1.42\\\hline
$0.05$&$1/160$&$1/560$&$1.655*10^{-3}$&1.42&$4.430*10^{-2}$&1.53\\\hline
$0.05$&$1/320$&$1/1120$&$1.094*10^{-3}$&1.51&$2.538*10^{-2}$&1.75\\\hline
$0.05$&$1/320$&$1/12800$&$6.012*10^{-4}$&-&$9.303*10^{-3}$&-\\\hline
\end{tabular}
\end{center}\caption{Example 1. $T=0.1$, the $L^2$ norm of the relative
error between the reference solution which is calculated with a very
fine mesh $\Delta x=1/1280,\Delta t=1/128000$ and the numerical
results for different $\epsilon$ with different $\Delta x,\Delta t$
are displayed.}\label{table1_2}\end{table}
\item[(v)] We emphasize the AP property in this final part.
For $\epsilon=0.005$, the numerical results at $T=0.01$ with
unresolved mesh $\Delta x=1/20,\Delta t=1/500$ and resolved mesh
$\Delta x=1/2000,\Delta t=1/5000$ are displayed in Figure
\ref{figure1_3}, while the fully explicit Lax-Fridrich scheme is not
stable with the same mesh size. We do capture the incompressible
limit when $\Delta x,\Delta t$ do not resolve $\epsilon$.
\begin{figure}
\begin{center}
\includegraphics[width=0.4\textwidth]{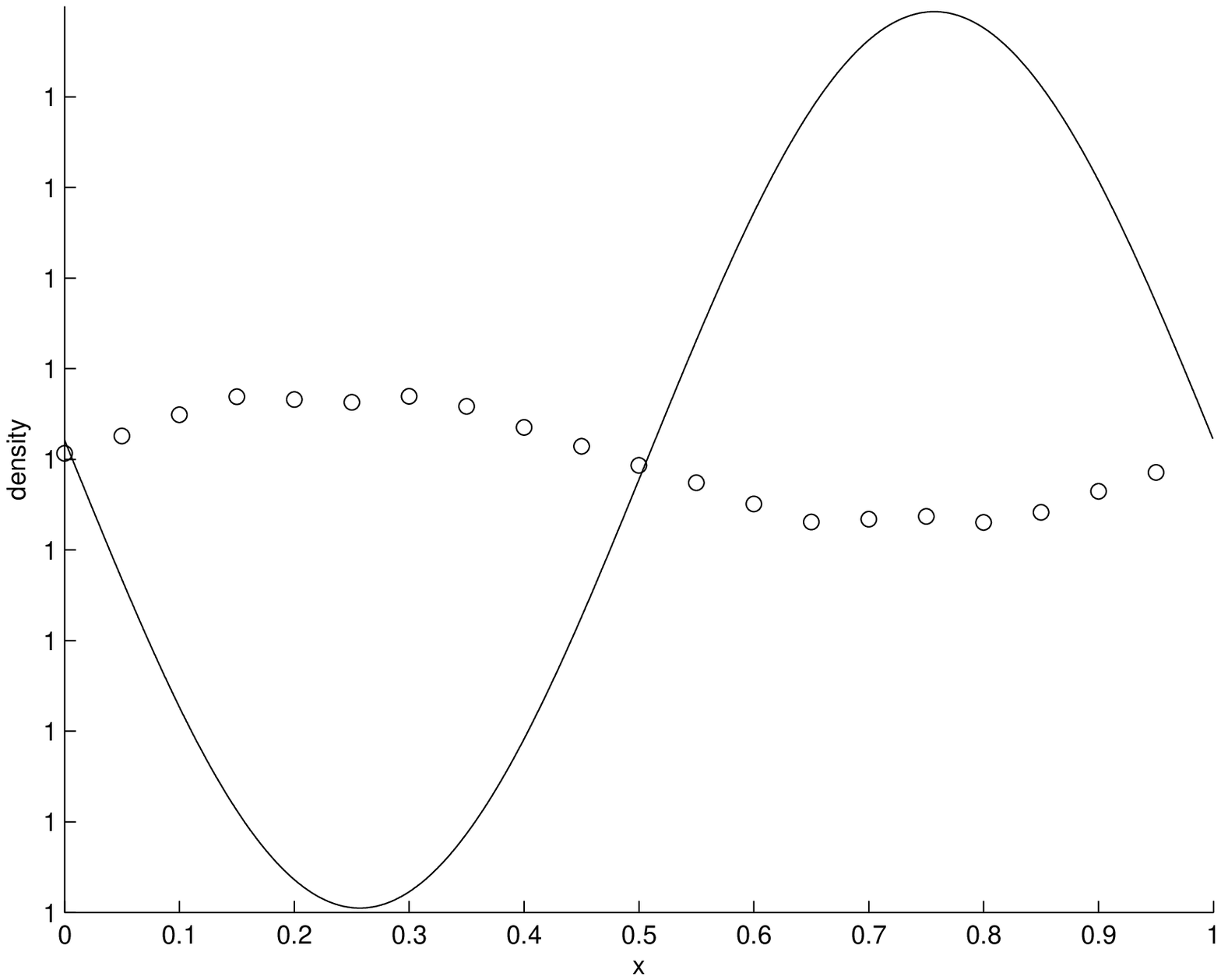}
\includegraphics[width=0.4\textwidth]{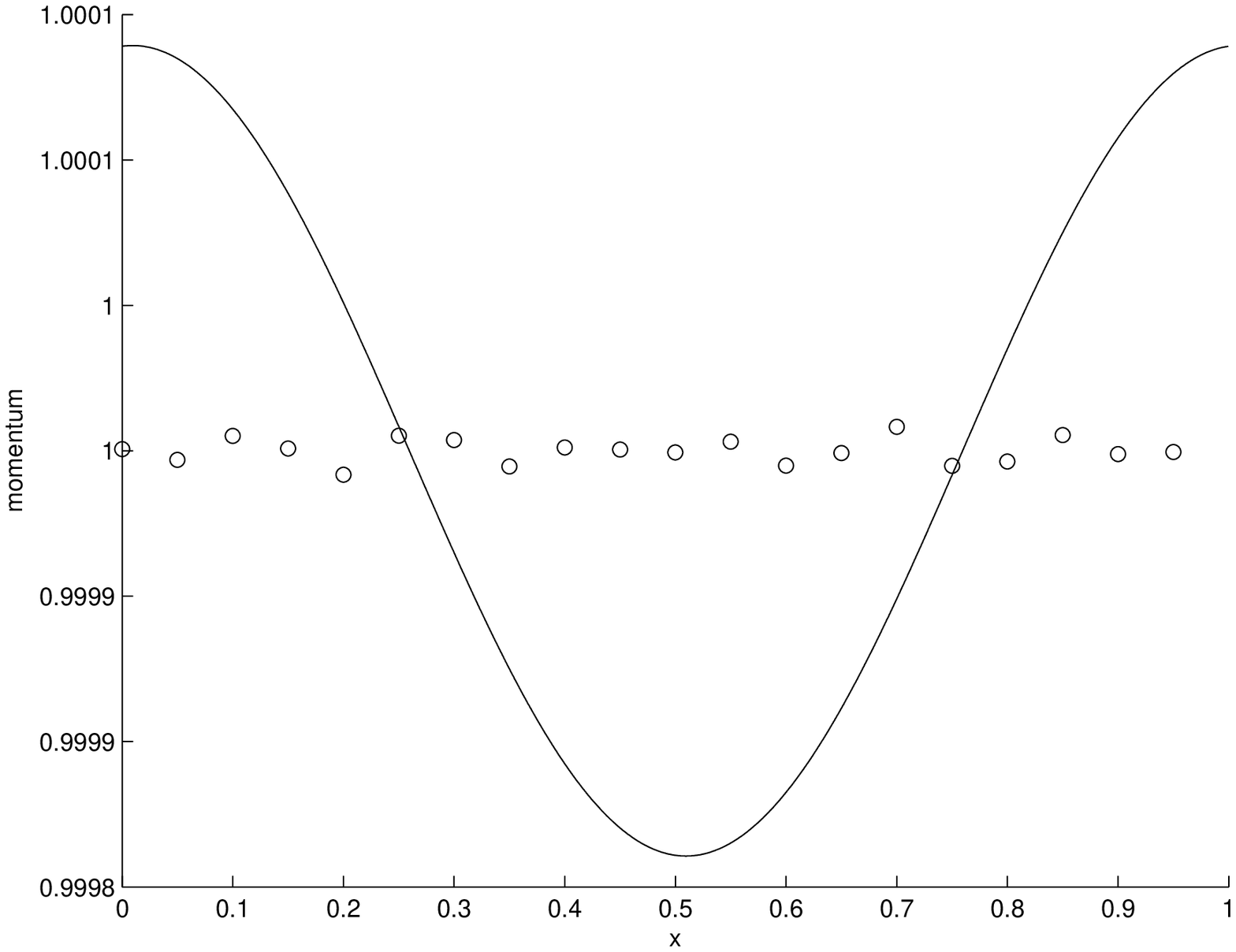}
 \caption{Example 1. By
using the "LD" scheme, the density (left) and momentum (right) for
$\epsilon=0.005$ at $T=0.01$ are represented. The circles are the
results for $\Delta x=1/20,\Delta t=1/500$ and the solid line is
calculated with $\Delta x=1/2000,\Delta t=1/5000$.}\label{figure1_3}
\end{center}
\end{figure}
\end{itemize}
\[\]
\textbf{Example 2:} In this example we simulate the evolution of two
collision acoustic waves by the "LD" scheme and test the
convergence. We choose $\alpha=1,\epsilon=0.1, \Delta x=1/100,\Delta
t=1/1000$. Here $\Delta t$ is chosen to stabilize the scheme and
decreasing $\Delta t$ alone will not improve much the numerical
accuracy. Similar to Klein's paper \cite{Klein}, $P(\rho_\epsilon)$
and the initial conditions are chosen as
\[\begin{array}{cc}P(\rho_\epsilon)=\rho_\epsilon^{1.4},&\mbox{for } x\in[-1,1]\\
\rho_\epsilon(x,0)=0.955+\frac{\epsilon}{2}\big(1-\cos(2\pi
x)\big),& u_\epsilon(x,0)=-\mbox{sign}(x)\sqrt{1.4}\big(1- \cos(2\pi
x)\big).
\end{array}\] The
initial density and momentum are displayed in Figure \ref{figure2}.
\begin{figure}
\begin{center}
\includegraphics[width=0.4\textwidth]{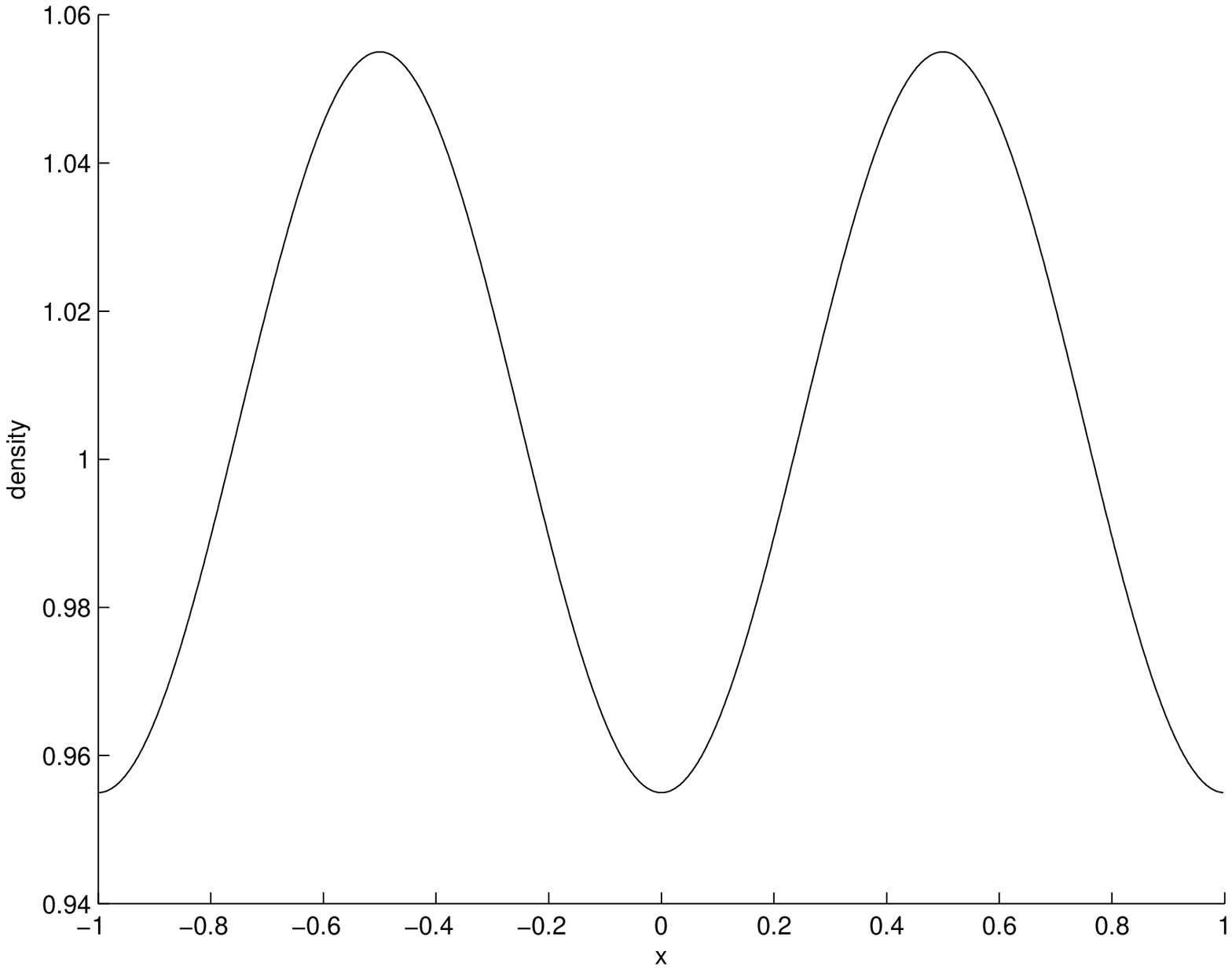}
\includegraphics[width=0.4\textwidth]{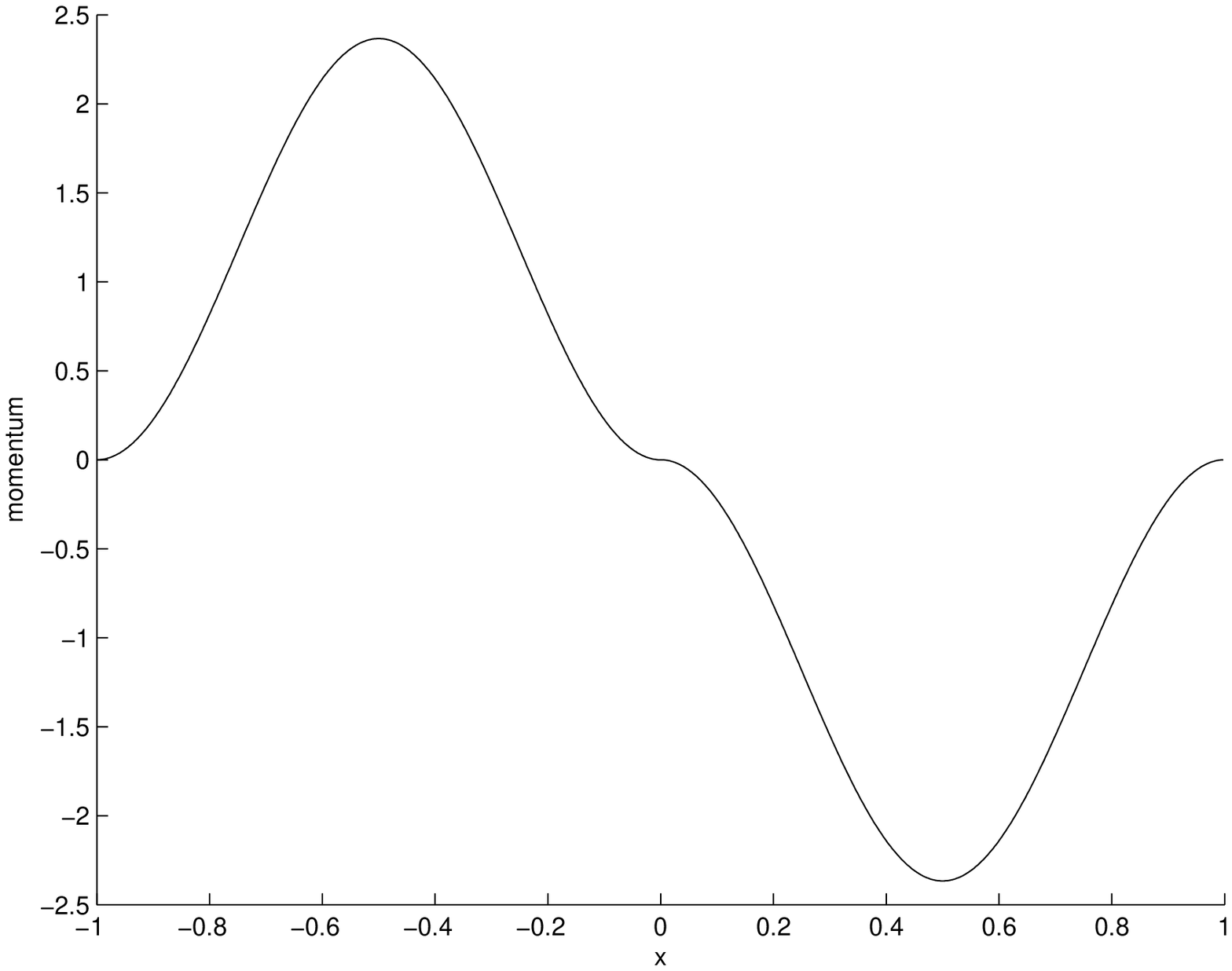}
\caption{Example 2. When $\epsilon=0.1$, the initial density and
momentum are displayed.}\label{figure2}
\end{center}
\end{figure}

\begin{figure}
\begin{center}
a)\includegraphics[width=0.4\textwidth]{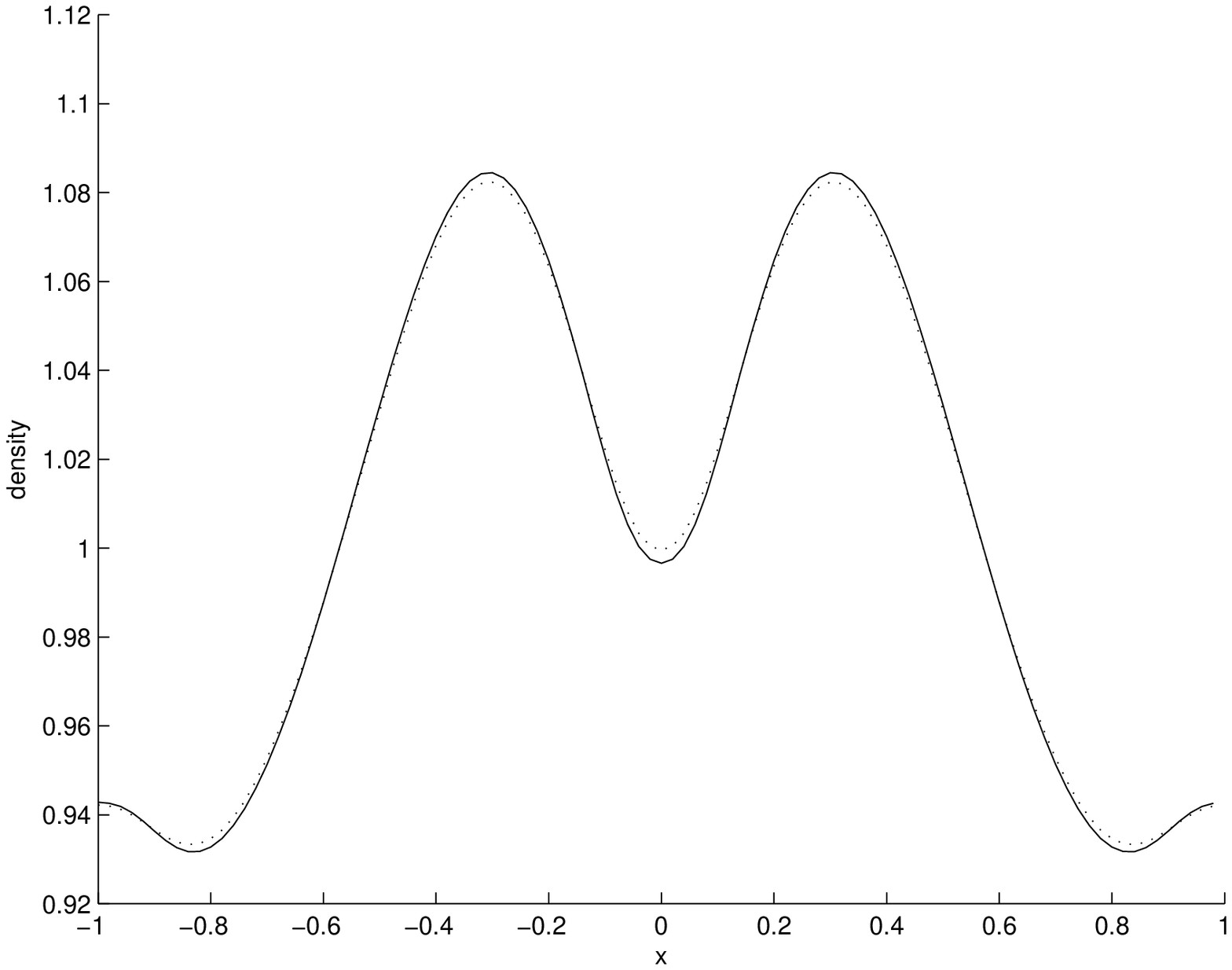}
\includegraphics[width=0.4\textwidth]{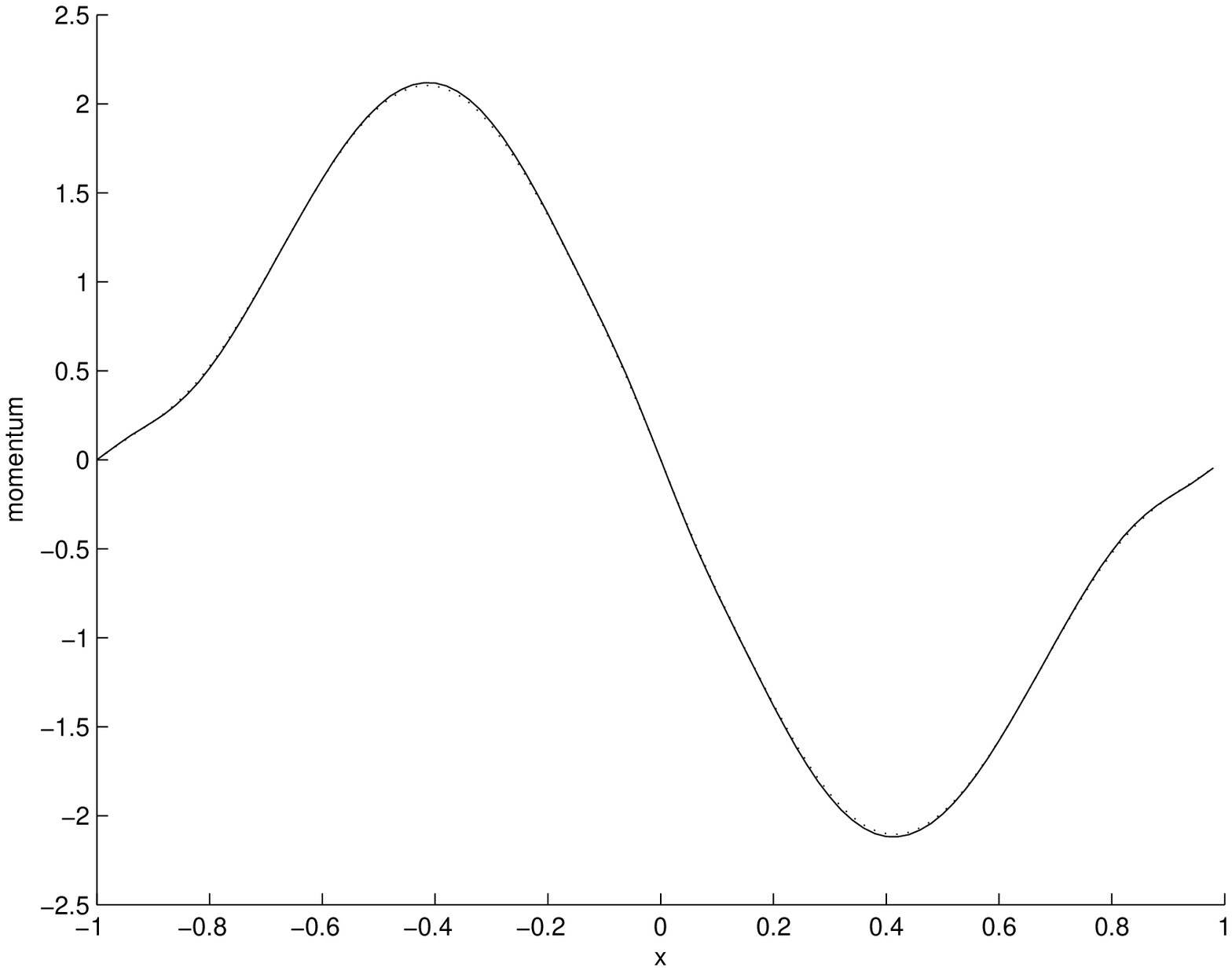}
b)\includegraphics[width=0.4\textwidth]{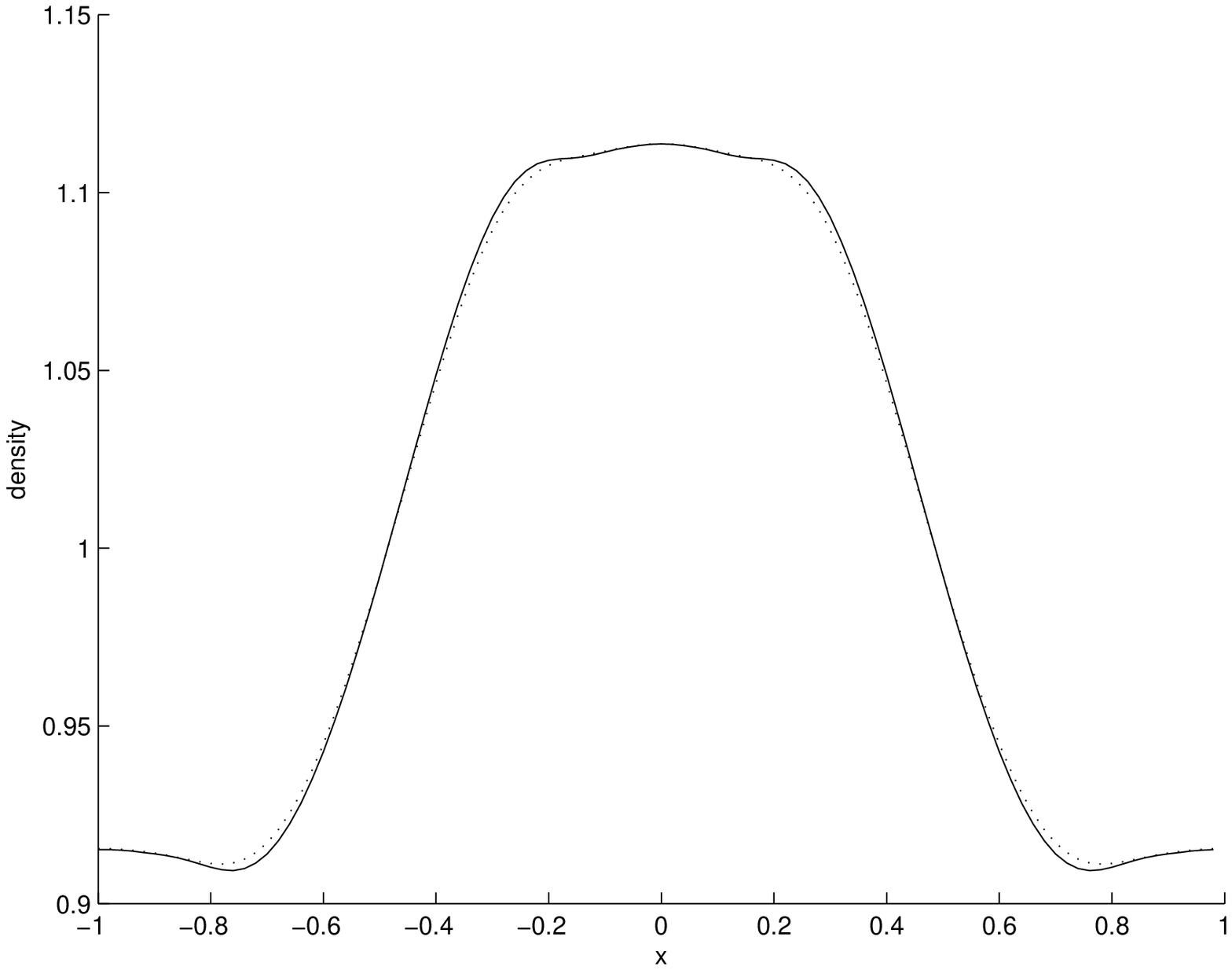}
\includegraphics[width=0.4\textwidth]{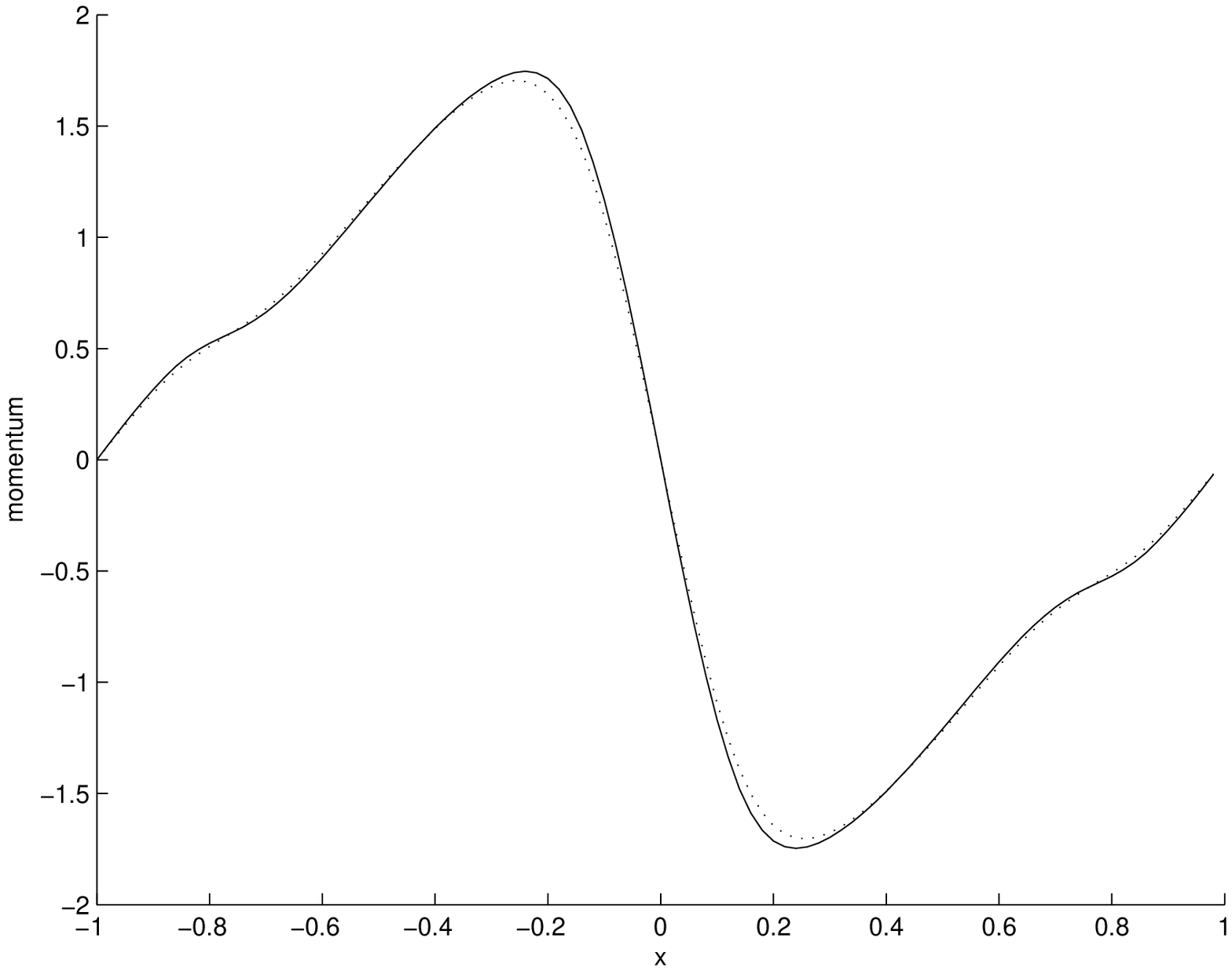}
c)\includegraphics[width=0.4\textwidth]{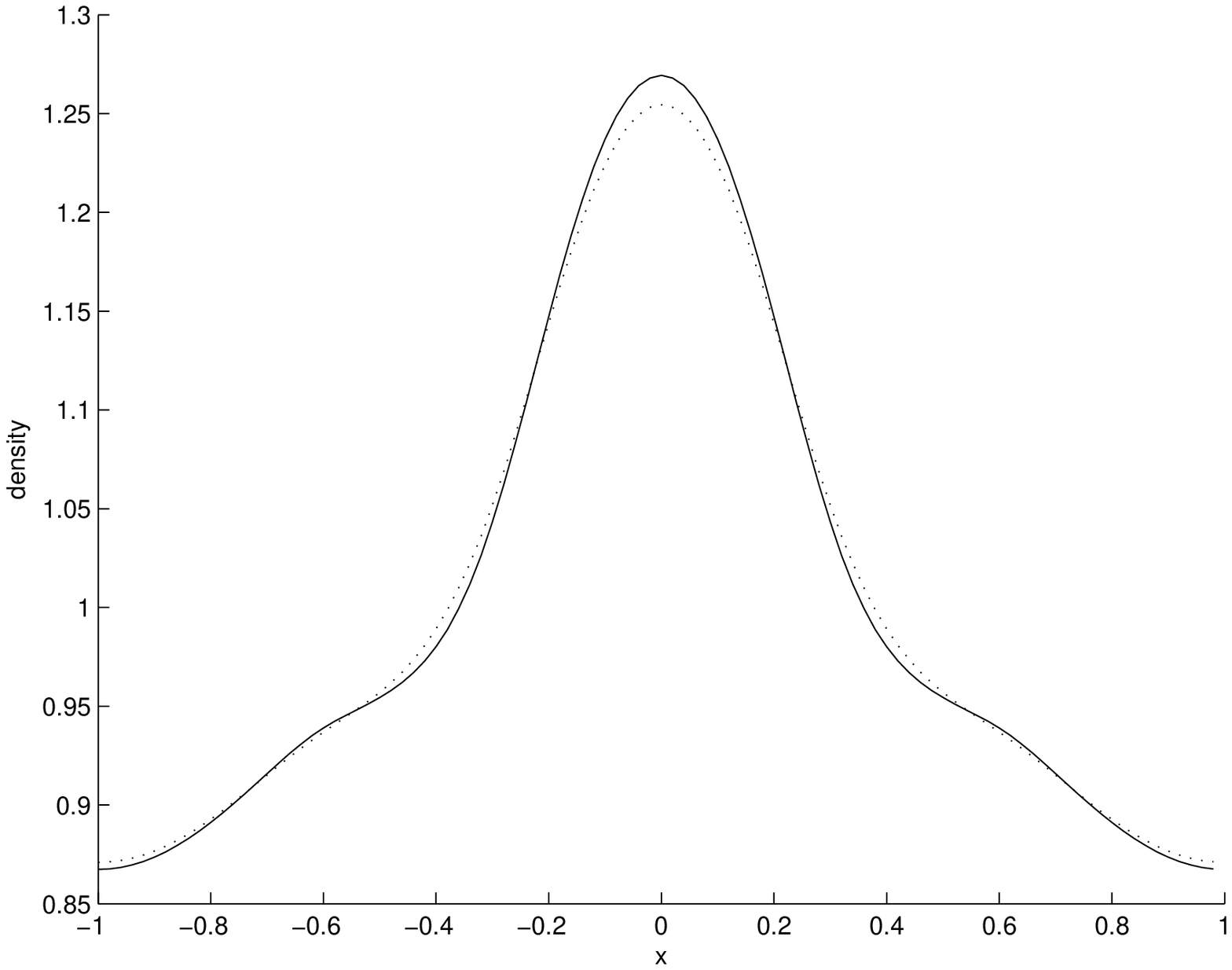}
\includegraphics[width=0.4\textwidth]{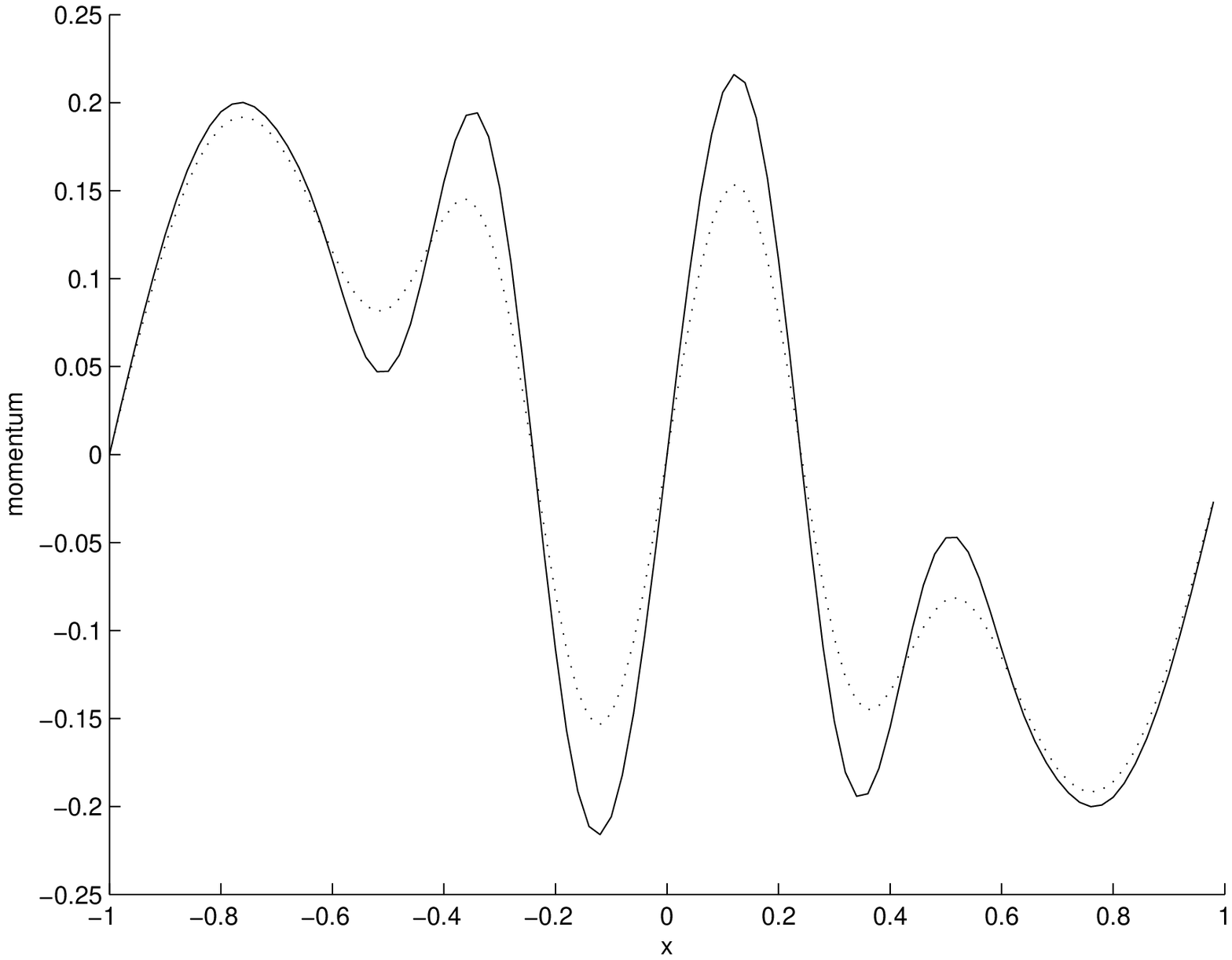}
d)\includegraphics[width=0.4\textwidth]{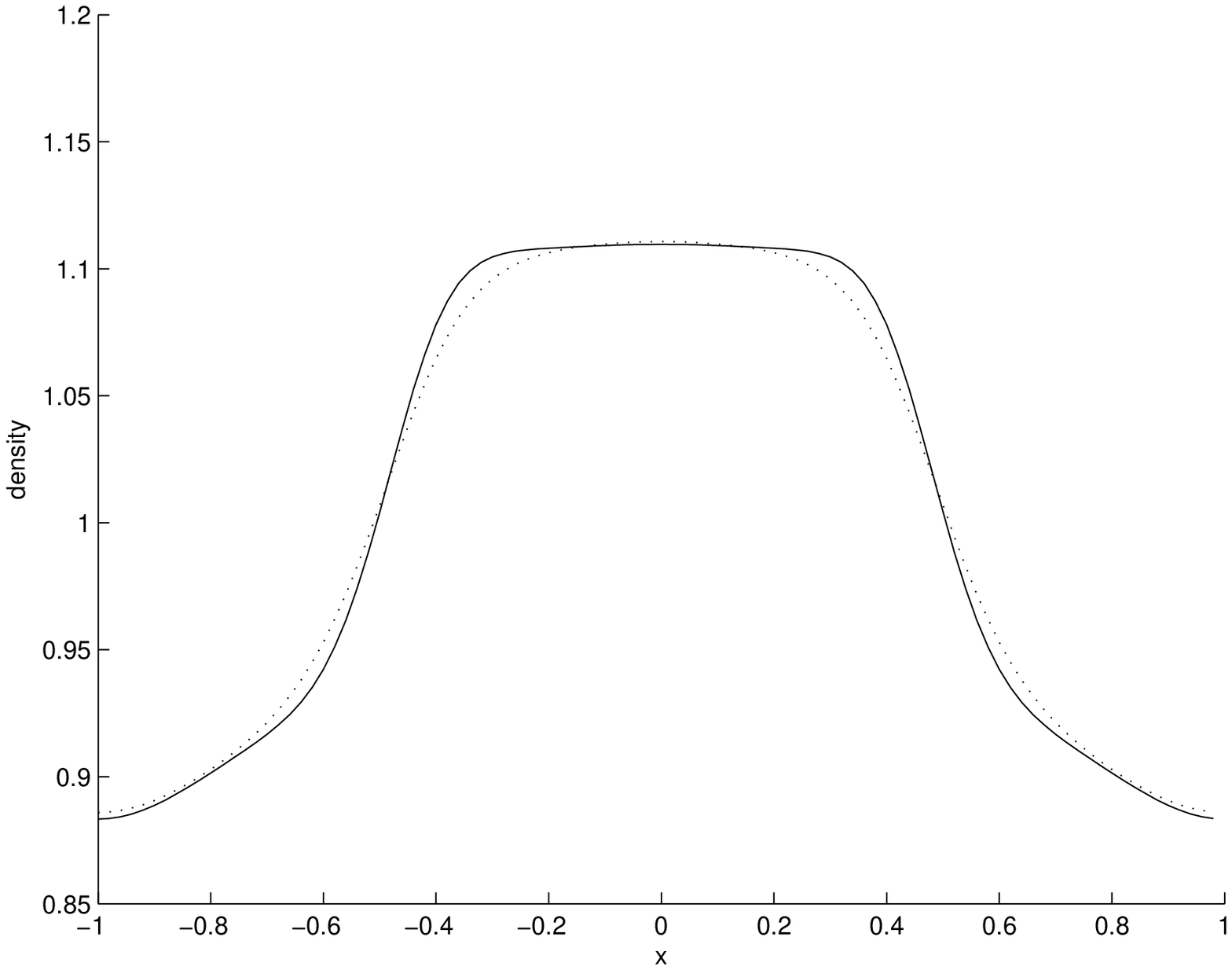}
\includegraphics[width=0.4\textwidth]{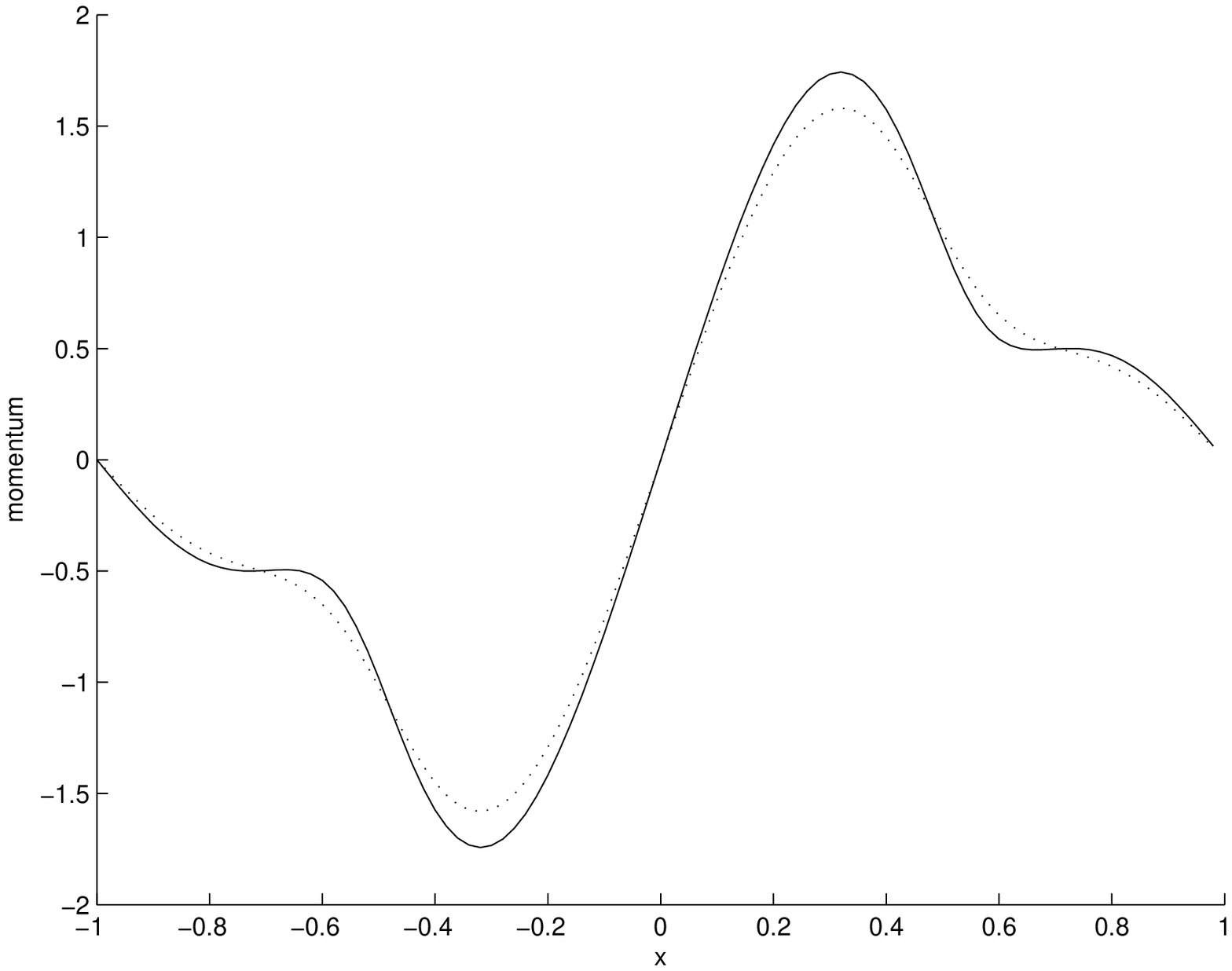}
e)\includegraphics[width=0.4\textwidth]{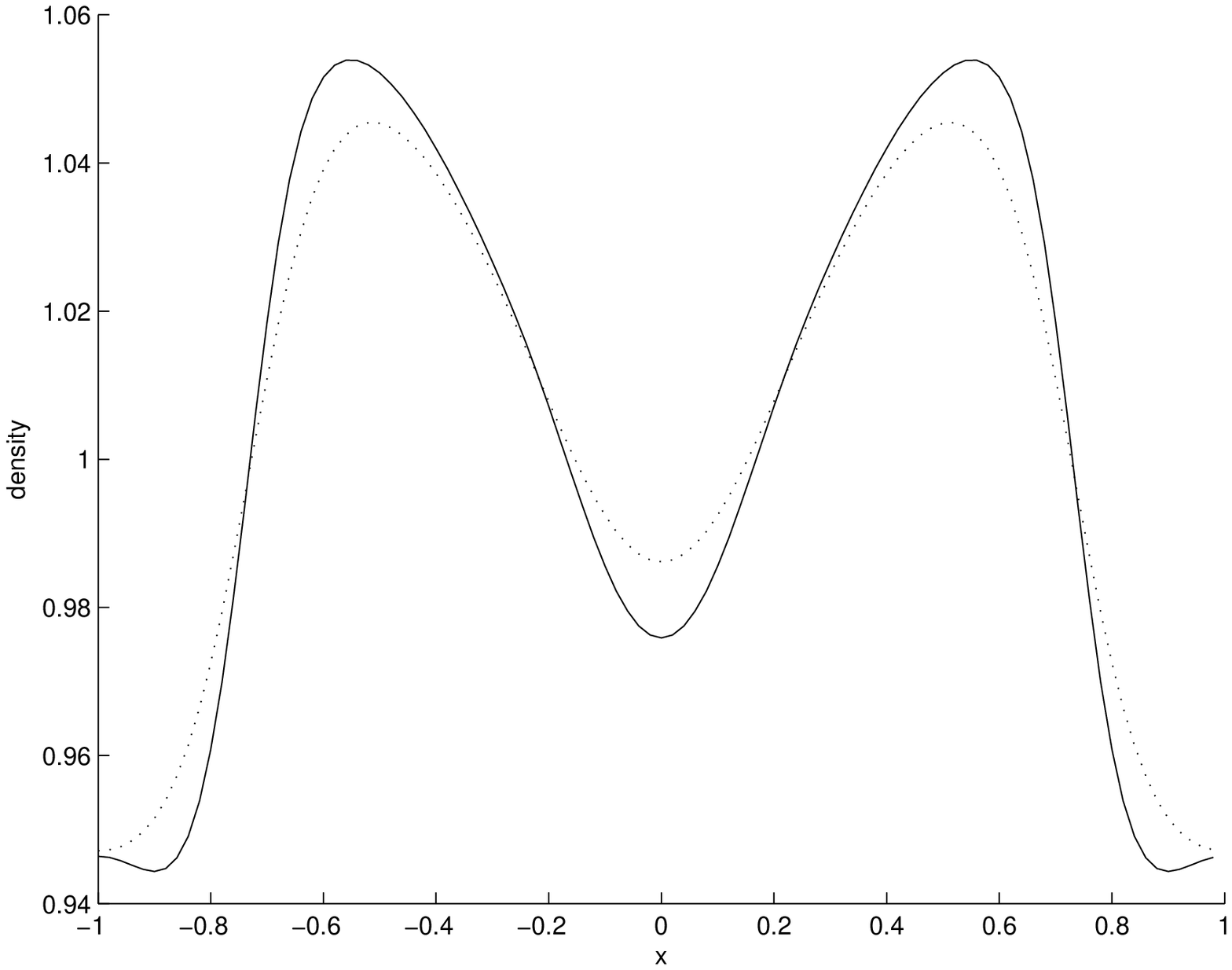}
\includegraphics[width=0.4\textwidth]{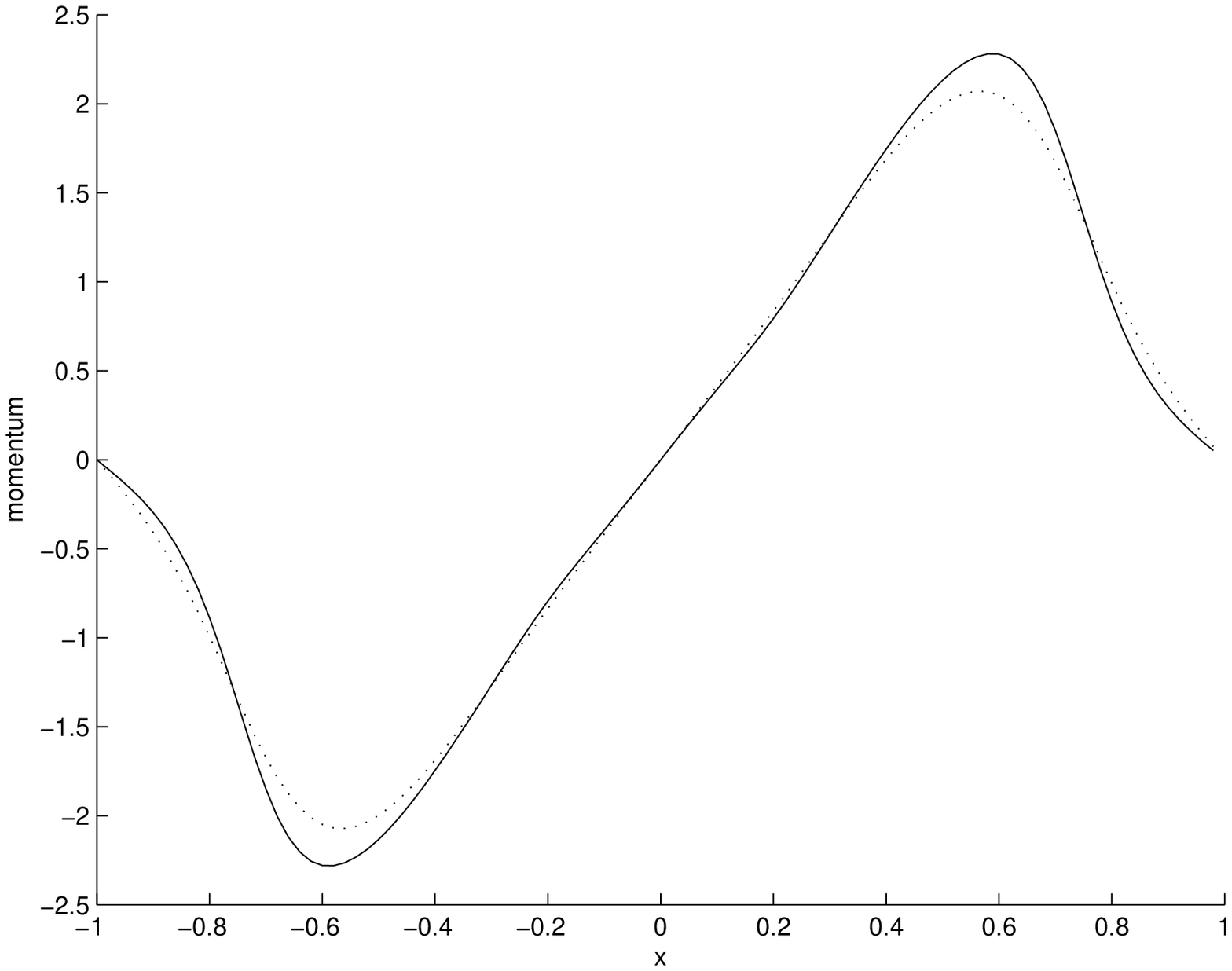}
\caption{Example 2. When $\epsilon=0.1$, the density and momentum of
the "LD" scheme at different times are represented: a) $T=0.01$; b)
$T=0.02$; c) $T=0.04$; d) $T=0.06$. e) $T=0.08$. All these pictures
correspond to $\Delta x=1/50,\Delta t=1/1000$.}\label{figure2_2}
\end{center}
\end{figure}

\begin{figure}
\begin{center}
a)\includegraphics[width=0.45\textwidth]{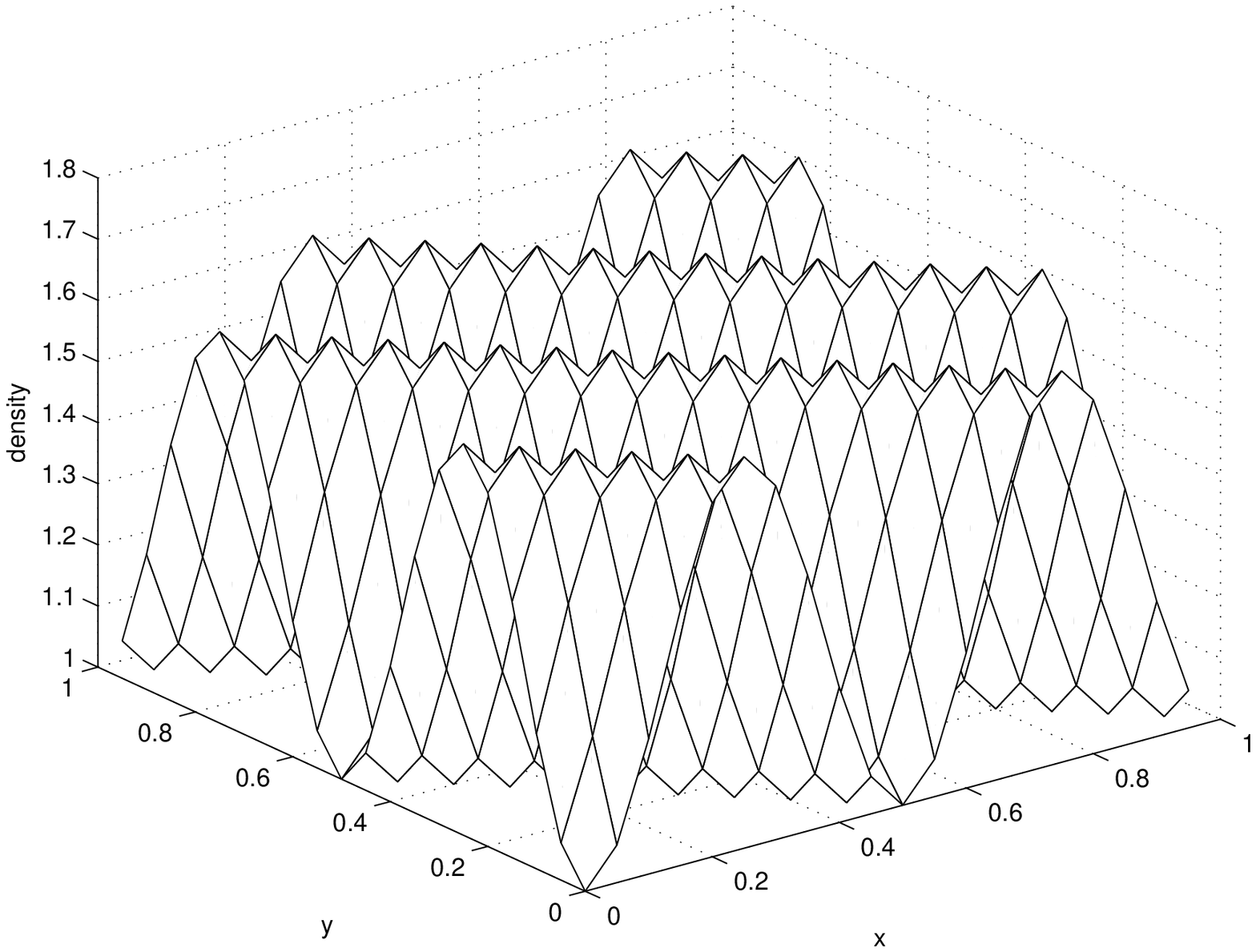}
\includegraphics[width=0.45\textwidth]{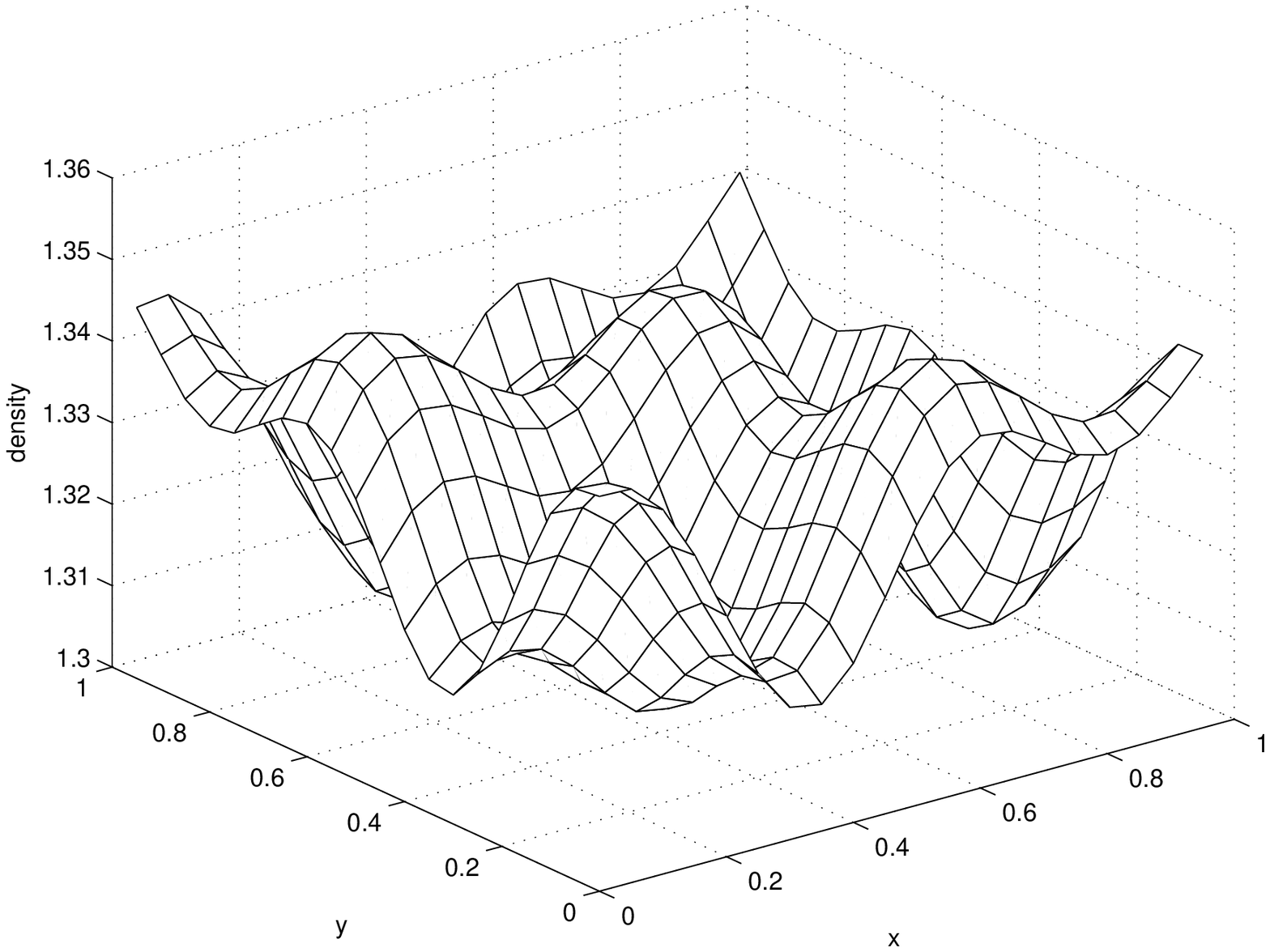}
b)\includegraphics[width=0.45\textwidth]{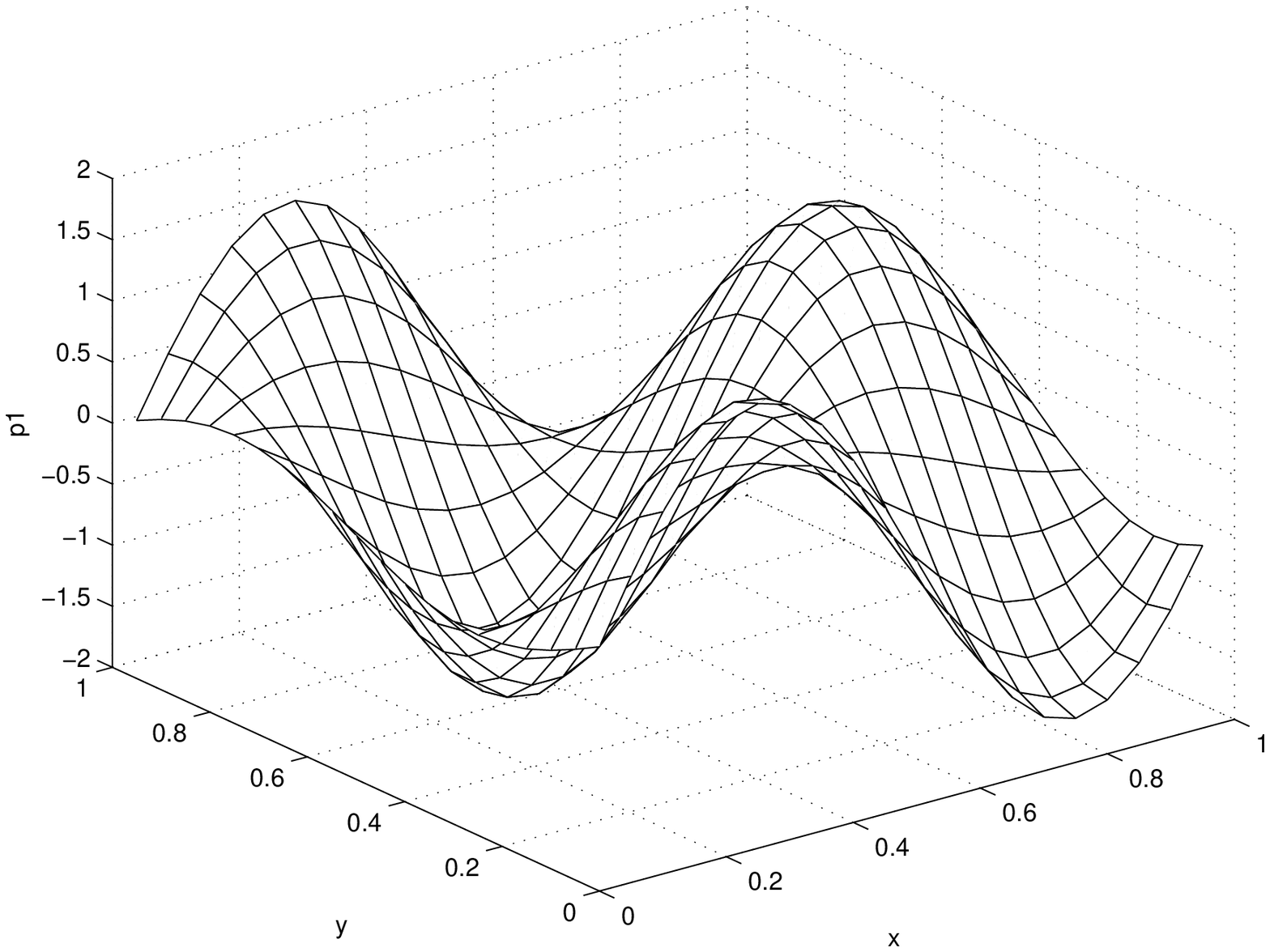}
\includegraphics[width=0.45\textwidth]{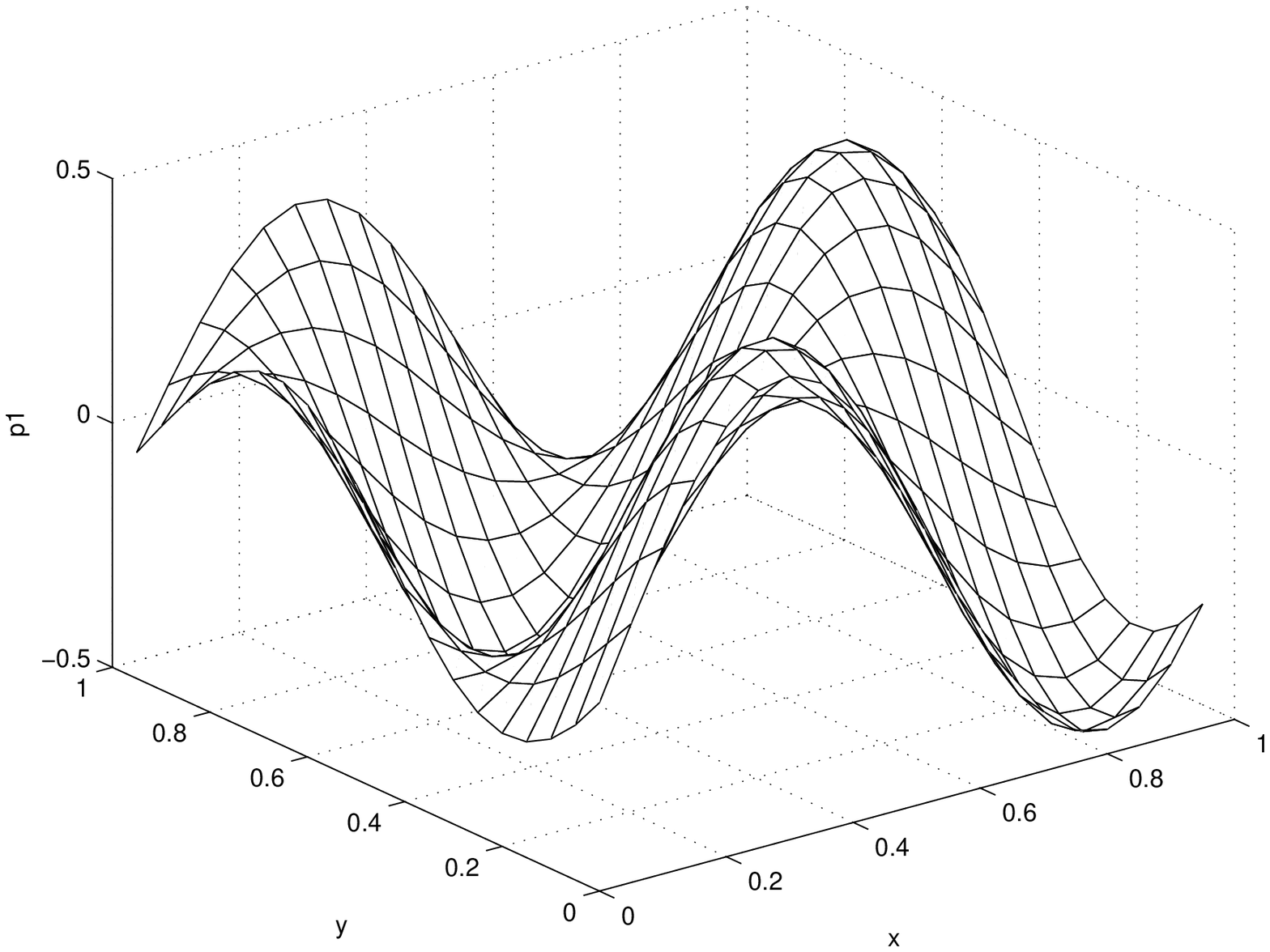}
c)\includegraphics[width=0.45\textwidth]{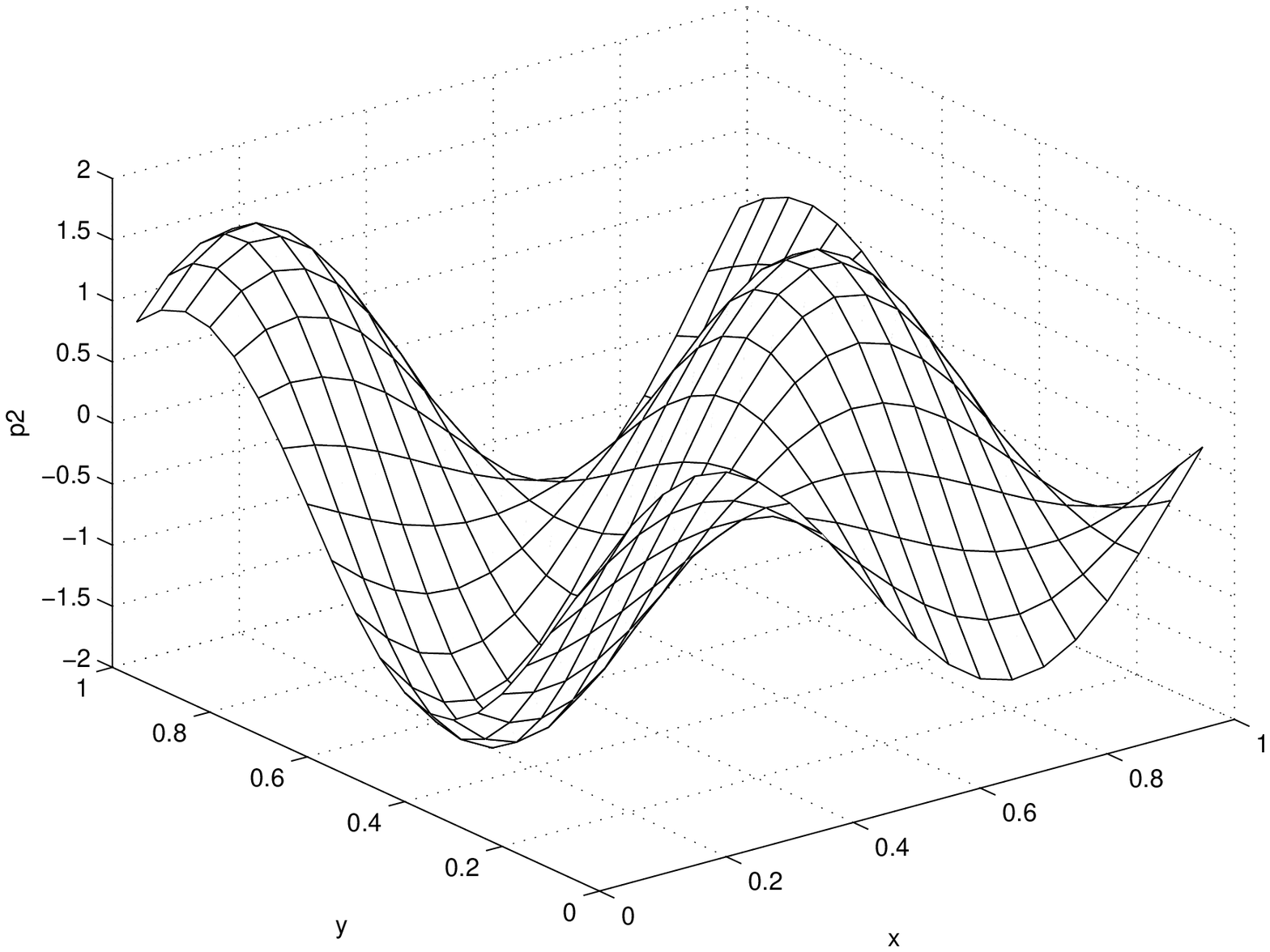}
\includegraphics[width=0.45\textwidth]{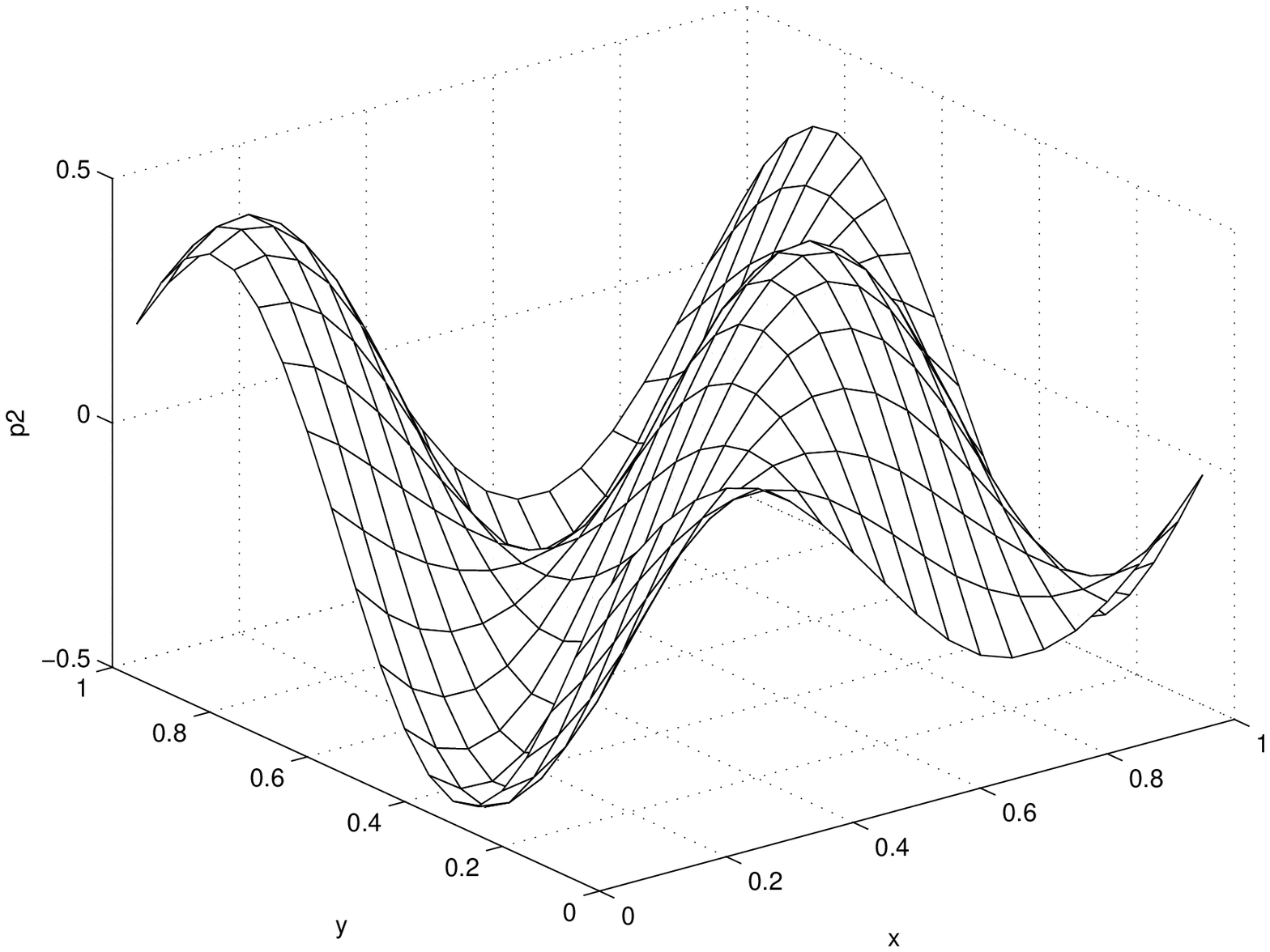}
\caption{Example 3. When $\epsilon=0.8$, the initial density and
momentum (left) and the numerical result with $\Delta x=1/20,\Delta
t=1/80$ at time $T=1$ (right) are represented. a) $\rho_\epsilon$;
b) $\mathbf{p}_{1\epsilon}$; c)
$\mathbf{p}_{2\epsilon}$.}\label{figure3}
\end{center}
\end{figure}

\begin{figure}
\begin{center}
a)\includegraphics[width=0.45\textwidth]{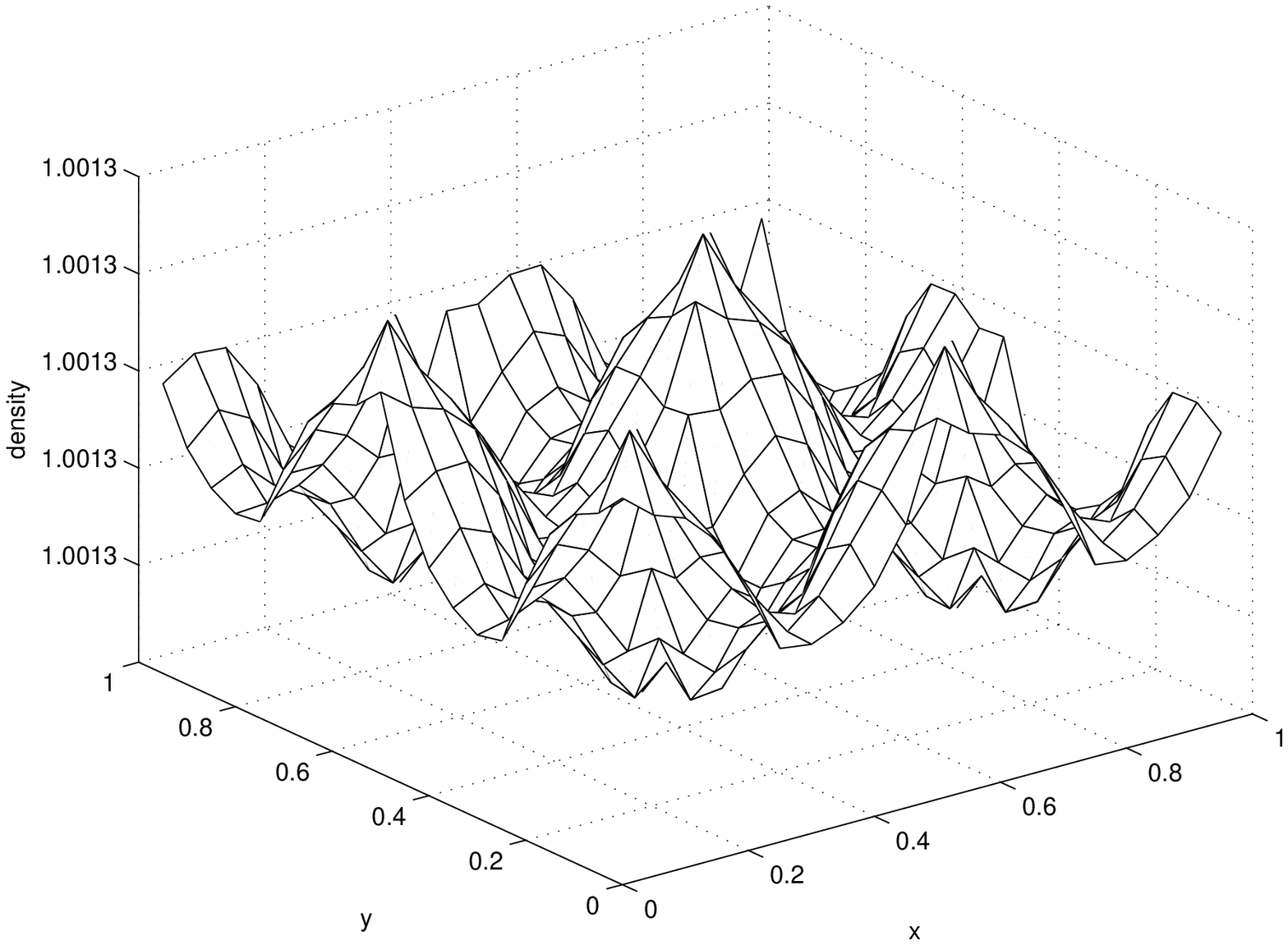}
\includegraphics[width=0.45\textwidth]{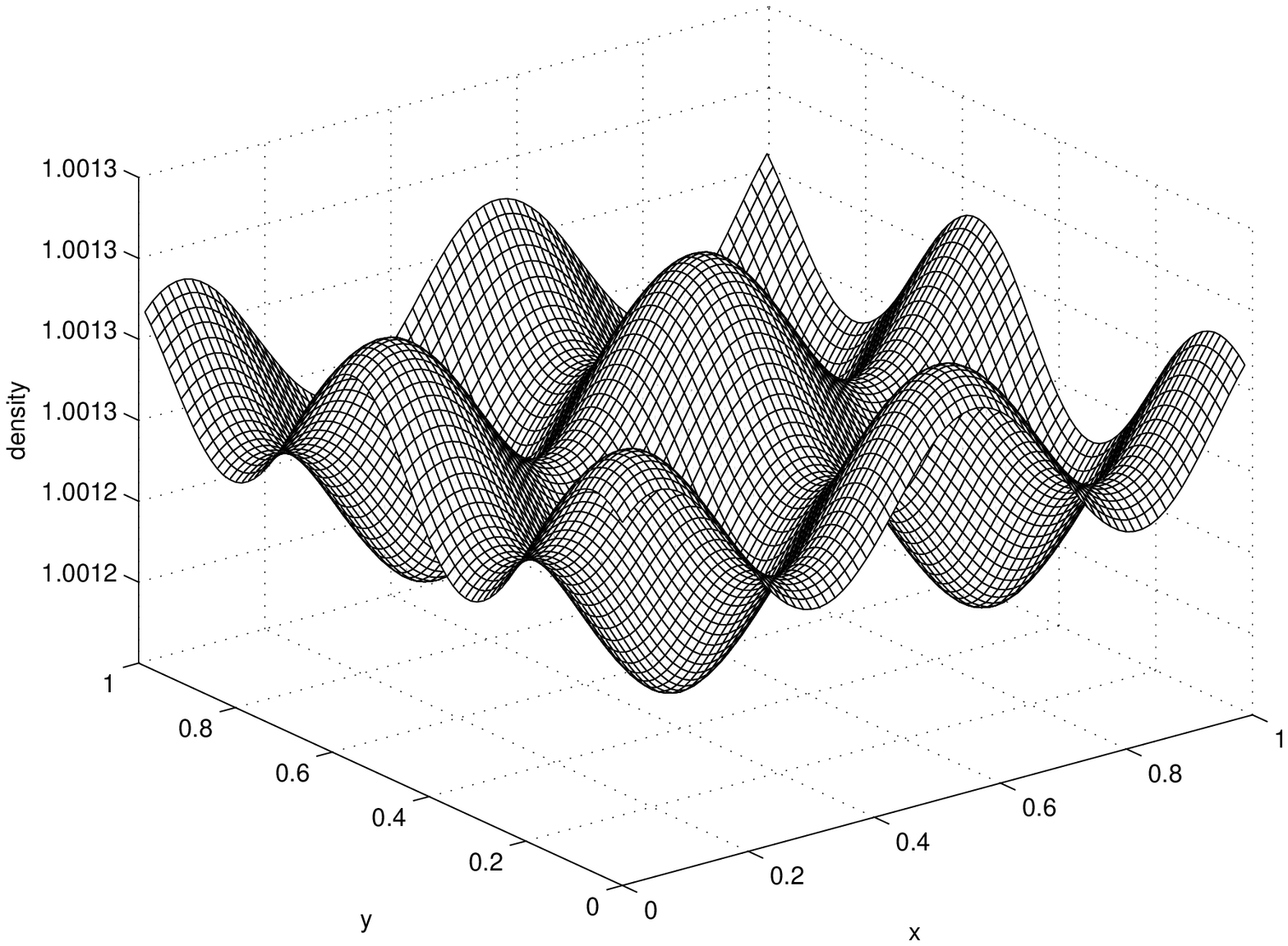}
b)\includegraphics[width=0.45\textwidth]{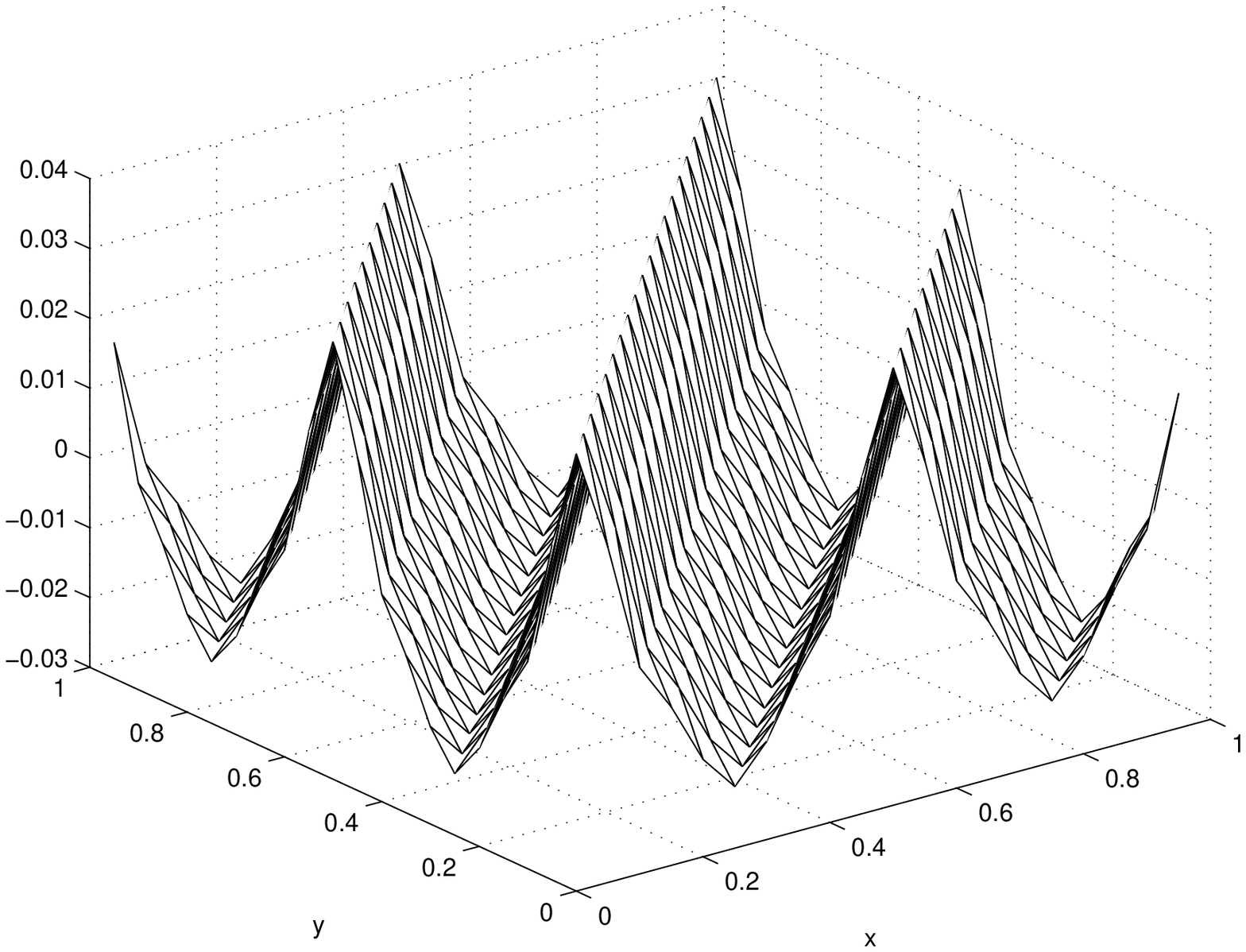}
\includegraphics[width=0.45\textwidth]{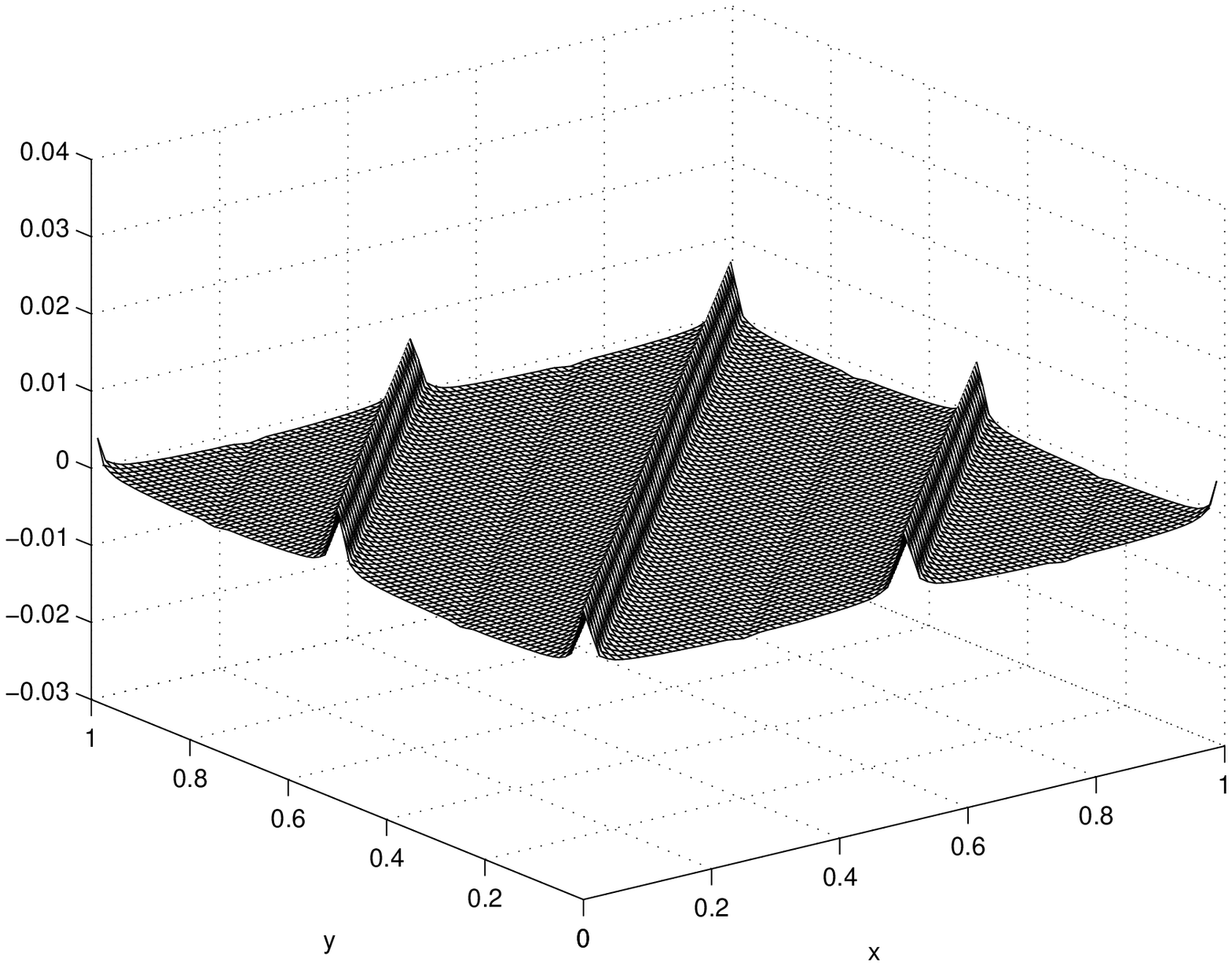}
c)\includegraphics[width=0.45\textwidth]{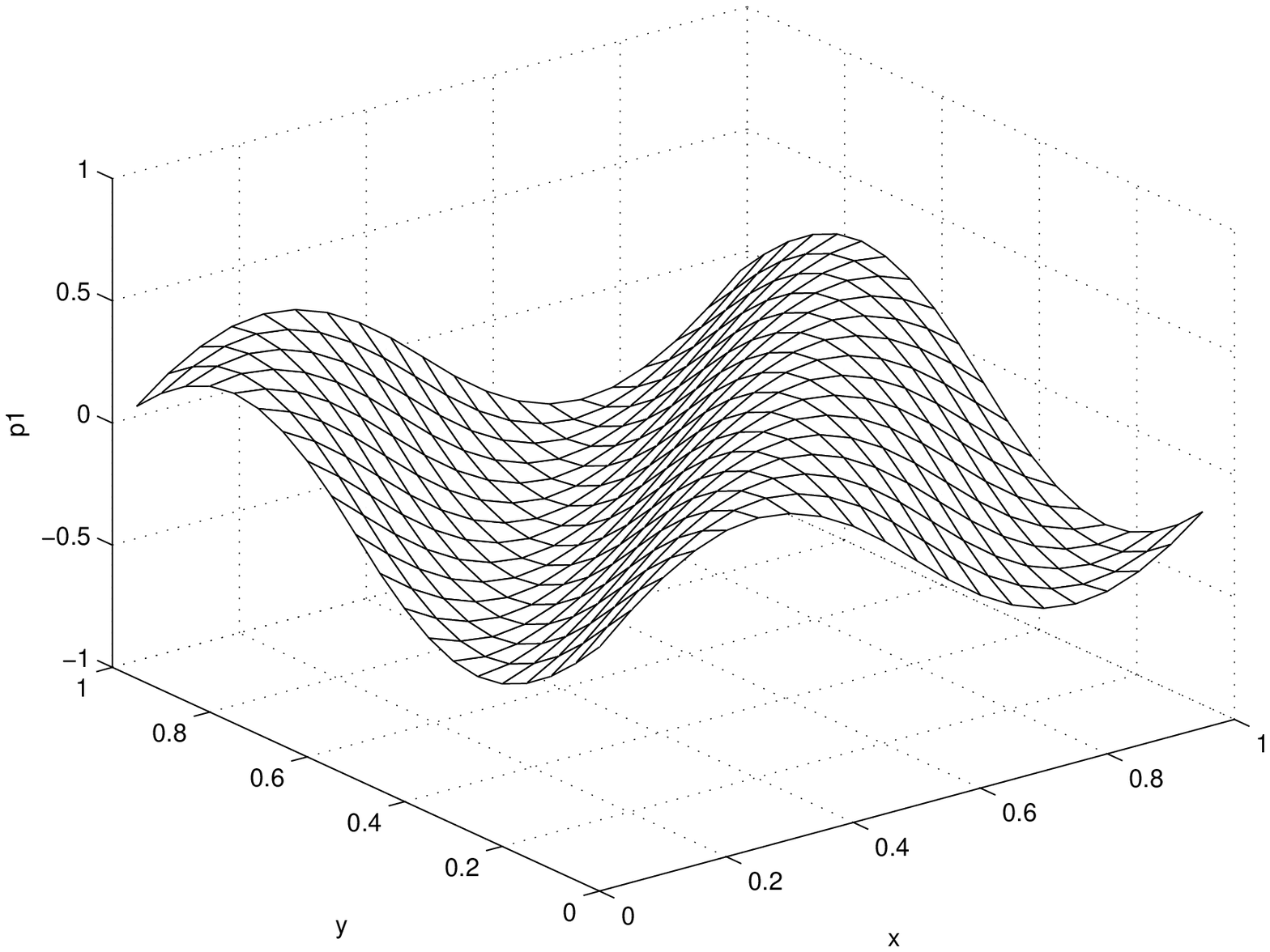}
\includegraphics[width=0.45\textwidth]{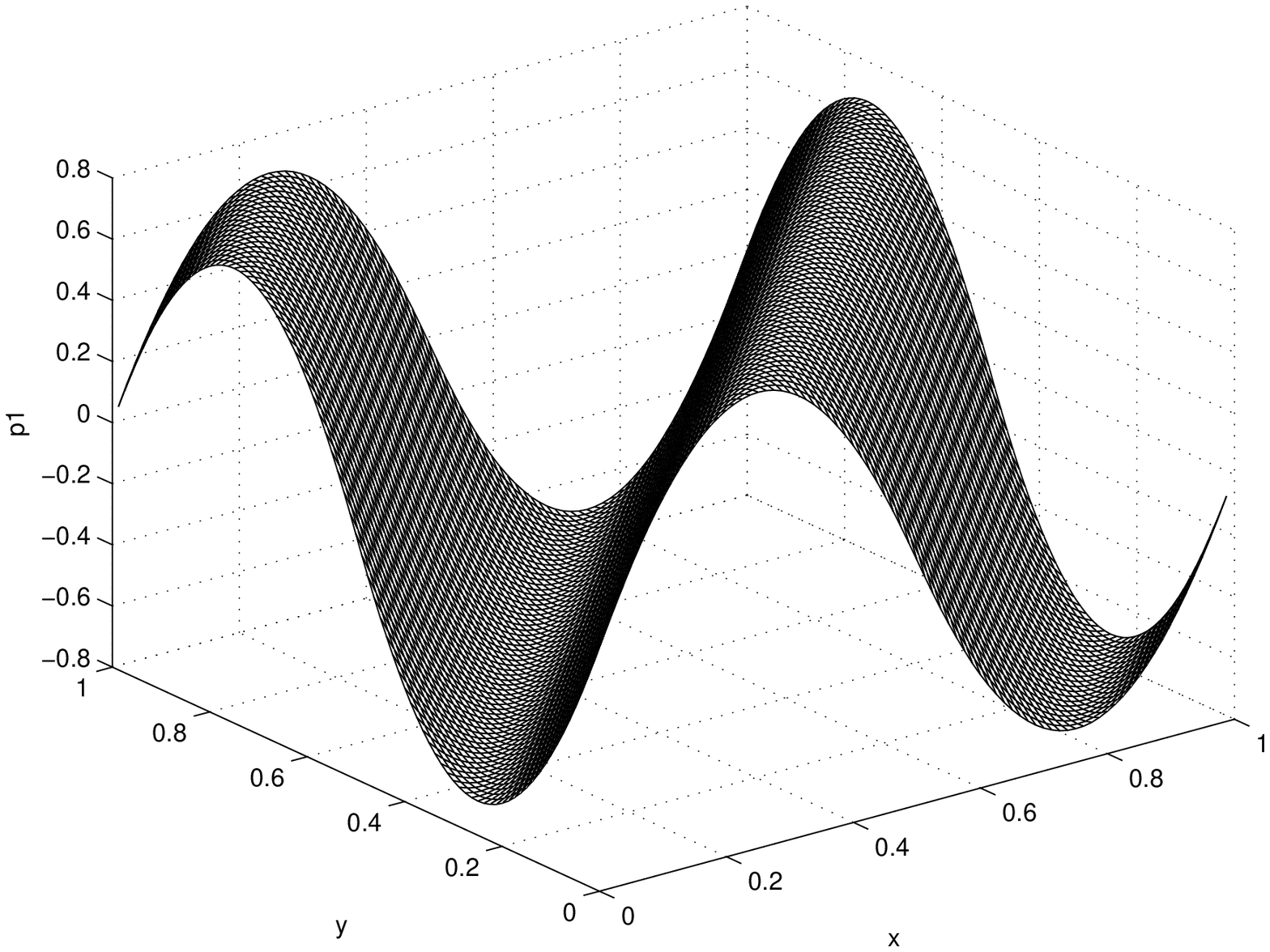}
d)\includegraphics[width=0.45\textwidth]{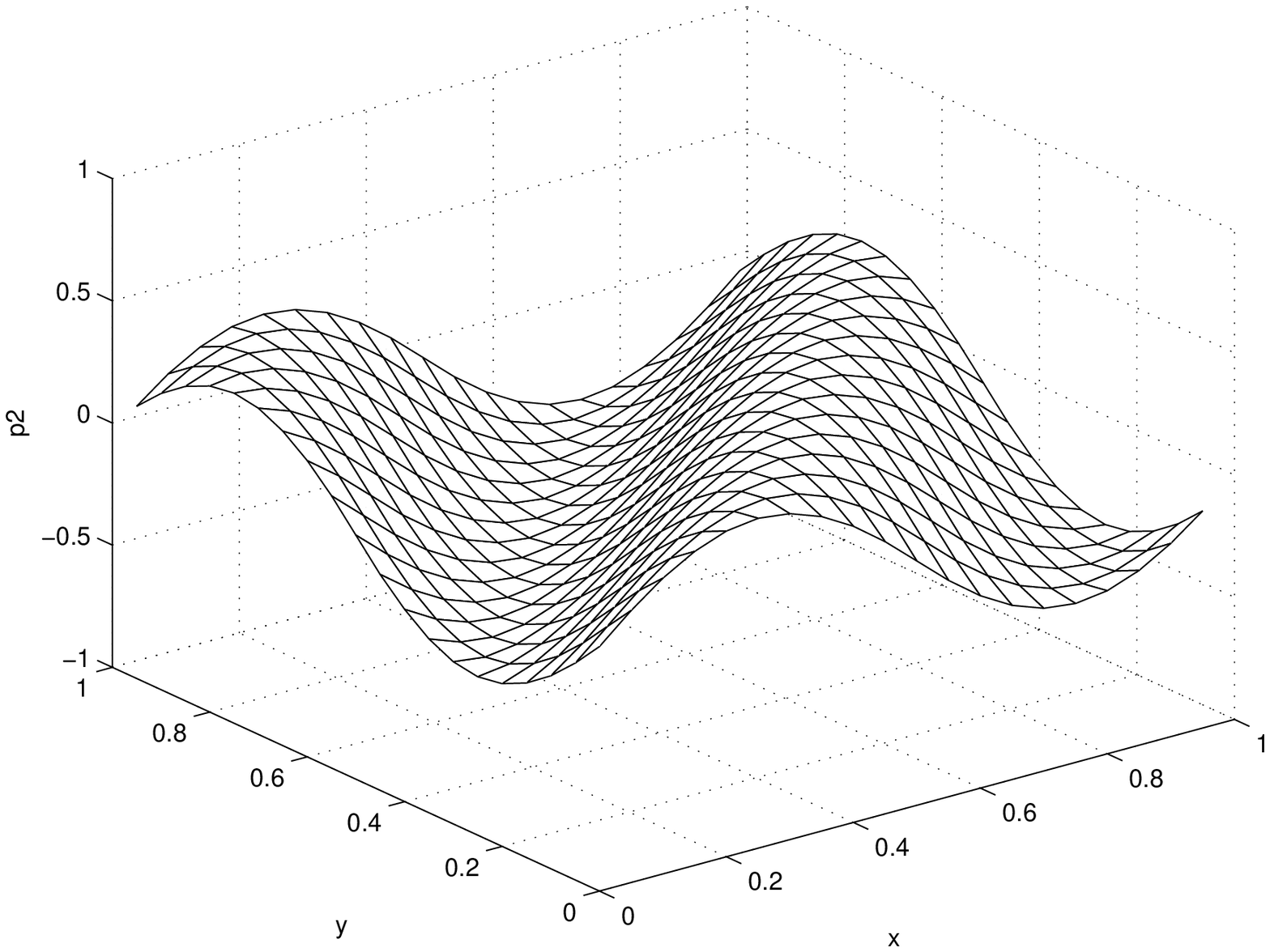}
\includegraphics[width=0.45\textwidth]{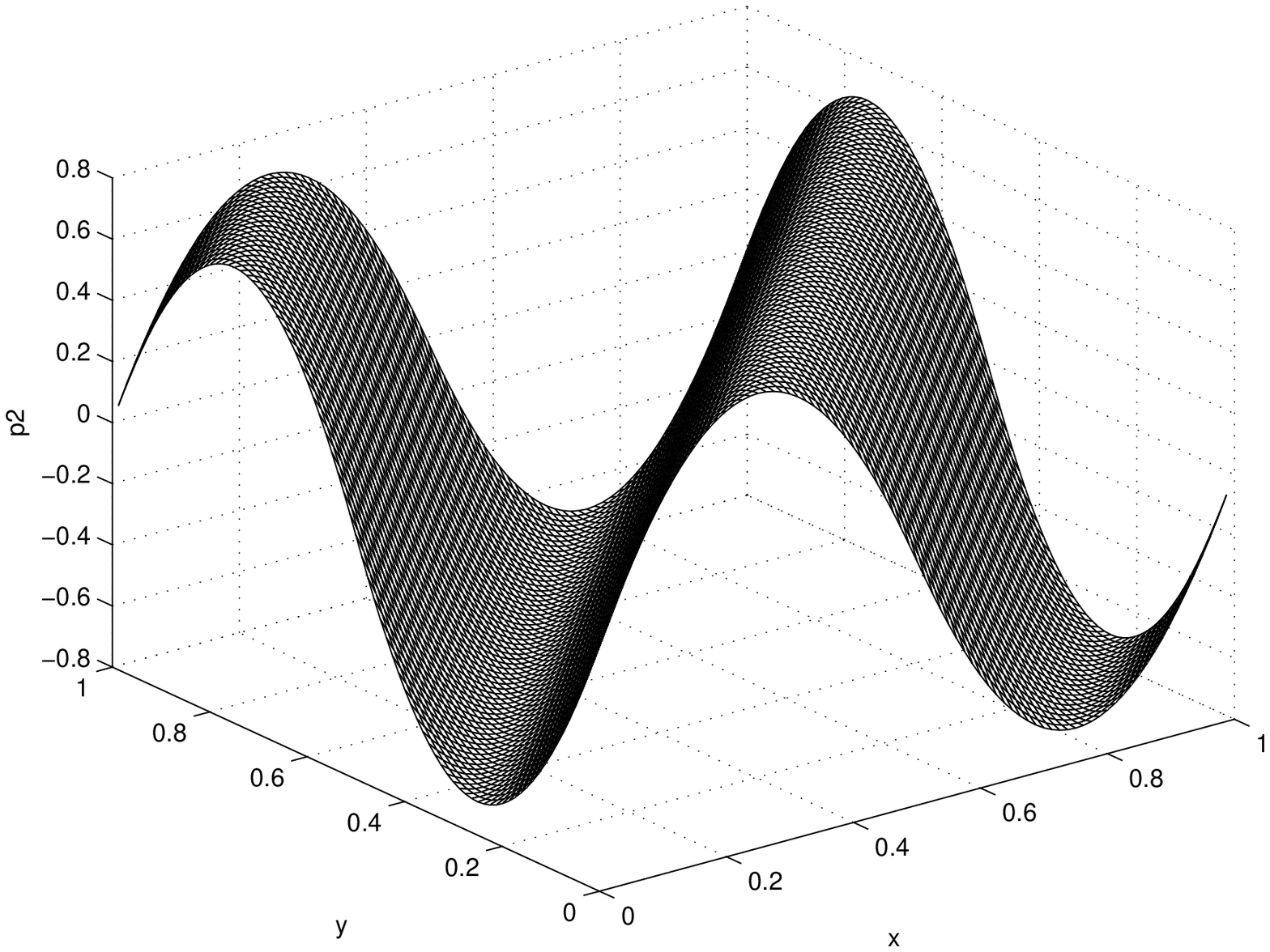}
\caption{Example 3. When $\epsilon=0.05$, the numerical result with
$\Delta x=1/20,\Delta t=1/80$ (left) and $\Delta x=1/80,\Delta
t=1/320$ (right) at time $T=1$ are represented respectively. a):
$\rho_\epsilon$; b):
$D^x\mathbf{p}_{1\epsilon}+D^y\mathbf{p}_{2\epsilon}$; c):
$\mathbf{p}_{1\epsilon}$; d):
$\mathbf{p}_{2\epsilon}$.}\label{figure3_1}
\end{center}
\end{figure}

For $\epsilon=0.1$, the numerical results of the "LD" scheme at
different times $T$ are shown in Figure \ref{figure2_2}. The initial
data approximate two acoustic pulses, one right-running and one
left-running. They collide and their superposition gives rise to a
maximum in the density. Then the pulses separate again. This
procedure is demonstrated clearly in Figure \ref{figure2_2}.
\[\]
\textbf{Example 3 }In this example, we show numerical results for
the two dimensional case. Let $P(\rho)=\rho^2$ and the computational
domain be $(x,y)\in[0,1]\times[0,1]$. Because no shock will form in
this example, we choose $\alpha=0$ and the initial condition as
follows:
\[
 \rho(x,y,0)=1+\epsilon^2
\sin^2(2\pi(x+y)),\]\[
\mathbf{q}_1(x,y,0)=\sin(2\pi(x-y))+\epsilon^2\sin(2\pi(x+y)),\]\[
\mathbf{q}_2(x,y,0)=\sin(2\pi(x-y))+\epsilon^2\cos(2\pi(x+y)).
\] The initial conditions for $\epsilon=0.8$ and numerical results
at $T=1$ calculated with $\Delta x=1/20,\Delta t=1/80$ are shown in
Figure \ref{figure3}. Numerical tests show that a similar CFL
condition is required as for the one-dimensional case. When
$\epsilon=0.05$ at time $T=1$, the numerical results with an
unresolved mesh $\Delta x=1/20,\Delta t=1/80$ and a resolved mesh
$\Delta x=1/80,\Delta t=1/320$ are displayed in Figure
\ref{figure3_1}. We can see that the results using the coarse mesh
are much 'smoother' than the one using the refined mesh. In this
example the amplitude decay due to numerical diffusion cannot be
ignored. When a coarse mesh is used, the first order method is known
to have dissipation. This is mainly due to the numerical diffusion
term, which smoothes out the solution. This phenomenon not only
happens when $\epsilon$ is small but also when $\epsilon$ is $O(1)$.
We can also see from Figure \ref{figure3_1} that when $\epsilon$ is
small, $D^x\mathbf{p}_{1\epsilon}+D^y\mathbf{p}_{2\epsilon}$ is
close to $0$.

\begin{figure}
\begin{center}
a)\includegraphics[width=0.4\textwidth]{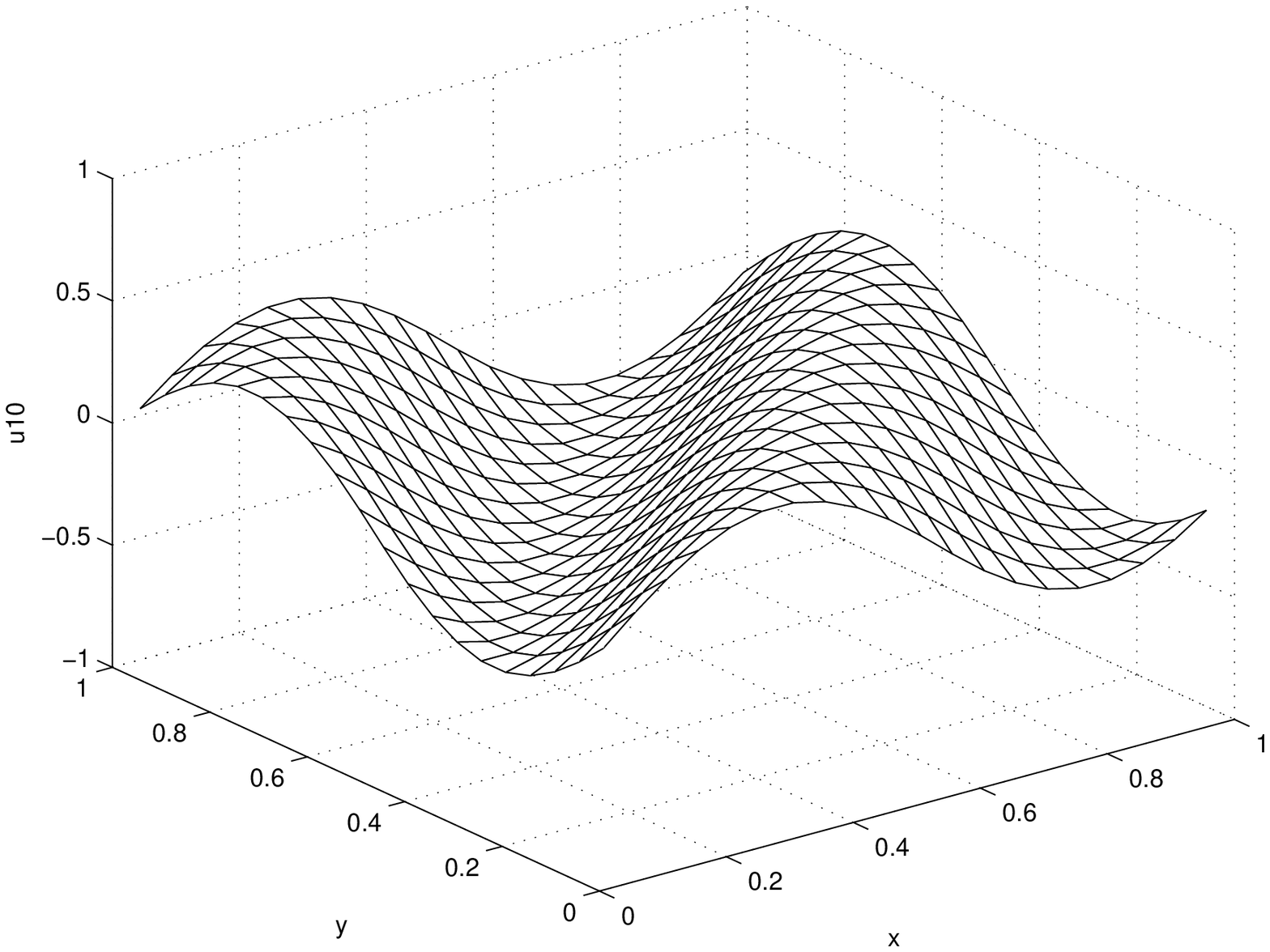}
\includegraphics[width=0.4\textwidth]{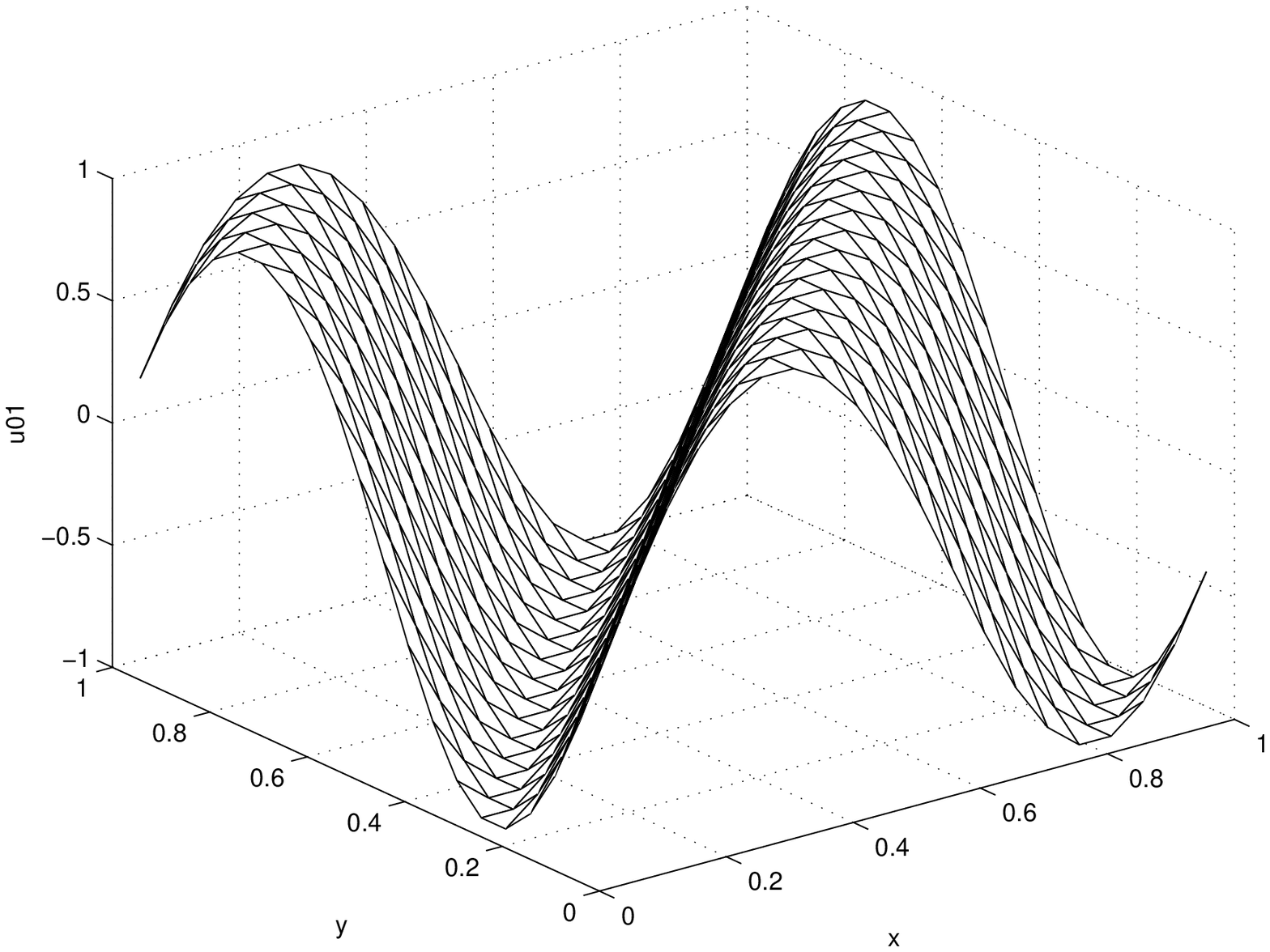}
b)\includegraphics[width=0.4\textwidth]{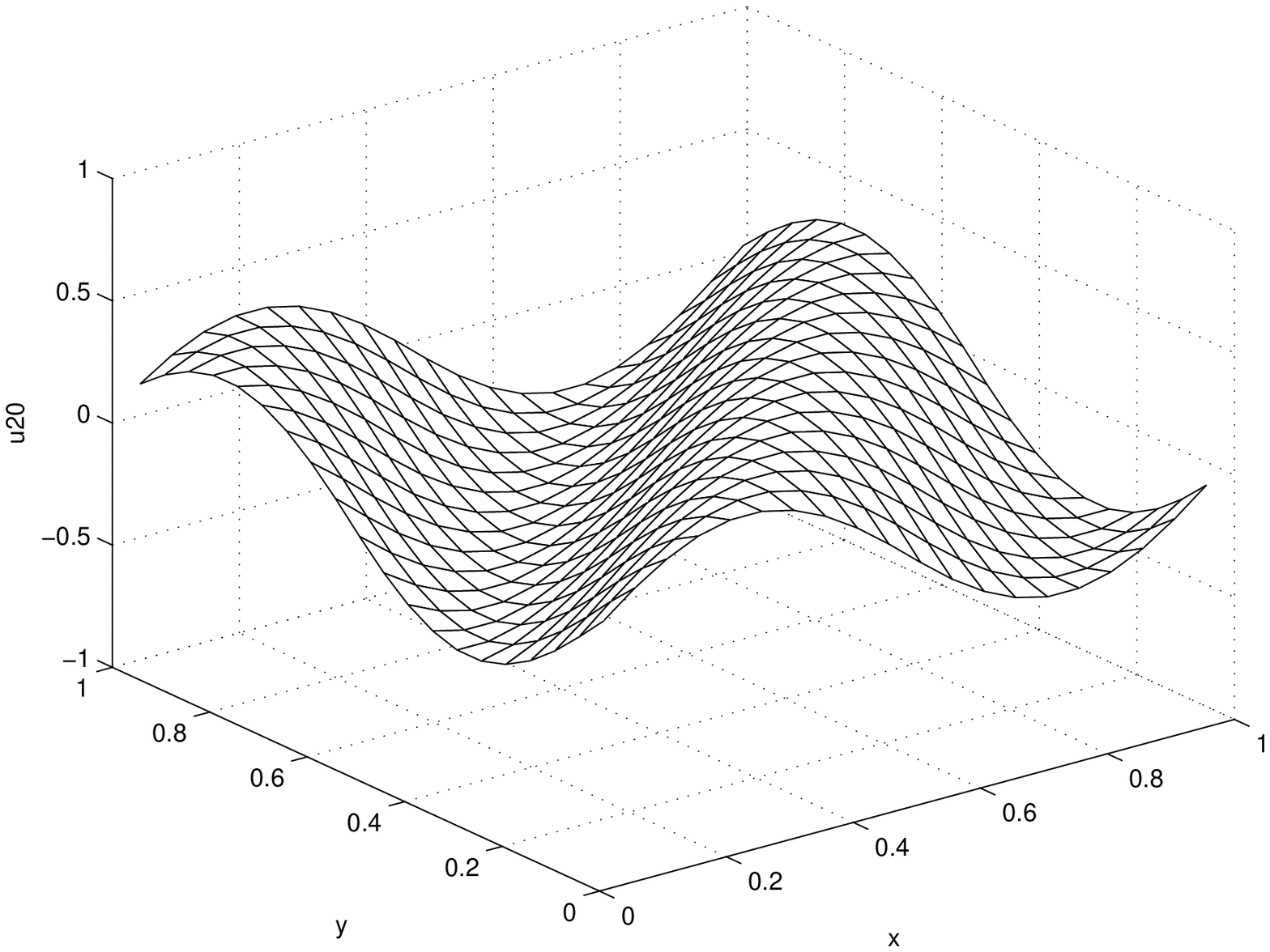}
\includegraphics[width=0.4\textwidth]{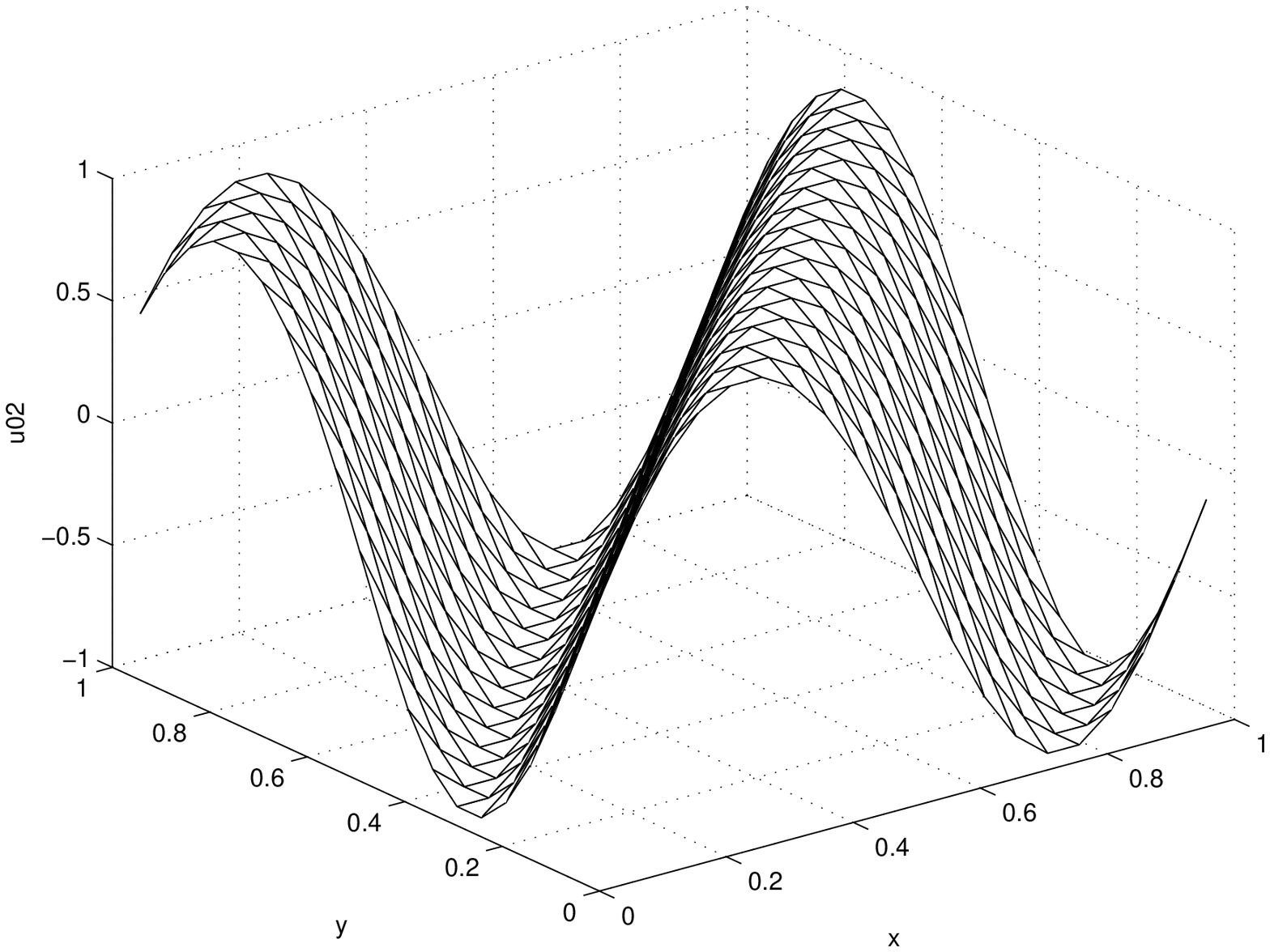}
\caption{Example 3. The numerical results of the incompressible
Euler limit using $\Delta x=1/20,\Delta t=1/80$ with (left) and
without (right) viscosity at time $T=1$ are represented. a): the
first element of the velocity $\mathbf{u}_{01}$; b): the second
element of the velocity $\mathbf{u}_{02}$.}\label{figure3_2}
\end{center}
\end{figure}

As a comparison, the numerical solutions of the incompressible limit
(\ref{lim}) with and without numerical viscosity are shown in Figure
\ref{figure3_2}. The latter is obtained by a difference method based
on a staggered grid configuration \cite{HW}. This staggered
difference method is attractive for incompressible flows, since no
artificial terms are needed to obtain stability and suppress the
oscillations. Because of the stable pressure-velocity coupling,
solutions with almost no viscosity can be obtained. The viscosity
introduced here is of the form $\frac{A}{2}\Delta x$ where $A$ is
given by (\ref{A2}). We can see that the amplitude of the wave
decays as time evolves even though the viscosity is only $O(\Delta
x)$. In the limit of $\epsilon\to 0$, (\ref{momentum2d}) generates a
discretization of the incompressible limit with $O(\Delta x)$
numerical diffusion terms. This is why the results for $\Delta
x=1/20,\Delta t=1/80$ in Figure \ref{figure3_1} are close to those
with viscosity in Figure \ref{figure3_2}. When the meshes are
refined, less diffusion is introduced and the solution becomes
closer to the solution with no viscosity. The scheme indeed catches
the incompressible Euler limit and good numerical approximations can
be obtained without resolving $\epsilon$, which confirms the AP
property that is proved in section 4. However we need to take care
of the numerical diffusion when coarse meshes are used. One possible
way to improve this is to use less diffusive shock capturing schemes
at first order or higher order schemes using the MUSCL strategy for
instance \cite{DPUV,KT, KT2}, or to use staggered grid
discretizations.

\section{Conclusion}
   We propose an all speed scheme for the Isentropic Euler equation.
   The key idea is the semi-implicit time
   discretization, in which the low Mach number
   stiff pressure term is divided into two parts, one being treated explicitly and the
   other one implicitly. Moreover, the flux of the density equation is also treated
   implicitly. The parameter which tunes the explicit-implicit decomposition of the pressure term
   allows to
   suppress the nonphysical oscillations. The numerical results show that the
   oscillations around shocks of $O(\epsilon^2)$ strength can be suppressed by
   choosing $\alpha=1$. The low Mach number
   limit of the time semi-discrete scheme becomes an elliptic
   equation for the pressure term, so that the density becomes a
   constant when $\epsilon\to 0$. In this way, the incompressible property
   is recovered in the limit $\epsilon\to 0$. Implemented with proper space
   discretizations, we can propose an AP
   scheme which can capture the incompressible limit without the need
   for $\Delta t,\Delta x$ to resolve $\epsilon$.

   In this paper we demonstrate the potential of this idea by using the first
   order Lax-Friedrich scheme with local evaluation of the wave speeds.
    Though this first order method is quite
   dissipative, we can observe that the scheme is
   stable independently of $\epsilon$ and that the CFL condition is
    $\Delta t=O(\Delta x)$ uniformly in $\epsilon$.
   It can also capture the right incompressible limit
   without resolving the mach number. Higher order space discretizations like
   the MUSCL method \cite{DPUV, KT, KT2} can be built into this
    framework. This is the subject of current work.

   This paper provides a framework for the design of a class of all
   speed schemes. Compared with the ICE method \cite{HW,HA} and some recent
   work by Jin, Liu and Hauck \cite{Jeff}, the idea is simpler and more natural.
   This framework can also be easily extended to the full Euler equation
   and flows with variable densities and temperatures. These extensions and
   applications \cite{MDR} will be the subject of future work.

 \section*{Acknowledgments}
This work was supported by the french 'Commissariat \`a l'Energie
Atomique (CEA)' (Centre de Saclay) in the frame of the contract
'ASTRE', \# SAV 34 160.

%%%% Bibliography  %%%%%%%%%%


\begin{thebibliography}{99}
\bibitem{Bijl}H. Bijl, P. Wesseling, A unified method for computing
incompressible and compressible flows in boundary-fitted
coordinates, J. Comput. Phys., 141: 153-173, (1998)

\bibitem{Bonner}M. P. Bonner, Compressible subsonic flow on a
staggered grid, Master thesis, The University of British Columbia.
(2007)

\bibitem{Constantin}P. Constantin, On the Euler equations of
incompressible fluids, Bulletin of the American Mathematical
Society, Vol. 44, No. 4, 603-621, (2007)

\bibitem{DEV07}P. Crispel,
P. Degond, M-H. Vignal,  An asymptotic preserving scheme for the
two-fluid Euler-Poisson model in the quasineutral limit, J. Comput.
Phys., 223, 208-234, (2007)



\bibitem{DDSV09}P. Degond, F. Deluzet,
A. Sangam, M-H. Vignal,  An asymptotic preserving scheme for the
Euler equations in a strong magnetic field,  J. Comput. Phys, Vol.
228, No.10, 3540-3558, (2009)



\bibitem{DJL}P. Degond, S. Jin and J-G. Liu, Mach-number uniform
asymptotic-preserving gauge schemes for compressible flows, Bulletin
of the Institute of Mathematics, Academia Sinica, New Series, 2, No.
4, 851-892, (2007)
\bibitem{DLV08}P. Degond, J-G. Liu,
M-H. Vignal, Analysis of an asymptotic preserving scheme for the
Euler-Poison system in the quasineutral limit, SIAM J. Numer. Anal.,
46, 1298-1322, (2008)
\bibitem{DPUV}P. Degond, P. F. Peyrard, G. Russo and P. Villedieu,
Polynomial upwind schemes for hyperbolic systems, Partial
Differential Equations, Series 1: 479-483, (1999)

\bibitem{GJL}F. Golse, S. Jin and C.D. Levermore,
The Convergence of Numerical Transfer Schemes in Diffusive Regimes
I: The Discrete-Ordinate Method, SIAM J. Numer. Anal., 36,
1333-1369, (1999)

\bibitem{HH}J. R. Haack and C. D. Hauck, Oscillatory Behavior of Asymptotic-Preserving
Splitting Methods for a Linear Model of Diffusive Relaxation, Los
Alamos Report LA-UR 08-0571, to appear in Kinetic and Related
Models, (2008)

\bibitem{Jeff}J. Haack, S. Jin and J. G. Liu, All speed asymptotic
preserving schemes for compressible flows. in preparation.

\bibitem{HA}F. H. Harlow, and A. Amsden, A numerical fluid dynamics
calculation method for all flow speeds, J. Comput. Phys, 8, 197-213,
(1971)

\bibitem{HW} F. H. Harlow and J. E. Welch, Numerical calculation of
time-dependent viscous incompressible flow of fluid with free
surface, Phys. Fluid, 8, No.12, 2182-2189, (1965)

\bibitem{IGW}R. I. Issa, A. D. Gosman, A. P. Watkins, The
computation of compressible and incompressible flow of fluid with a
free surface. Phys. Fluids, 8, 2182-2189, (1965)

\bibitem{KM1}S. Klainerman, A. Majda, Singular limits of quasilinear
hyperbolic systems with large parameters and the incompressible
limit of compressible fluids, Communication on Pure and Applied
Mathematics, 34: 481-524, (1981)

\bibitem{KM2}S. Klainerman, A. Majda, Compressible and incompressible
fluids, Communication on Pure and Applied Mathematics, 35: 629-653,
(1982)

\bibitem{Klein} R. Klein, Semi-implicit extension of a Godunov-type
scheme based on low Mach number asymptotics I: one-dimensional flow,
J. Comput. Phys. 121: 213-237, (1995)

\bibitem{KBSM}R. Klein, N. Botta, T. Schneider, C. D. Munz, S. Roller,
A. Meister, L. Hoffmann, T. Sonar, Asymptotic adaptive methods for
multi-scale problems in fluid mechanics, J. Eng. Math., 83:261-343,
(2001)

\bibitem{KT}A. Kurganov and E. Tadmor,
New high-resolution central schemes for nonlinear conservation laws
and convection-diffusion equations, J. Comput. Phys., 160:
214-282,(2000)

\bibitem{KT2}A. Kurganov and E. Tadmor
Solution of two-dimensional Riemann problems for gas dynamics
without Riemann problem solvers, Numerical Methods for Partial
Differential Equations, 18:548-608,(2002)

\bibitem{Leveque}R. J. Leveque, Numerical methods for conservation laws,
Lectures in Mathematics ETH Z¨¹rich, (1992)

\bibitem{MDR} C. D. Munz, M. Dumbser and S. Roller, Linearized acoustic
perturbation equations for low Mach number flow with variable
density and temperature, J. Comput. Phys. 224: 352-364, (2007)

\bibitem{MRKG}C. D. Munz, S. Roller, R. Klein and K. J. Geratz. The
extension of incompressible folw solvers to the weakly compressible
regime, Comp. Fluid, 32: 173-196, (2002)

\bibitem{PM} J. H. Park and C. D. Munz, Multiple pressure variables methods for fluid flow
at all Mach numbers ,Int. J. Numer. Meth. Fluid, 49: 905-931, (2005)

\bibitem{P}S. V. Patankar, Numerical heat transfer and fluid flow,
New York: McGraw-Hill, (1980)

\bibitem{RB}F. Rieper and G. Bader, The influence of cell geometry
on the accuracy of upwind schemes in the low mach number regime, J.
Comput. Phys., 228: 2918-2933, (2009)

\bibitem{HWconserve}D. R. van der Heul, C. Vuik and P. Wesseling, A
conservative pressure-correction method for flow at all speeds,
Comptuters and Fluids, 32, 1113-1132, (2003)

\end{thebibliography}
\end{document}